\documentclass[aps,pra,twocolumn]{revtex4-1}
\usepackage{amsmath}
\usepackage{amssymb}
\usepackage{graphicx}
\usepackage[utf8]{inputenc}
\usepackage[T1]{fontenc}
\usepackage[export]{adjustbox}
\usepackage{color}
\usepackage{bm}
\usepackage[colorlinks]{hyperref}
\usepackage{tikz}
\usepackage{booktabs}

\usetikzlibrary{calc}

\hypersetup{colorlinks = false}

\newcommand{\oumk}{\affiliation{School  of Physical Sciences, The Open University, Walton Hall, MK7 6AA Milton Keynes, UK}}

\newcommand{\cuni}{\affiliation{Institute of Theoretical Physics, Faculty of Mathematics and Physics, Charles University, V~Hole\v{s}ovi\v{c}k\'{a}ch~2, Prague~8, 180~00, Czech Republic}}

\begin{document}

\title{Analysis of RABITT time delays using the stationary multi-photon molecular R-matrix approach}

\author{J.~Benda}
\email[]{jakub.benda@matfyz.cuni.cz}
\cuni

\author{J.~D.~Gorfinkiel}
\oumk

\author{Z.~Ma\v{s}\'{\i}n}
\cuni

\date{\today}
\begin{abstract}
We employ the recently developed multi-photon R-matrix method for molecular above-threshold photoionization to obtain second-order ionization amplitudes that govern the interference in RABITT experiments. This allows us to extract RABITT time delays that are in better agreement with non-perturbative time-dependent simulations of this process than the typically used combination of first-order (Wigner) delays and asymptotic corrections. We calculate molecular-frame as well as orientation-averaged RABITT delays for H$_2$, N$_2$, CO$_2$, H$_2$O and N$_2$O and analyze the origin of various structures in the time delays including the effects of partial wave interference, shape resonances and orientation-averaging. Time-delays for B and C states of CO$_2^{+}$ are strongly affected by absorption of the second (IR) photon in the ion. This effect corresponds to an additional contribution, \(\tau_{\text{coupl}}\), to the asymptotic approximation for the RABITT delays \(\tau \approx \tau_{\text{mol}} +\tau_{cc} + \tau_{\text{coupl}}\). Applicability of the asymptotic theory depends on the target and IR photon energy but typically starts at approximately 30 -- 35~eV of XUV photon energy.
\end{abstract}

\maketitle

\section{Introduction}

%MAKE SURE TO INCLUDE: 
%HASE (S. Eckart, Phys. Rev. Res. 2, 033248 (2020))
%S. Eckart, Nat. Comm 12, (2021)
%Biswas et al, NATure PHySIcS, VOL 16, JULy 2020, 778–783

% Electron correlation effects in attosecond photoionization of CO2
%Andrei Kamalov, Anna L. Wang, Philip H. Bucksbaum, Daniel J. Haxton, and James P. Cryan Phys. Rev. A 102, 023118 (2020) https://journals.aps.org/pra/abstract/10.1103/PhysRevA.102.023118
%<-THEY HAVE PROBLEMS WITH LOW-LYING AUTOIONIZING RESONANCES THAT THEY CANNOT REPRESENT WELL.

The experimental method of reconstruction of attosecond beating by interference of two-photon transitions (RABITT, \cite{Paul,RABITT}) for measurement of the intrinsic attosecond photoionization time delay celebrates two decades since its invention. Since its conception, it has been applied to a variety of systems, including noble gasses~\cite{ArHeNe, Compar, ArgonCooper}, isolated molecules H$_2$~\cite{cattaneo2018}, H$_2$O and N$_2$O~\cite{Huppert}, N$_2$~\cite{LoriotN2,Nandi}, CO~\cite{vos_2018} or CO$_2$~\cite{KamalovCO2}, and has been the subject of many theoretical works aiming at more or less accurate numerical simulation of the process~\cite{Dahlstrom,Dahlstrom_JPB,SerovKheifets,BaykushevaWorner}. Related methods have been used for time-resolved spectroscopy in liquid water~\cite{LiquidDelays}, on surfaces~\cite{Gallmann} and in other contexts~\cite{HASE,Eckart,Biswas}. Angularly dependent RABITT spectra have been studied too, see e.g.~\cite{joseph2020,fuchs2020}.

\begin{figure}[htbp]
    \centering
    \includegraphics[width=0.98\columnwidth]{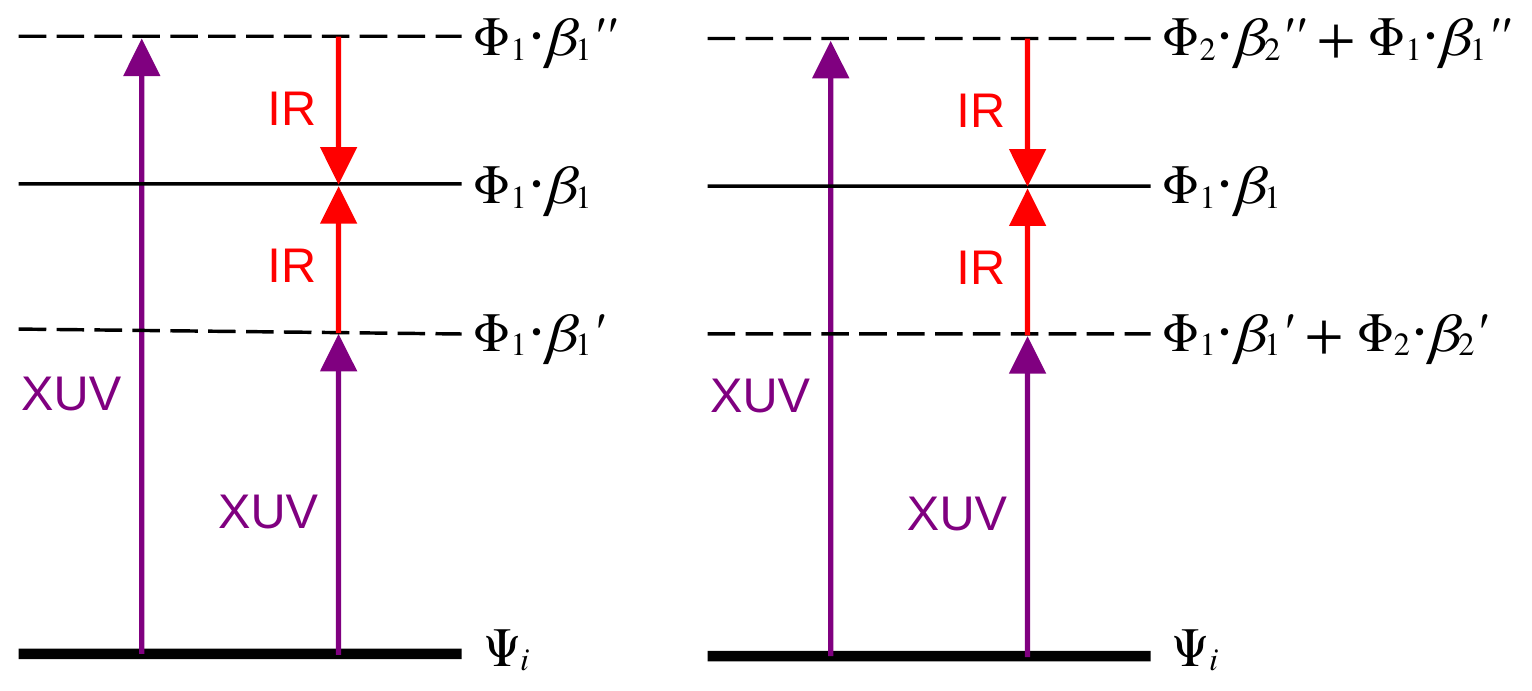}
    \caption{Single- (left) and multi-channel RABITT (right). \(\Psi_i\) represents the initial state of the molecule, \(\Phi_j\) is the \(j\)-th residual ion state, \(\beta_j\) the associated continuum wave function of the photoelectron.}
    \label{fig:multichannel}
\end{figure}

RABITT is a two-photon process where two different absorption pathways interfere, though more complex setups have been considered as well~\cite{Loriot2017,Multisideband}. An XUV photon is absorbed, with energy \(\Omega_<\) or \(\Omega_>\), releasing the photoelectron, which subsequently either absorbs another IR photon, or---stimulated by the ambient IR field---emits one, with energy \(\omega\). The two ionization pathways leading to the same photoelectron energy, \(\Omega_< + \omega = \Omega_> - \omega\), interfere. Their phase difference depends on many factors, including the phase of the ionizing field, the intrinsic photoionization time delay and the relative temporal delay \(\Delta \tau\) between the XUV and IR field~\cite{Dahlstrom_JPB,Dahlstrom}. When observing the photoionization yield in some direction, its magnitude \(I\) periodically fluctuates with varying \(\Delta \tau\),
\begin{equation}
    \Delta I(\Delta\tau) \sim \cos \left(2\omega (\Delta\tau + \tau_{sb})\right)
    \label{eq:sbcos}
\end{equation}
giving rise to the ``beating'' RABITT sideband~\cite{Dahlstrom}. In experiments the delay $\Delta\tau$ is varied and the signal fitted to Eq.~\eqref{eq:sbcos} thus determining the sideband delay $\tau_{sb}$. The sideband delay \(\tau_{sb}\), has been commonly expressed as a sum of the one-photon Wigner ionization delay \(\tau_W\) and the continuum-continuum delay \(\tau_{cc}\) arising from the additional interaction with the second (absorbed or emitted) IR photon~\cite{IvanovSmirnova}, see Fig.~\ref{fig:multichannel}, and possibly other additional phases~\cite{Pazourek}. However, the separation of the time delay into \(\tau_W\) and \(\tau_{cc}\) is only an approximation.

% One arm of the RABITT interferometer corresponding to the XUV+IR absorption in the single- (left) and multi-channel RABITT (right). The latter is included in this work.

Direct calculation of the RABITT delay without the aforementioned approximation is complicated since it is a two-photon above-threshold ionization process (i.e. involving  absorption or emission of the second photon by the photoelectron \textit{in the continuum}) for which specialized treatments are necessary. For this reason time-dependent theoretical studies~\cite{Haessler,SerovKheifets,ChaconCO,KheifetsBray} have prevailed over time-independent ones~\cite{Hockett,BaykushevaWorner}. Moreover, the latter mostly consisted of approximate methods that extended various asymptotic forms of the photoelectron wave functions all the way to the origin, as in~\cite{Dahlstrom}, where they are necessarily inaccurate. This led to various approximate forms for the continuum-continuum correction \(\tau_{cc}\)~\cite{Dahlstrom,IvanovSmirnova}, all of which share the same feature that they are incapable of providing correct results in the low-energy limit. At the same time, access to a theoretical method both computationally efficient and accurate is invaluable for analysis of the observed experimental results.

While there have been a few successful attempts to obtain two-photon above-threshold ionization amplitudes for atoms using time-independent approaches~\cite{Feng,Shakeshaft,vinbladh2019}, these have been rarely applied to the RABITT process so far~\cite{ArgonCooper}. In molecules there hasn't been any such option available until recently. To overcome this limitation, we developed an efficient R-matrix-based time-independent method for the calculation of multi-photon above-threshold ionization amplitudes of many-electron atoms and molecules~\cite{multiphoton}, which can be directly applied to RABITT and similar multi-photon processes.

In this article we apply the time-independent multi-photon R-matrix method to calculation of RABITT sideband delays in molecular hydrogen, nitrogen, carbon dioxide, water and nitrous oxide. To obtain reference results for a specific combination of an IR field and a time-dependent XUV attosecond pulse train we also employ the time-dependent ``R-matrix with time dependence'' (RMT, \cite{RMT,MRMT}). We compare the accurate two-photon results to the one-photon Wigner delays alone as well as augmented with various forms of the continuum-continuum correction \(\tau_{cc}\) reported in the literature. Where available, we contrast the calculations with experimental data.

%Finally, we investigate the origin and properties of several conspicuous structures in the time delays at specific photon energies, showing how interference of partial waves can dramatically affect the RABITT measurement without actually revealing much about the photoionization process itself.

In Section~\ref{sect:theory} we briefly summarize the notation used throughout the paper, define the observable RABITT delays and outline the theoretical method used for calculation of the multi-photon amplitudes. In Section~\ref{sect:results} we present the calculated delays for all studied molecules and discuss the effects of electron correlation revealed by the use of various molecular models. In the last two sections we analyze in detail two special aspects of the calculated time delays: In Section~\ref{sect:chancoupl} we show that the channel coupling mediated by the probing IR field significantly affects the calculated (and measurable) sideband delays in ionization of CO$_2$ into higher excited states (B~$^2\Sigma_u^+$ and C~$^2\Sigma_g^+$) of CO$_2^+$. This is due to the energy separation of these two states being close to resonance with the IR field. In the asymptotic treatment of molecular RABITT time delays, the resonant transition leads to an additional delay, \(\tau_{\text{coupl}}\), for which we provide explicit formula. The separate treatment of the channel coupling contribution enables us to isolate time delay structures that are caused by more complex electron correlation effects. In Section~\ref{sect:wpacket} the origin of a structure in time delays for oriented H$_2$ molecule and parallel emission direction is tracked down to the interference of partial waves. A simple time-dependent wave-packet model is introduced for ease of interpretation, based on the time-independent amplitudes.

\section{Theory}
\label{sect:theory}

We use Hartree atomic units throughout the text; $a_0$ denotes the Bohr radius.
The one-photon ionization time delay for a specific orientation of the molecule, fixed photoelectron emission direction and known field polarization direction \(\bm{\epsilon}\) is defined using the asymptotic continuum wave function \(\psi\) of the photoelectron as~\cite{Hockett}
\begin{equation}
    \tau_W = -\frac{\mathrm{d}}{\mathrm{d}E} \arg \psi
           = +\frac{\mathrm{d}}{\mathrm{d}E} \arg d^{(1)} \,.
    \label{eq:tau1}
\end{equation}
Here \(d^{(1)} = \bm{\epsilon} \cdot \bm{d}^{(1)} = \langle \Psi_{f\bm{k}}^{(-)} | D(\bm{\epsilon}) | \Psi_i \rangle\) is the one-photon ionization amplitude into the final residual ion state \(\Phi_f\), \(E\) is the photoelectron kinetic energy and
\begin{equation}
    D(\bm{\epsilon}) = \bm{\epsilon} \cdot \sum_{i = 1}^{N} \bm{r}_i
\end{equation}
is the projection of the electronic dipole operator along the polarization direction, expressed as a sum over coordinates of all \(N\) electrons. The stationary photoionization state \(\Psi^{(-)}\) is asymptotically resolved into channels of the final residual ion states. Due to the phase freedom, we choose that the initial state \(\Psi_i\) as well as all residual ion states are real, which justifies the second equality in Eq.~\eqref{eq:tau1}.

In the RABITT experiment, the total ionization signal in a given direction depends on two interfering 2-photon amplitudes \(d_+^{(2)}\) and \(d_-^{(2)}\) corresponding to pathways \(\Omega_< + \omega\) and \(\Omega_> - \omega\), respectively:
\begin{align}
    I &\sim |d_+^{(2)} + d_-^{(2)}|^2 \nonumber \\
    &= |d_+^{(2)}|^2 + |d_-^{(2)}|^2 + 2 |d_+^{(2)}| |d_-^{(2)}| \cos \arg d_+^{(2)*} d_-^{(2)} \,.
    \label{eq:signal}
\end{align}
The magnitude of the ionization signal depends on the relative phase of the two-photon amplitudes due to the last, interference, term. Assuming zero time delay \(\Delta \tau = 0\) between the XUV and IR pulses, the RABITT sideband time delay can be obtained by matching Eq.~\eqref{eq:signal} to Eq.~\eqref{eq:sbcos} as
\begin{equation}
    \tau_{sb} = \frac{1}{2\omega} \arg d_+^{(2)*} d_-^{(2)} \,.
    \label{eq:tau2}
\end{equation}
The required two-photon amplitudes are given by the leading-order perturbation theory expressions~\cite{cohen-tannoudji1998}
\begin{align}
    d_+^{(2)} &= \langle \Psi_{f\bm{k}}^{(-)} | D(\bm{\epsilon}) \frac{1}{E_i + \Omega_< - H + \mathrm{i}0} D(\bm{\epsilon}) | \Psi_i \rangle \,, \label{eq:melp} \\
    d_-^{(2)} &= \langle \Psi_{f\bm{k}}^{(-)} | D(\bm{\epsilon}) \frac{1}{E_i + \Omega_> - H + \mathrm{i}0} D(\bm{\epsilon}) | \Psi_i \rangle \,. \label{eq:melm}
\end{align}
Here \(E_i\) is the energy of the initial bound neutral state \(\Psi_i\), \(H\) is the field-free Hamiltonian and the many-electron final stationary photoionization state \(\Psi_{f\bm{k}}^{(-)}\) with the total energy \(E_f = E_i + \Omega_< + \omega = E_i + \Omega_> - \omega\) is obtained using the standard R-matrix photoionization method~\cite{Harvey}.

\begin{figure}[htbp]
    \centering
    \includegraphics[width=\columnwidth]{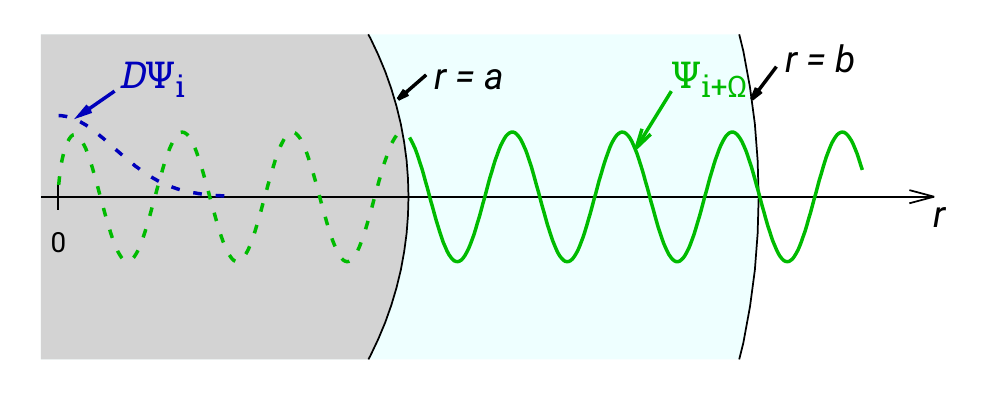}
    \caption{Physical space division into the inner region (left, dark shaded), and the numerically and analytically integrated sections of the outer region (central, light shaded and right, in white, respectively). The symbols \(D\Psi_i\) and \(\Psi_{i+\Omega}\) illustrate the right-hand side and the solution of Eq.~\eqref{eq:schrinterm}.}
    \label{fig:regions}
\end{figure}

In the R-matrix approach, space is split into two parts: (a) The spherical surroundings of the target up to some radius \(r = a\) (the ``R-matrix radius''), where multi-electron wave functions and quantum-chemical methods are used, and (b) the outer region, where only the uncoupled one-electron ionization channels are considered. The two-photon ionization amplitudes in Eqs.~\eqref{eq:melp} and~\eqref{eq:melm} are evaluated in two steps: First, the intermediate state \(|\Psi_{i+\Omega_\lessgtr}^{(+)}\rangle\) is calculated from the equation
\begin{equation}
    (E_i + \Omega_\lessgtr - H) | \Psi_{i+\Omega_\lessgtr}^{(+)} \rangle = D(\bm{\epsilon}) | \Psi_i \rangle
    \label{eq:schrinterm}
\end{equation}
with outgoing wave boundary condition imposed at the boundary between the two regions. Then, the remaining free-free matrix element can be written in the molecular frame by means of the partial wave expansion in terms of the photoelectron emission direction,
\begin{equation}
    d_{\pm}^{(2)} = \langle \Psi_{f\bm{k}}^{(-)} | D(\bm{\epsilon}) | \Psi_{i+\Omega_\lessgtr}^{(+)} \rangle
                  = \sum_{l m p q} d_{\pm, f l m p q}^{(2)} X_{lm}(\hat{\bm{k}}) \epsilon_p \epsilon_q \,.
    \label{eq:dip}
\end{equation}
The function \(X_{lm}(\hat{\bm{k}})\) is a real spherical harmonic. This dipole integral is evaluated numerically in the inner region, where expansion in many-electron Hamiltonian eigenstates is used, and analytically using the asymptotic method from~\cite{AymarCrance} in the outer region. This approach is supplemented with an efficient integration using the numerical Levin quadrature~\cite{Levin} up to a large enough radius $r=b$, where the asymptotic approach is applicable, see Fig.~\ref{fig:regions}. This numerical method can take advantage of the known recurrence relations for Coulomb functions~\cite{Powell} and offers significant speed-up compared to classical quadrature methods. Details of the R-matrix multi-photon calculation procedure as well as its generalization to photon transitions of higher order were explained in~\cite{multiphoton}. The application of the Levin quadrature, which was not used in the original presentation of the R-matrix multi-photon method~\cite{multiphoton}, is discussed in detail in Appendix~\ref{sect:levin}.

%\subsection{Reference RABITT delays from time-dependent calculations}
%Additionally, the RABITT experiment can be directly simulated using a solution of the time-dependent Schrödinger equation in the molecular frame. To provide such benchmark results, we used the molecular RMT~\cite{MRMT}. In all time-dependent calculations performed as part of this work we used the same combination of an attosecond XUV pulse train (APT, consisting of odd harmonics of the IR field frequency) and a probe IR pulse as given in~\cite{SerovKheifets}. The time evolution was calculated for 19 uniformly spaced relative temporal offsets of the two pulses to cover one full period of the IR field. The extraction of the RABITT sideband time delay was done by transforming the evolved photoelectron wave packet to momentum space. In momentum space, the sidebands were located. These correspond to even harmonics not contained in the APT. The oscillatory behaviour of the sidebands with the relative XUV/IR delay \(\Delta \tau\) was fitted to the expected form given by Eq.~\eqref{eq:sbcos} and, finally, the absolute delay of the oscillations \(\tau_{sb} \approx \tau_W + \tau_{cc}\) corresponding to \(\Delta \tau = 0\) was read out. This is the time delay given by Eq.~\eqref{eq:tau2}, associated with the XUV photon energy \(\Omega = (\Omega_> + \Omega_<)/2\).

\subsection{Orientation averaging}

The following sections discuss time delays for specific orientation of the polarization and direction of emission in the molecular frame, as well as fully emission-integrated and orientation-averaged time delays. The latter can be obtained by substituting Eqs.~\eqref{eq:tau2} and~\eqref{eq:dip} into Eq.~\eqref{eq:signal} and averaging over all orientations. The result is
\begin{equation}
    \tau^{(2)} =
    \frac{1}{2\omega} \arg \sum_{\substack{lm\\ q_1 q_2 q_1' q_2'}} d_{+, f l m q_1 q_2}^{(2)*} d_{-, f l m q_1' q_2'}^{(2)} A_{q_1 q_2 q_1' q_2'} \,,
    \label{eq:oavg2}
\end{equation}
where the factor~\cite{Zamastil}
\begin{align}
    A_{q_1 q_2 q_1' q_2'}
    &= \int \hat{\epsilon}_{q_1} \hat{\epsilon}_{q_2} \hat{\epsilon}_{q_1'} \hat{\epsilon}_{q_2'} \frac{\mathrm{d\hat{\bm{\epsilon}}}}{4\pi} \nonumber \\
    &= \frac{1}{15} (\delta_{q_1 q_2} \delta_{q_1' q_2'} + \delta_{q_1 q_1'} \delta_{q_2 q_2'} + \delta_{q_1 q_2'} \delta_{q_1' q_2})
    \label{eq:Aabcd}
\end{align}
arises due to orientation averaging. Here we assume Cartesian basis (real spherical harmonics) for indices \(q_1\), \(q_2\), \(q_1'\) and \(q_2'\), which correspond to molecular-frame polarization directions of the two absorbed photons in the first and the second pathway. In the derivation of Eq.~\eqref{eq:oavg2} we first integrated over emission directions of the photoelectron in the molecular frame and then over all possible relative orientations of the linear polarization vector with respect to the molecular axis.

\subsection{Reference RABITT delays from time-dependent calculations}
The RABITT experiment can be directly simulated using a solution of the time-dependent Schrödinger equation in the molecular frame. To provide such benchmark results, we used the molecular RMT~\cite{MRMT}. In all time-dependent calculations performed as part of this work we used the same combination of an attosecond XUV pulse train (APT, consisting of odd harmonics of the IR field frequency) and a probe IR pulse as given in~\cite{SerovKheifets}. The time evolution was calculated for 19 uniformly spaced relative temporal offsets of the two pulses to cover one full period of the IR field. The extraction of the RABITT sideband time delay was done by transforming the evolved photoelectron wave packet to momentum space. In momentum space, the main bands and the sidebands were identified as local maxima of the momentum probability distribution. The sidebands correspond to even harmonics not contained in the APT. The oscillatory behaviour of the sidebands with the relative pulse delay \(\Delta \tau\) was fitted to the expected form given by Eq.~\eqref{eq:sbcos} and, finally, the absolute delay of the oscillations \(\tau_{sb} \approx \tau_W + \tau_{cc}\) corresponding to \(\Delta \tau = 0\) was read out. This is the time delay given by Eq.~\eqref{eq:tau2}, associated with the XUV photon energy \(\Omega = (\Omega_> + \Omega_<)/2\).

With realistic XUV pulses of finite duration the maxima of the momentum distribution are not infinitely narrow, so some ambiguity of data extraction may arise. Nevertheless, the synthetic pulse used in this work was sufficiently long (and the molecules sufficiently simple) to generate well-separated, non-overlapping peaks in the momentum space, allowing unambiguous identification of centres of the individual bands.

\subsection{Molecular delay vs RABITT delay}

The second-order RABITT results can be approximated using one-photon delays, where the amplitudes for ionization after absorption of two photons in Eq.~\eqref{eq:tau2} are replaced by amplitudes corresponding to absorption of the XUV photon only (\(\Omega_<\) and \(\Omega_>\)), i.e.\ neglecting the effect of absorption of the second photon. This makes the formula a discrete approximation of Eq.~\eqref{eq:tau1}. Then, an expression similar to Eq.~\eqref{eq:oavg2} can be obtained for the averaged one-photon time delay:
\begin{align}
    \tau^{(1)} =
    \frac{1}{2\omega} \arg \sum_{lm q} d_{+, f l m q}^{(1)*} d_{-, f l m q}^{(1)} \,.
    \label{eq:oavg1}
\end{align}
Here the one-photon partial wave transition dipoles \(d_{\pm,flmq}^{(1)}\) correspond to absorption of one of the two interfering harmonics with frequencies \(\Omega_\lessgtr\) and Cartesian component \(q\) of the polarization vector.

Neglecting the interaction with the IR field can be partly compensated for by inclusion of the continuum-continuum correction \(\tau_{cc}\), which is accurate at high energies~\cite{Dahlstrom} as we explicitly demonstrate below by comparing to our two-photon delays which don't use this splitting.

We would like to stress  that the two-photon delay \(\tau^{(2)}\) considered in this work is not the same quantity as the so-called ``two-photon molecular delay'' discussed in the literature (\(\tau_{\text{mol}}\) in \cite{BaykushevaWorner}, \(\tau_{PI}\) in \cite{KamalovCO2}). The latter can be expressed in the present definitions as \(\tau_{\text{mol}} \approx \tau^{(2)} - \tau_{cc}\) and, despite its name, is actually closer to approximating \(\tau^{(1)}\) than \(\tau^{(2)}\) (see Appendix~\ref{sect:taumol}), which was also confirmed by calculations of Kamalov \textit{et al.} in the supplemetary material to Ref.~\cite{KamalovCO2} ``\textit{in stark contrast to the previous observation by Baykusheva and Wörner}~\cite{BaykushevaWorner}''.

\section{Time delay calculations and results}
\label{sect:results}

\subsection{Details of the calculations}

In all calculations presented in this article, the calculated energy of the neutral ground state was manually shifted to recover the experimental vertical first ionization potential of the molecule. This is a common practice in the R-matrix method, where the same molecular orbitals and active space are typically used for both the initial neutral molecule and the residual ion states. As a consequence, the initial and the final states are described with a different degree of accuracy but their energies can be manually adjusted. Such a correction does not alter excitation thresholds within the ion. Comparison of the original calculated, the corrected, and the experimental vertical ionization thresholds for molecules discussed in this article is shown in Tab.~\ref{tab:ionization-potentials}.

\begin{table}[htbp]
    \centering
    \begin{tabular}{llcl}
        \toprule
         molecule & ion state          & calculated (shifted) & measured \\
        \midrule
         N$_2$    & X \(^2 \Sigma_g^+\)& 16.67      (15.60)   & 15.6 \cite{N2IP} \\
                  & A \(^2 \Pi_u\)     & 18.12      (17.05)   & 17.0 \\
                  & B \(^2 \Sigma_u^+\)& 19.87      (18.80)   & 18.8 \\
         CO$_2$   & X \(^2 \Pi_g\)     & 14.85      (13.78)   & 13.8 \cite{CO2IP} \\
                  & A \(^2 \Pi_u\)     & 18.82      (17.75)   & 17.6 \\
                  & B \(^2 \Sigma_u^+\)& 19.27      (18.19)   & 18.1 \\
                  & C \(^2 \Sigma_g^+\)& 20.59      (19.52)   & 19.4 \\
         H$_2$O   & X \(^2 B_1\)       & 12.80      (12.60)   & 12.6 \cite{H2OIP} \\
                  & A \(^2 A_1\)       & 15.18      (14.96)   & 14.7 \\
                  & B \(^2 B_2\)       & 19.35      (19.12)   & 18.5 \\
         N$_2$O   & X \(^2 \Pi\)       & 13.66      (12.89)   & 12.9 \cite{Truesdale} \\
                  & A \(^2 \Sigma\)    & 17.18      (16.41)   & 16.4 \\
                  & B \(^2 \Pi\)       & 19.33      (18.56)   & 18.3 \\
                  & C \(^2 \Sigma\)    & 21.31      (20.54)   & 20.1 \\
        \bottomrule
    \end{tabular}
    \caption{Calculated, corrected (in brackets) and experimental vertical ionization potentials in eV for the final cationic states of the molecules analyzed in this work.}
    \label{tab:ionization-potentials}
\end{table}

Some plots below present smoothed energy dependencies of the time delays. They were obtained transforming often highly oscillatory raw calculated results to data suitable for comparison to experimentally observable average trends, including effects like vibrational averaging, finite bandwidth and similar. In Eqs.~\eqref{eq:oavg2} and~\eqref{eq:oavg1}, time delays are calculated as complex phases of the differential cross section interference term. These interference terms were smoothed using a convolution with a Gaussian distribution \(g(x)\). The smoothing is also very useful for practical purposes. Extremely close to thresholds, the multi-photon R-matrix method is extremely sensitive to channel energies. Due to the finite precision of a computer calculation, this often translates into very narrow but extremely high spikes of numerical origin in the calculated dipole energy dependence. To avoid distortion of the results by these spurious spikes, we included an additional weighting factor in the smoothing procedure, which is approximately inversely proportional to the distance between the raw calculated value and the smoothed value at the same energy:
\begin{equation}
    f_{\text{smooth}}(x_i) = \sum_j \frac{g(x_i - x_j) f(x_j)}{\sqrt{1 + |f(x_j) - f_{\text{smooth}}(x_j)|^2}}
\end{equation}
The additional factor penalizes strongly outlying values. This self-consistent smoothing procedure is performed for a few iterations until it converges.

Where both time-dependent and time-independent calculations were performed, we always used the same molecular models. In all cases fixed geometry was assumed.

\subsection{H$_2$ molecule}

We first calculated the photoionization amplitudes and time delays for the H$_2$ molecule. We used two different models: ionization of a single Hartree-Fock orbital from a single Slater determinant (static exchange model, SE) and full CI. In both cases we employed the basis set cc-pVDZ and Hartree-Fock orbitals of H$_2^+$ obtained from Psi4~\cite{PSI4}. The size of the inner region was set to 150$a_0$ (200 in the full CI calculation) and the partial wave expansion was truncated at the angular momentum \(\ell = 4\). We performed the time-independent calculations with the molecular scattering suite UKRmol+~\cite{UKRmolp} for a range of photoelectron energies from the threshold up to 80~eV. To obtain reference data, we also ran a time-dependent simulation in RMT for the combination of pulses detailed in the previous section. All calculations on H$_2$ were done for two specific configurations: with the linear polarization direction parallel or perpendicular to the molecular axis. Only electron emission in the direction of the polarization vector was studied. The same configurations were studied by Serov and Kheifets~\cite{SerovKheifets}.

\begin{figure}[htbp]
    \centering
    \includegraphics[width=0.95\columnwidth]{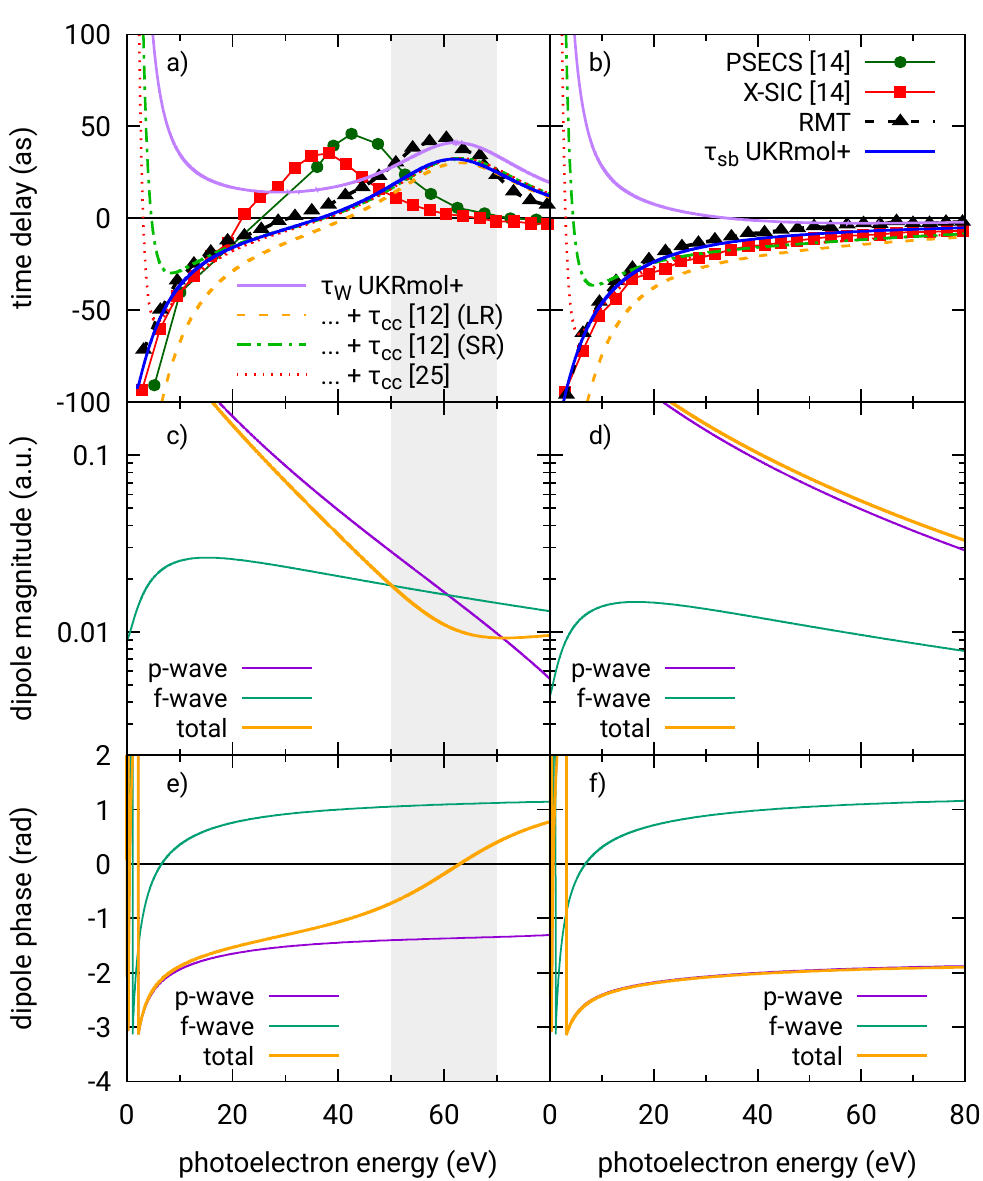}
    \caption{Calculated RABITT sideband delays and other computed reference time delays for photoionization of H$_2$ in the static exchange model into the ground state of H$_2^+$  by a pulse linearly polarized parallel (left column) or perpendicular (right column) to the molecular axis and photoelectron emission in the same direction. The top row shows the time delays calculated from the time-dependent (RMT) and time-independent method (one-photon \(\tau_W\), two-photon \(\tau_{sb}\)) compared to calculations of Serov and Kheifets~\cite{SerovKheifets}. Different asymptotic corrections of Dahlström et al.~\cite{Dahlstrom} and Ivanov and Smirnova~\cite{IvanovSmirnova} are used to complement \(\tau_W\). The middle row shows the magnitude of the one-photon ionization dipole per partial wave, and the bottom row contains the phases of these dipoles, partial and total. The shaded region, 50--70~eV, highlights the interval of energies where the two contributing partial wave amplitudes exchange in magnitude and their interference gives rise to the feature in the time delays.}
    \label{fig:H2-delays-SE}
\end{figure}

\begin{figure}[htbp]
    \centering
    \includegraphics[width=0.95\columnwidth]{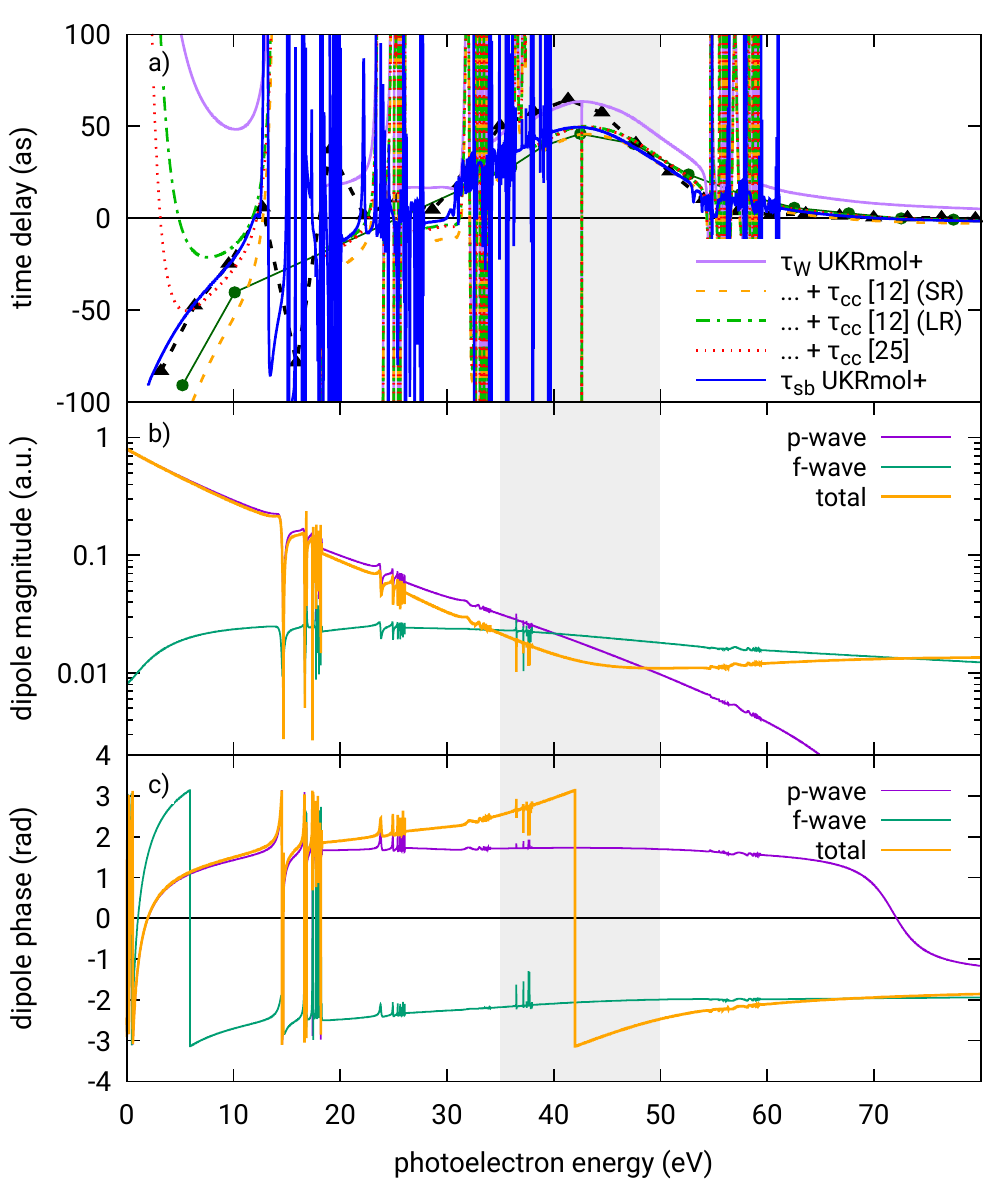}
    \caption{Calculated RABITT sideband delays (\(\tau_{sb}\)) and other computed reference time delays (\(\tau_W\), \(\tau_{cc}\)) for photoionization of H$_2$ in the full CI model into the ground state of H$_2^+$ by a pulse linearly polarized parallel with the molecular axis and photoelectron emission in the same direction. The results are compared to the same data as in~Fig.~\ref{fig:H2-delays-SE}: the full geen line with circles (barely visible) corresponds to the PSECS calculation~\cite{SerovKheifets} and the black dashed line with triangles to RMT. The shaded region highlights the region of partial wave interference as in Fig.~\ref{fig:H2-delays-SE}.}
    \label{fig:H2-delays-FCI-z}
\end{figure}

The results of the SE model are shown in Fig.~\ref{fig:H2-delays-SE} for both perpendicular and parallel direction. The plots include both the 1- and 2-photon delays as defined by Eqs.~\eqref{eq:tau1} and~\eqref{eq:tau2}, respectively. On top of that, for both polarizations in Figs.~\ref{fig:H2-delays-SE}a--b we also add to the 1-photon (Wigner) ionization delays several high-energy corrections from the literature~\cite{IvanovSmirnova, Dahlstrom}, approximating the complete 2-photon picture. We see that only our full 2-photon method is able to correctly reproduce the decreasing behaviour of the time delays towards low energies. In the long-wavelength limit all of the corrections either over-estimate the decrease, or eventually are dominated by the large positive Coulomb delay coming from the 1-photon delay. Finally, it is obvious that the wide hill-like structure in the time delays between 50 and 70~eV, that is highlighted by the shaded background, is present both in the first-order and second-order results. In Section~\ref{sect:wpacket} we discuss its origin in the context of one-photon ionization.

The results for the full CI model, Fig.~\ref{fig:H2-delays-FCI-z} (parallel polarization only), are qualitatively equal to those of the SE model in Fig.~\ref{fig:H2-delays-SE}a, only the position of the broad structure is shifted from approximately 60~eV to around 40~eV due to an improved accuracy of the dipole matrix elements which shifts the interference structure to lower energies, see Section~\ref{sect:wpacket}. The wild oscillations in the full CI results are caused by Feshbach resonances and are also apparent in the dipole magnitudes, which are directly linked to the differential cross section. As before, the second-order calculation agrees best with the time-dependent results from RMT.

\subsection{N$_2$ molecule}

\begin{figure}[htbp]
    \centering
    \includegraphics[width=0.95\columnwidth]{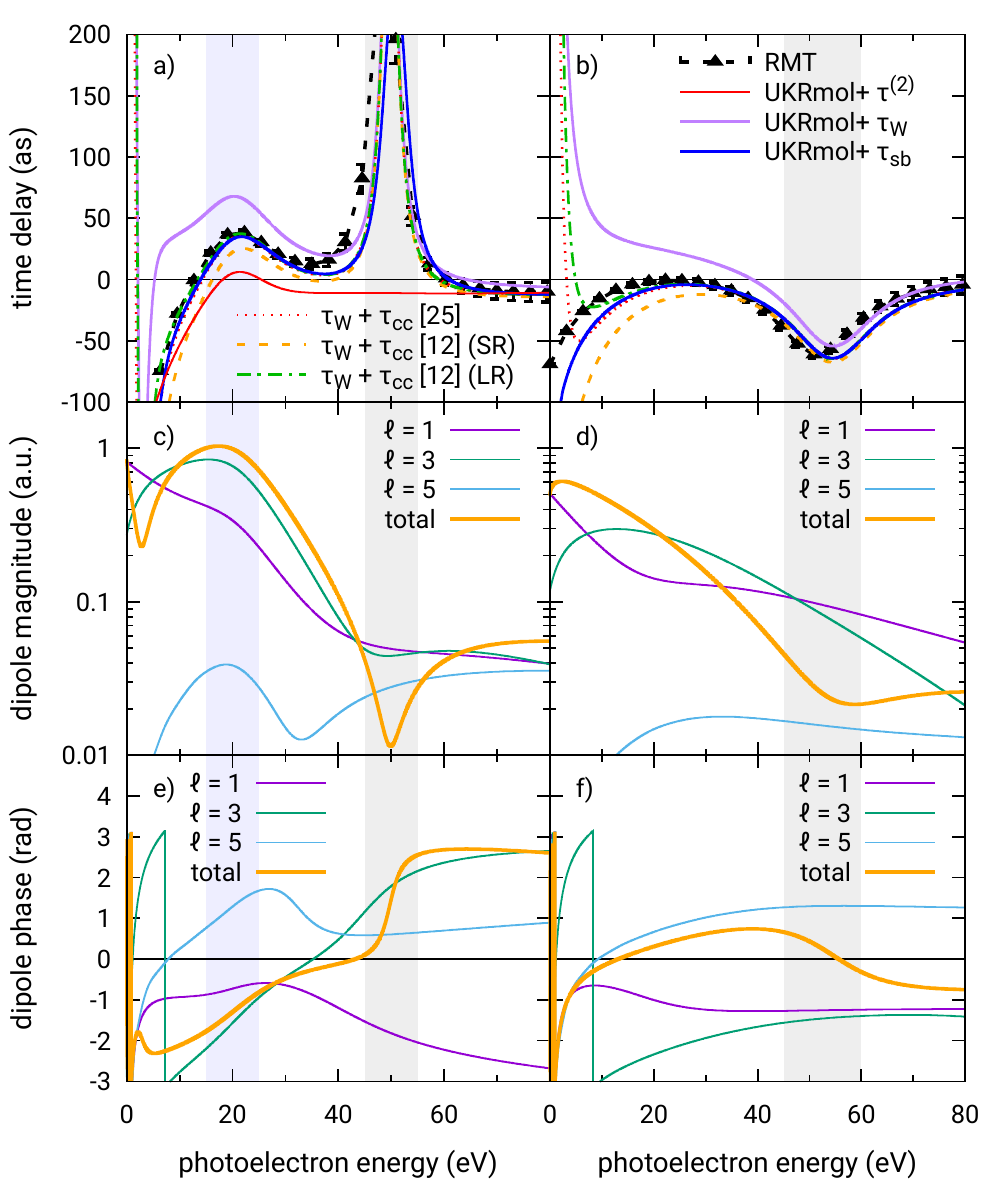}
    \caption{RABITT sideband delays and other reference time delays for photoionization of N$_2$ in the static exchange model into the ground state of N$_2^+$  by a pulse linearly polarized parallel (left column) or perpendicular (right column) to the molecular axis and photoelectron emission in the same direction. The top row shows the time delays, the middle row shows the magnitude of the one-photon ionization dipole per partial wave, and the bottom row contains the phases of these dipoles, partial and total. The grey shaded regions 45--60~eV in both columns highlight the interval of energies where the contributing partial wave amplitudes exchange in magnitude and their interference gives rise to the feature in the time delays. The light blue shaded region in the left column, around 20~eV, is where the dominant \(f\)-wave is the one most affected by the nitrogen's broad shape resonance.}
    \label{fig:N2-delays-SE}
\end{figure}

For N$_2$ we first used the same basis set and the same SE model as for H$_2$. That is the cc-pVDZ atomic basis set was used together with all N$_2$ HF molecular orbitals of the neutral ground state. The partial wave expansion was extended to \(\ell = 6\), R-matrix radius was set to 15$a_0$, while other parameters were left unchanged. The calculated results, both time-dependent and time-independent are in Fig.~\ref{fig:N2-delays-SE}. As before, there is an energy interval around 50~eV where the magnitudes of the leading partial photoionization dipole elements become comparable, causing a large variation of the total phase. On top of that, the \(f\)-wave (\(l = 3\)) exhibits another feature related to a broad shape resonance around 20~eV. Unlike the time delay feature caused by the partial wave interference, this latter one is also visible in the results for an unoriented sample of molecules (red curve in Fig.~\ref{fig:N2-delays-SE}a).

The various asymptotic forms of the continuum-continuum delay perform similarly as in H$_2$. For low energies they diverge away from the time-dependent and second-order calculations, but for photoelectron kinetic energies above 20~eV they generally perform very well, the short-range variant of~\cite{Dahlstrom} being the least accurate.

\begin{figure}[htbp]
    \centering
    \includegraphics[width=0.95\columnwidth]{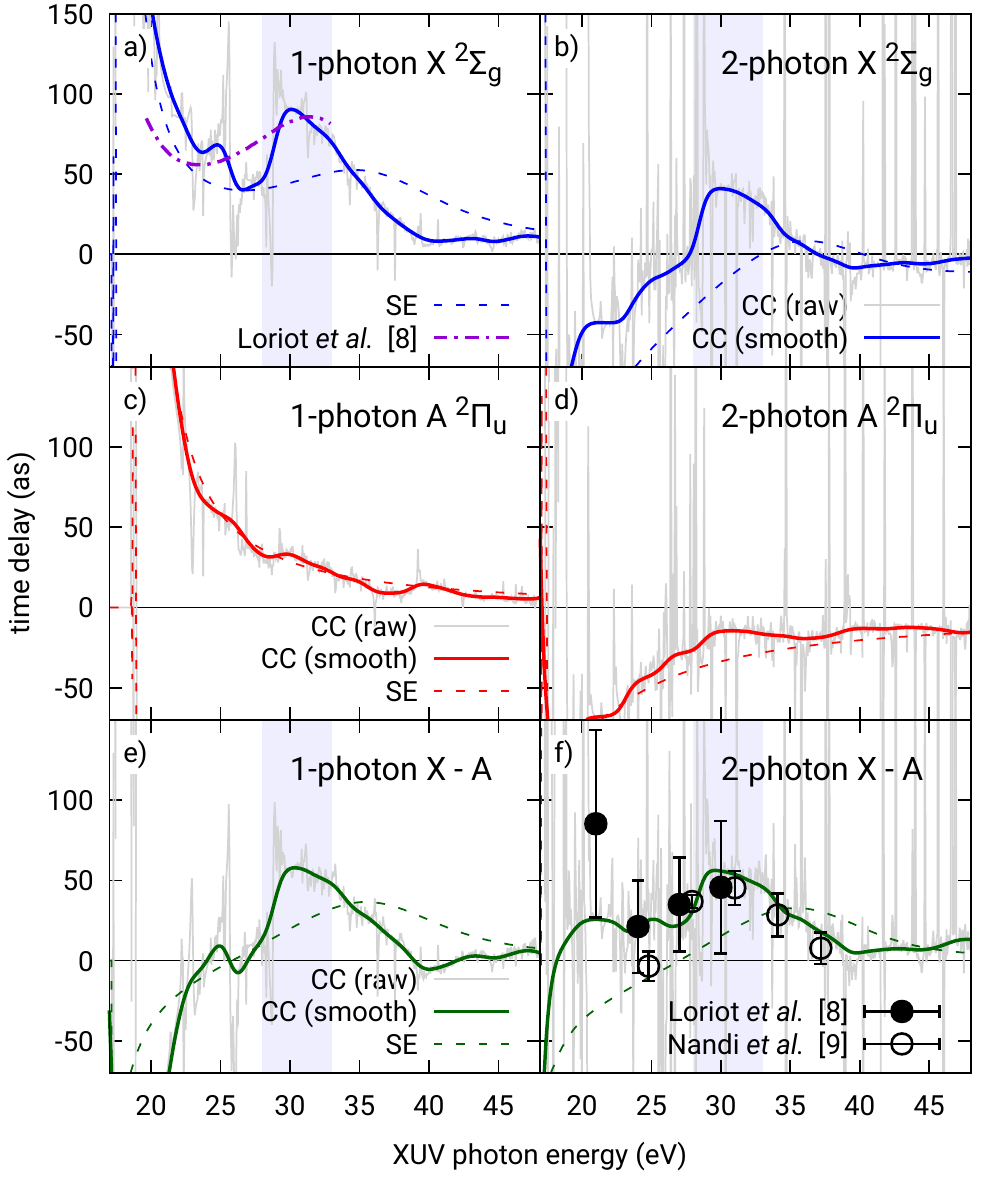}
    \caption{Unsmoothed (grey) and smoothed RABITT sideband delays for photoionization of N$_2$ into the lowest two states of N$_2^+$ in a larger close-coupling model, averaged over orientations of the molecule. Dashed curves correspond to the static exchange results, solid to close-coupling. Left column: raw and smoothed discrete one-photon delays \(\tau^{(1)}\). Right column: 2-photon delays \(\tau^{(2)}\). The experimental and theoretical results of Loriot \textit{et al.}~\cite{LoriotN2} are shown using the filled circles and the purple curve, respectively. The experimental data of Nandi \textit{et al.}~\cite{Nandi} are plotted using the empty circles. Experimental data for the transition between the lowest vibrational states are plotted. The resonance in the X state discussed in text is highlighted by the shaded region. }
    \label{fig:N2-delays-oavg}
\end{figure}

\begin{figure}[htbp]
    \centering
    \includegraphics[width=0.95\columnwidth]{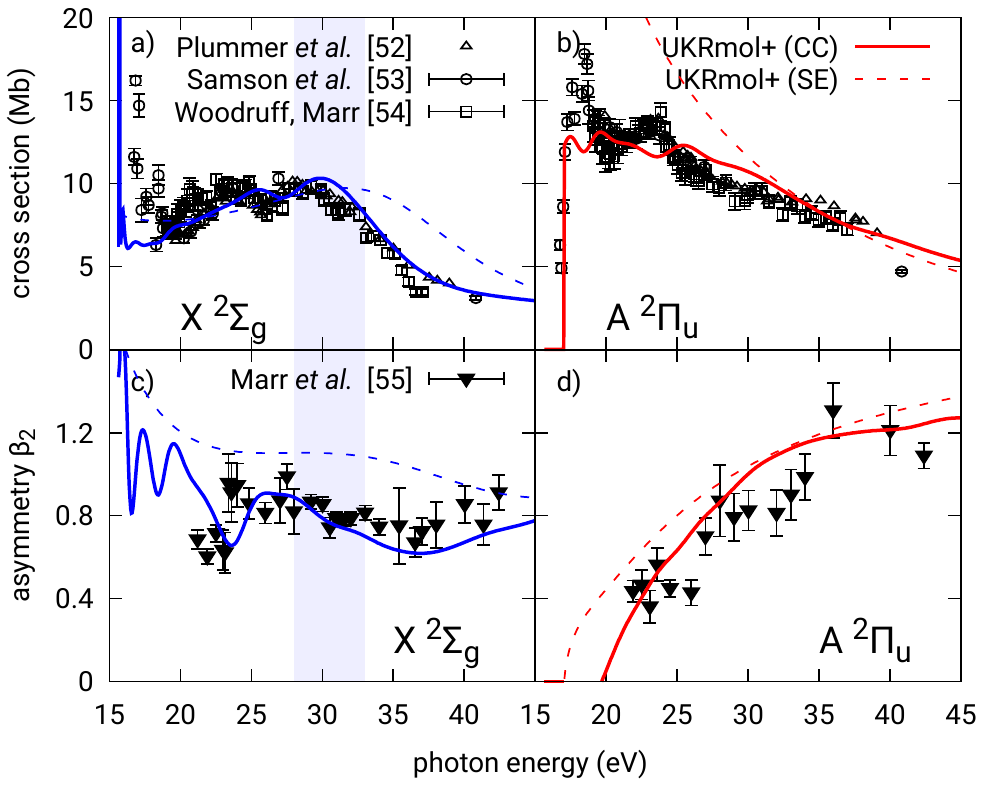}
    \caption{Smoothed isotropic cross sections and asymmetry parameters for one-photon ionization of N$_2$ into the lowest two states of N$_2^+$. Results are compared to measurements of Plummer \textit{et al.}~\cite{PlummerN2}, Samson \textit{et al.}~\cite{SamsonN2}, Woodruff and Marr~\cite{WoodruffMarrN2} and Marr \textit{et al.}~\cite{MarrN2}. The peak of the resonance in the X state discussed in the text is highlighted. Labels ``SE'' and ``CC'' distinguish the static exchange and close coupling models.}
    \label{fig:N2-cs-CAS}
\end{figure}

Loriot \textit{et al.}~\cite{LoriotN2} and Nandi \textit{et al.}~\cite{Nandi} measured relative RABITT time delays of the ground ionic state X~$^2\Sigma_g^+$ with respect to the excited ionic state A~$^2\Pi_u$ for an unoriented sample of N$_2$ molecules in the vicinity of a broad shape resonance located at 30~eV of photon energy. Having access to molecular-frame partial wave transition elements in the present R-matrix method, we are easily able to calculate the laboratory-frame observables using Eq.~\eqref{eq:oavg2} and~\eqref{eq:oavg1}. In Fig.~\ref{fig:N2-delays-oavg} we present the results of such calculation, where we used a larger molecular fixed-nuclei model with internuclear separation \(d = 1.1\,\)\AA, based on the cc-pVQZ basis set as employed in an accurate N$_2$ structure study~\cite{DuncanN2a,DuncanN2b}. We used complete active space self-consistent field (CASSCF) orbitals of N$_2^+$ obtained in Molpro~\cite{Molpro} by optimizing with respect to the sum of energies of the three lowest states of N$_2^+$ in the D$_{2h}$ point group, corresponding to the two states X and A in D$_{\infty h}$. We constructed the molecular model from 2 frozen orbitals, 9 active orbitals and 300 ionic states, employing R-matrix radius \(r = 15a_0\), partial wave expansion up to \(\ell = 6\) and a B-spline basis consisting of 30 equally spaced functions for construction of the radial continuum functions. This model gives sufficiently accurate one-photon cross sections and asymmetry parameters, see Fig.~\ref{fig:N2-cs-CAS}, particularly for the X state displaying the prominent resonance around 30~eV.

To explore electron correlation effects, we also calculated the same process with the SE model discussed before, see dashed lines in Figs.~\ref{fig:N2-delays-oavg} and~\ref{fig:N2-cs-CAS}. In that calculation we used HF orbitals generated from the same basis set, cc-pVQZ; 4 of these orbitals were included in the continuum basis as virtual orbitals. Other parameters were left unchanged.

For the RABITT calculation, the IR wavelength was 800~nm, which is different than the wavelength of 400~nm used in~\cite{LoriotN2}; however, in the experiment a frequency doubled fundamental pulse was used which resulted in a more complex interference of three ionization pathways. Despite the slightly different method, the calculated and orientation-averaged molecular two-photon delays agree very well with the experiment in the energy range 24--30~eV, see Fig.~\ref{fig:N2-delays-oavg}f. In Fig.~\ref{fig:N2-delays-oavg} our calculated results are given in their raw, unsmoothed form (grey), as well as with Gaussian smoothing as explained in the theory section. The experimental point at 21~eV is somewhat off the smoothed curve but the upward trend of the delays observed in the experiment is confirmed by our calculations. The agreement with both experiments is excellent, with only a minor deviation in the lowest-energy experimental point.

Figure~\ref{fig:N2-delays-oavg} also clearly demonstrates that only the large close-coupling (CC) model is able to reproduce the increase of the relative delay at low energies. This difference in the relative delays arises due to the structure in the absolute delays for the X state, see Fig.~\ref{fig:N2-delays-oavg}b, where in the region from 23~eV to 19~eV the smoothed CC results maintain constant delays at variance with the SE results which drop down monotonously, while the absolute delays pertaining to the A state drop down in the same interval in both types of calculation. This clearly demonstrates the important role of electron correlation at low energies in ionization into the X state.

\subsection{H$_2$O and N$_2$O molecules}

\begin{figure}[htbp]
    \centering
    \includegraphics[width=0.95\columnwidth]{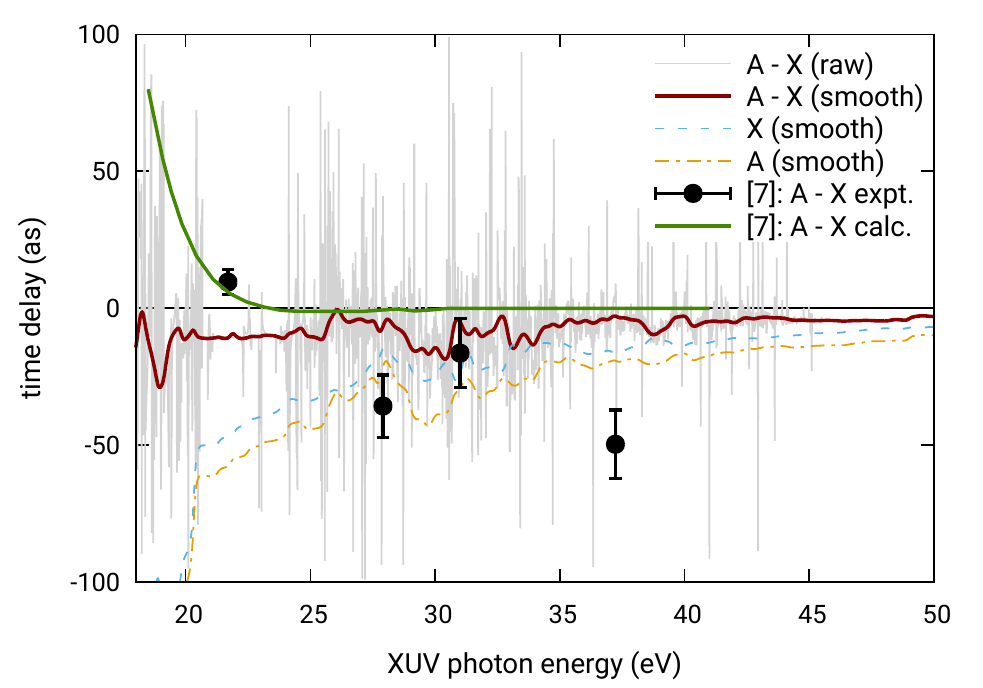}
    \caption{Orientation-averaged two-photon time delays \(\tau^{(2)}\) for ionization of H$_2$O into the first two states of its ion and their difference. Comparison is done to the measurement and calculation of Huppert \textit{et al.}~\cite{Huppert}.}
    \label{fig:H2O-XA-oavg}
\end{figure}

Huppert \textit{et al.}~\cite{Huppert} measured relative RABITT time delays in unoriented H$_2$O and N$_2$O molecules, exploring the effect of photoionization shape resonances. They observed that the calculated relative delays between the ionization of H$_2$O molecule into states A\(\,{}^2A_1\) and X\(\,{}^2B_1\) are essentially featureless. Our fully two-photon calculation confirms their conclusion, see Fig.~\ref{fig:H2O-XA-oavg}, even though neither our nor their calculation is able to fully reproduce the measured data within the reported experimental uncertainty. In this calculation we used the large molecular model from~\cite{MRMT}, which was shown to provide one-photon cross sections in a good agreement with experiment. The wavelength of the IR field was assumed to be 800~nm.

\begin{figure}[htbp]
    \centering
    \includegraphics[width=0.95\columnwidth]{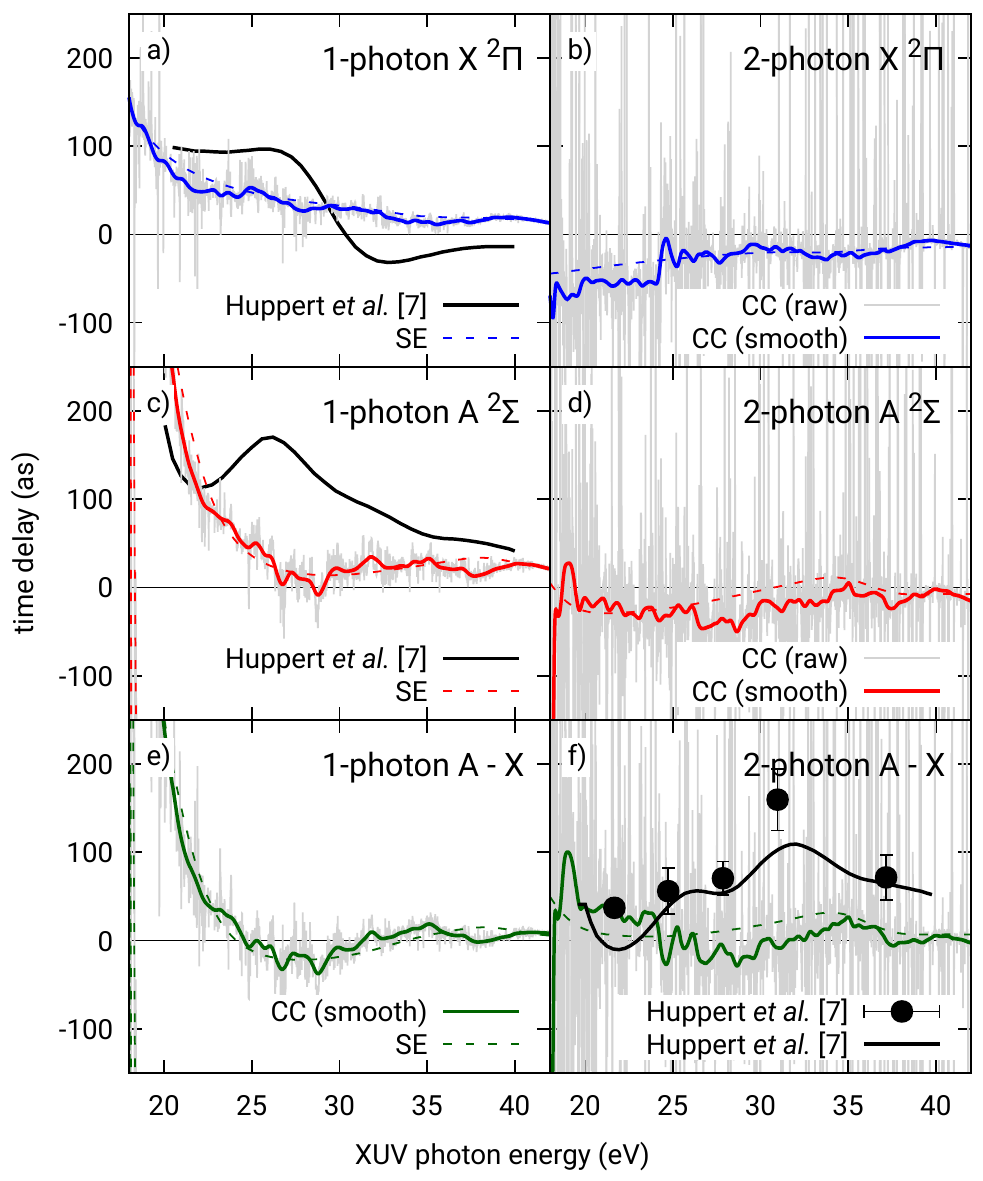}
    \caption{Smoothed and unsmoothed RABITT sideband delays for photoionization of N$_2$O into the lowest two states of N$_2$O$^+$ in a small static exchange (dashed lines) and large close-coupling (solid lines) model, averaged over orientations of the molecule. Left column: raw and smoothed 1-photon delays \(\tau^{(1)}\). Right column: 2-photon delays \(\tau^{(2)}\). The experimental data (black points) as well as the additional calculations (black curves) are from Huppert \textit{et al.}~\cite{Huppert}.}
    \label{fig:N2O-delays-CAS}
\end{figure}

\begin{figure}[htbp]
    \centering
    \includegraphics[width=0.95\columnwidth]{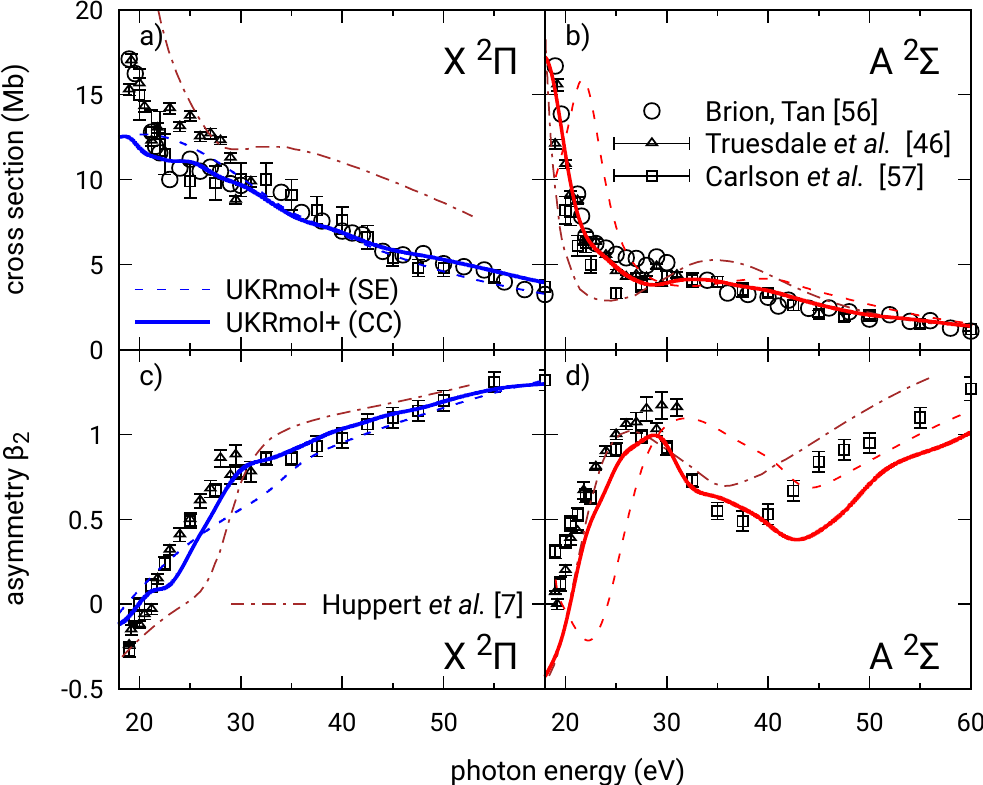}
    \caption{Smoothed cross sections and asymmetry parameters of one-photon ionization of N$_2$O into the lowest two states of N$_2$O$^+$ calculated for the static-exchange and close-coupling molecular model, compared to the calculation of Huppert \textit{et al.}~\cite{Huppert} and measurements of Brion and Tan~\cite{BrionTan}, Truesdale \textit{et al.}~\cite{Truesdale} and Carlson \textit{et al.}~\cite{Carlson}.}
    \label{fig:N2O-cs-CAS}
\end{figure}

Figure~\ref{fig:N2O-delays-CAS} compares our one- and two-photon delays to measurement and calculation of time delays in N$_2$O~\cite{Huppert}, while Fig.~\ref{fig:N2O-cs-CAS} shows the one-photon observables for this molecule. Figures~\ref{fig:N2O-delays-CAS} and~\ref{fig:N2O-cs-CAS} also compare two different molecular models: SE and CC. In the former case we used Hartree-Fock orbitals of the neutral N$_2$O molecule at its experimental equilibrium geometry (N--N 1.128~\AA, N--O 1.184~\AA), obtained from Psi4~\cite{PSI4} using the basis set cc-pVTZ. The B-spline continuum basis was the same as in the above N$_2$ calculations and included 10 additional virtual HF orbitals of the molecule. In contrast, for the CC model we used CASSCF molecular orbitals of N$_2$O obtained in Molpro~\cite{Molpro} by state-averaging including the neutral ground state and the eight lowest states of the ion (in D$_{2h}$). The larger molecular model consisted of 3 frozen orbitals, 11 active orbitals, 7 virtual orbitals added to the continuum basis and 200 ionic states. The IR wavelength was 800~nm.

Judging by the agreement with photoionization measurements of Brion and Tan~\cite{BrionTan}, Truesdale \textit{et al.}~\cite{Truesdale} and Carlson \textit{et al.}~\cite{Carlson} in Fig.~\ref{fig:N2O-cs-CAS}, the CC model yields very good one-photon cross sections. Only the asymmetry parameter for the excited state (Fig.~\ref{fig:N2O-cs-CAS}d) somewhat deviates from the angularly resolved measurement~\cite{Carlson} above 40~eV. Generally, our one-photon cross sections and asymmetries seem to agree with the experiments better than the results of Huppert \textit{et al.}~\cite{Huppert}.

However, while we are confident that our one-photon observables are accurate, we were not able to reproduce the experimental and the theoretical results of Huppert \textit{et al.}~\cite{Huppert}.

Having achieved such a good agreement for an analogous experiment with N$_2$, Fig.~\ref{fig:N2-delays-oavg}f, we have no simple explanation that would account for the qualitative difference. In contrast to N$_2$, the partial one-photon ionization cross sections of N$_2$O do not exhibit as dramatic resonance features, cf.\ Figs.~\ref{fig:N2O-cs-CAS}a--b to Figs.~\ref{fig:N2-cs-CAS}a--b, even though some shape resonances are undoubtedly present. According to Braunstein and McKoy~\cite{BraunsteinMcKoy}, one of the shape resonances in the ionization into the A state affects the cross sections around the threshold. In our calculation we do observe a build-up of the time delay towards low energies, but this is almost out of the range of the RABITT experiment of Huppert \textit{et al}. The second shape resonance in the A state, in the range of photon energies approx. 30--40~eV, that Braustein and McKoy discuss, manifests solely as the dip in the asymmetry parameter. While there seems to be a very low and broad structure in our calculated time delays centered at 35~eV, see Fig.~\ref{fig:N2O-delays-CAS}f, roughly coinciding with the bottom of the dip in the asymmetry parameter, Fig.~\ref{fig:N2O-cs-CAS}d, we found no trace of a time delay feature as prominent as those calculated by Huppert \textit{et al.}~\cite{Huppert} (black curves in Figs.~\ref{fig:N2O-delays-CAS}a,c) by means of the molecular theory of Baykusheva and Wörner~\cite{BaykushevaWorner}.

\begin{figure}[htbp]
    \centering
    \includegraphics[width=0.95\columnwidth]{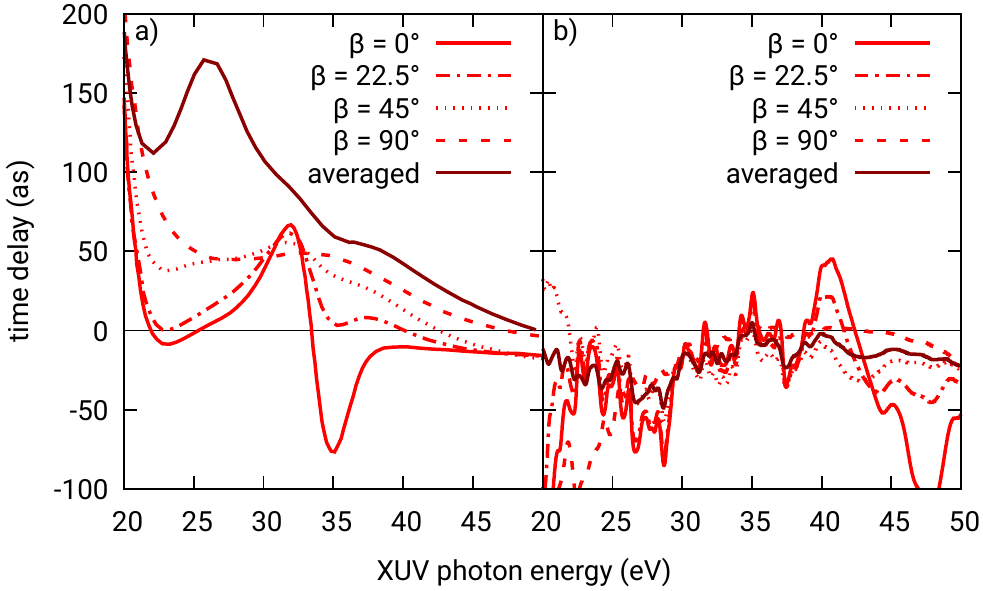}
    \caption{Photoelectron-emission-integrated sideband time delays \(\tau_{sb}(\hat{\bm{\epsilon}})\) of photoionization into the A state of N$_2$O$^+$ for several molecular alignment angles with respect to the field polarization. Left panel: Molecular delay \(\tau_{\text{mol}}\) of Baykusheva and Wörner~\cite{BaykushevaWorner}. Right panel: Present two-photon calculation in UKRmol+.}
    \label{fig:N2O-oriented-delays}
\end{figure}

Baykusheva and Wörner also report photoionization delays for the A state for several fixed alignment angles of the N$_2$O molecule with respect to the polarization \(\hat{\bm{\epsilon}}\) of the field, see Fig.~\ref{fig:N2O-oriented-delays}a. Since none of their fixed-alignment time delays for photon energies around 25~eV exceeds 75~as, it is quite surprising that their reported alignment-averaged time delay (darker curve in Fig.~\ref{fig:N2O-oriented-delays}a) exceeds 150~as in this energy range, even though we would expect the alignment-averaged photoelectron distribution to exhibit a smoother energy variance and smaller delays. In the present fully two-photon calculation we obtained the emission-integrated but alignment-resolved second-order time delay in the molecular frame and Cartesian basis from
\begin{equation}
    \tau_{sb}(\hat{\bm{\epsilon}}) = \frac{1}{2\omega} \arg \sum_{q_1 q_2 q_1' q_2'} d_{+,flm q_1 q_2}^{(2)*} d_{-,flm q_1' q_2'}^{(2)}
    \hat{\epsilon}_{q_1} \hat{\epsilon}_{q_2} \hat{\epsilon}_{q_1'} \hat{\varepsilon}_{q_2'} \,.
    \label{eq:emissionintegrated}
\end{equation}
The results calculated from Eq.~\eqref{eq:emissionintegrated} are in Fig.~\ref{fig:N2O-oriented-delays}b. The energy dependence of our molecular-orientation-averaged time delay is more in line with the above expectation, i.e.\ not deviating significantly from the fixed-alignment data. The continuum-continuum delay was considered separately in~\cite{BaykushevaWorner}, which is why our second-order data in Fig.~\ref{fig:N2O-oriented-delays}b, inherently containing the continuum-continuum component, are generally lower than the molecular delays in Fig.~\ref{fig:N2O-oriented-delays}a.

In~\cite{BaykushevaWorner} it is argued that the large change in magnitude of the molecular delay arising in the orientation averaging is a key result of their theoretical method; see also the discussion in Appendix~\ref{sect:taumol}. However, the results of the present second-order calculation do not exhibit such behaviour.

The disagreement between the experiment of Huppert~\textit{et al.} and our calculations may originate in dynamical factors that either cannot be included in our calculations, such as nuclear dynamics, or come from the experimental setup departing from the purely two-photon picture: finite bandwidth of the pulse and intensity-dependent effects including higher-order IR transitions, dynamical polarization of the target, etc. While the latter could be modeled using RMT, investigation of these effects lies well beyond the scope of the present work.

Finally, going back to Fig.~\ref{fig:N2O-delays-CAS}, we also see that the SE model yields time delays that are in fair agreement with the CC model for energies above approximately 25~eV. For lower energies the results are strongly affected by resonances close to the channel thresholds, requiring a more complex description. The overall agreement between the single-channel and the multi-channel calculations also to some degree dispels a concern that the lack of anticipated shape resonance features in the second-order time delays is caused by insensitive smoothing approach. Still, given the large density of resonances in the two-photon data, some uncertainty related to the smoothing persists. It also leaves open the question as to how the unsmoothed results would look like if nuclear vibration was included in the model. Below 25~eV the difference between the models increases, particularly due to the improved position of the low-energy shape resonance in the A state.

\subsection{CO$_2$ molecule}

\begin{figure}[htbp]
    \centering
    \includegraphics[width=0.95\columnwidth]{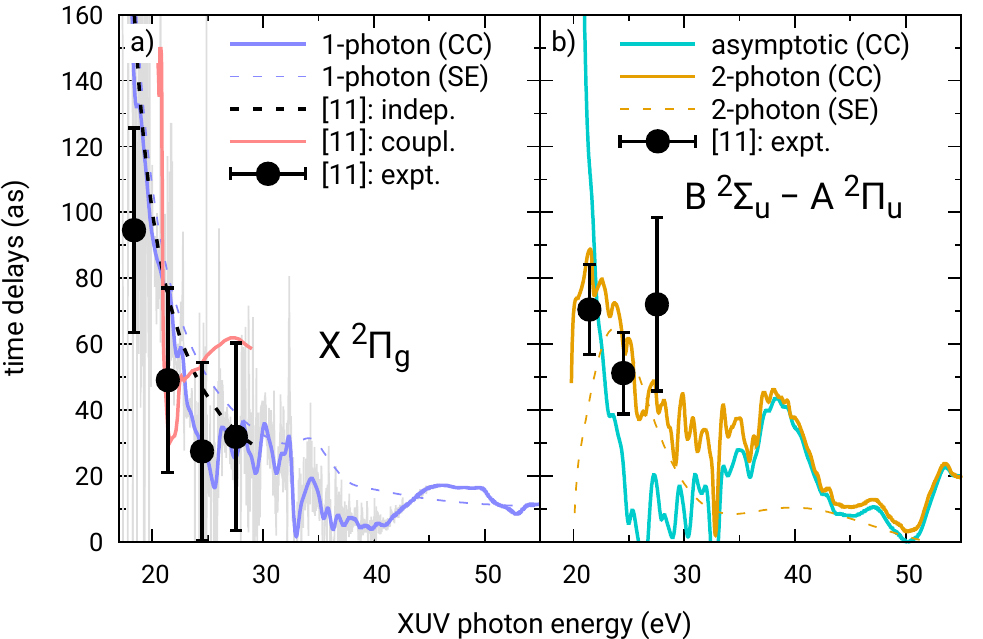}\\
    \includegraphics[width=0.95\columnwidth]{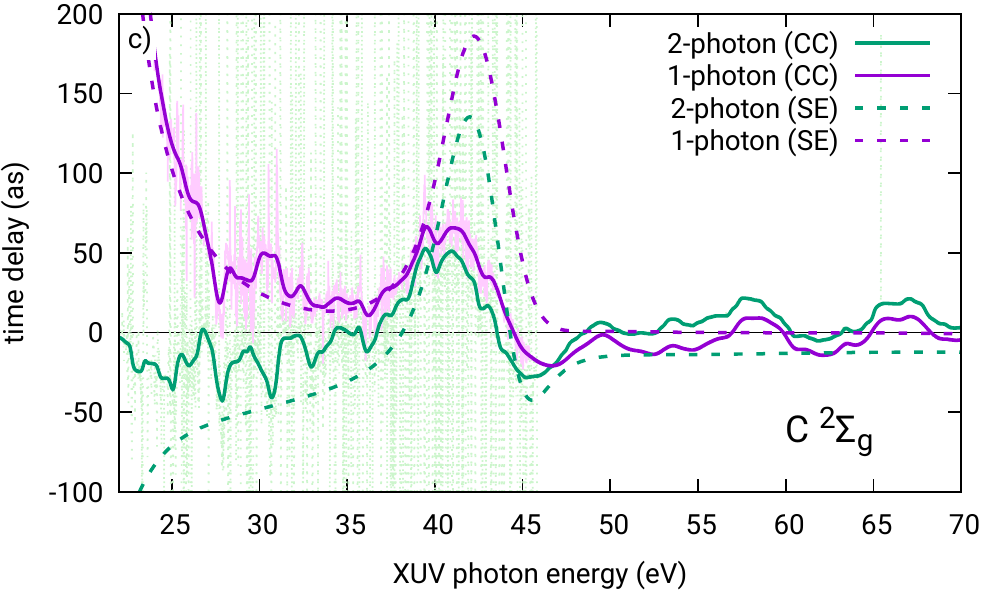}
    \caption{Orientation-averaged time delays for ionization of CO$_2$. (a) Smoothed one-photon delay \(\tau^{(1)}\) for ionization into the ground state of CO$_2^+$. (b) Difference of smoothed two-photon delays \(\tau^{(2)}\) for ionization into the first two excited states of CO$_2^+$; the asymptotic 2-photon results are obtained from Eq.~\eqref{eq:tau2asy}. Comparison is done to the calculations (independent and coupled channels, respectively) and the experiment of Kamalov \textit{et al.}~\cite{KamalovCO2}. (c) Ionization into C $^2\Sigma_g^+$. Light curves in (a) and (c) depict unsmoothed CC results.}
    \label{fig:CO2-X-oavg}
\end{figure}

Kamalov \textit{et al.}~\cite{KamalovCO2} measured molecular orientation-averaged time delays for photoionization of CO$_2$ into the three lowest states of its ion. In their work, they measured relative RABITT time delays with respect to ionization of krypton. Because the ionization thresholds of Kr and CO$_2$ are very close to each other, the uncertainty in the difference between \(\tau_{cc}\) in these two gasses was considered very small and the asymptotic forms for \(\tau_{cc}\) sufficiently accurate for this purpose. Subsequently, they made use of accurate calculations of atomic time delays in krypton to obtain the (1-photon) molecular delays in CO$_2$. We used the accurate molecular model of~\cite{CO2HHG} to calculate the 1-photon delays for the ground cation state, X~\({}^2\Pi_g\), according to Eq.~\eqref{eq:oavg1} and assumed an 800~nm IR field. In contrast to~\cite{CO2HHG}, though, here we replaced the Gaussian continuum basis with a B-spline-based one to maintain accuracy even at somewhat higher energies. The calculation results in an excellent agreement at higher energies, see Fig.~\ref{fig:CO2-X-oavg}a, though there is some divergence at lower energies, possibly related to the above-mentioned \(\tau_{cc}\)-compensating procedure. Even though the coupled-channel theory of Kamalov~\textit{et al.} seems to reproduce the experimental point at \(\sim21\)~eV very well compared to our calculation, we find their theoretical approach questionable due to their use of incorrect boundary conditions, see Appendix~\ref{sect:coupling} for details.

We further calculated two-photon relative ionization delays between the excited states A~\({}^2\Pi_u\) and B~\({}^2\Sigma_u^+\), Fig.~\ref{fig:CO2-X-oavg}b. Here the agreement is very good, too. Instead of using the post-processed molecular delay difference presented directly in~\cite{KamalovCO2} for comparison, we took the original measured (two-photon) delays from the supplementary material of that article and evaluated the B \(-\) A difference from those. This included the value at 21~eV otherwise omitted in the main text of that work due to the uncertainty in the \(\tau_{cc}\) used to extract the presented molecular delays. In our calculation we are not limited by the approximate forms of \(\tau_{cc}\) and we can compare directly to the measured relative two-photon delay. The experimental uncertainty for the relative delays in Fig.~\ref{fig:CO2-X-oavg}b was obtained by summing the experimental uncertainties for the measured krypton-referenced data pertaining to the two final excited states A and B.

Figure~\ref{fig:CO2-X-oavg} further demonstrates the effect of electron correlation on the photoionization delays. While the results for the X state are largely insensitive to the choice between the SE and CC model, indicating weak electron correlation, results for the excited states require the CC model for a good experimental agreement. In Fig.~\ref{fig:CO2-X-oavg}b, this is possibly caused by coupling of the C-channel core-excited shape resonance to the B channel, as suggested in the previous one-photon results~\cite{CO2HHG}. A similar shift of a shape resonance is visible in the C state of CO$_2^+$, Fig.~\ref{fig:CO2-X-oavg}c. We also see that while the difference between the two models' one-photon delays (magenta curves) is more or less limited to the resonant energy interval 40--50~eV, the two-photon results (green curves) differ from each other in a much broader range of energies, particularly in the long wavelengths. Analogous behaviour can be observed for the X state of N$_2^+$ in Fig.~\ref{fig:N2-delays-oavg}a,b. The different two-photon delays for the SE and CC models at low energies indicate that the interaction of the photoelectron with the second photon is significantly more complex than what the simple single-channel continuum-continuum correction would suggest. This is further analysed in Section~\ref{sect:chancoupl}.

Finally, we would like to remark that the very dense forest of narrow autoionizing resonances in Fig.~\ref{fig:CO2-X-oavg}c ends at approximately 45~eV, corresponding to the highest CO$_2$ ionization threshold included in the close-coupling expansion. This was a typical cutoff for all CC calculations in this article.

\section{Channel coupling in time delays}
\label{sect:chancoupl}

The conventional description of the RABITT two-photon process is compatible with the  single-channel picture: the photoelectron is released by the XUV photon, leaving a hole in its original orbital. This is followed by absorption or emission of the IR quantum by the very same electron.% In the single-channel picture, after all, there is no other object that could absorb any further energy.

However, this may not be a sufficient interpretation for multi-electron systems with multiple open channels. In this case, absorption of the IR photon can lead either to transition in the ion (i.e. between the various final channels) or to polarization (virtual excitation). Electron correlation is then responsible for redistributing the excitation among the different final photoionization channels. This means that while the absorption of the XUV photon results in a superposition of several ionic states coupled to the photoelectron wave function with the residual kinetic energy, interaction with the IR field further mixes these channels, proportionally to the dipole coupling between them. This is somewhat similar to the process investigated in~\cite{ResonantRABITT} (in the lithium atom), with the important distinction that in the latter case the resonant transition was occurring within the neutral target rather than in the residual ion.

\begin{figure}[htbp]
    \centering
    \includegraphics[width=0.75\columnwidth]{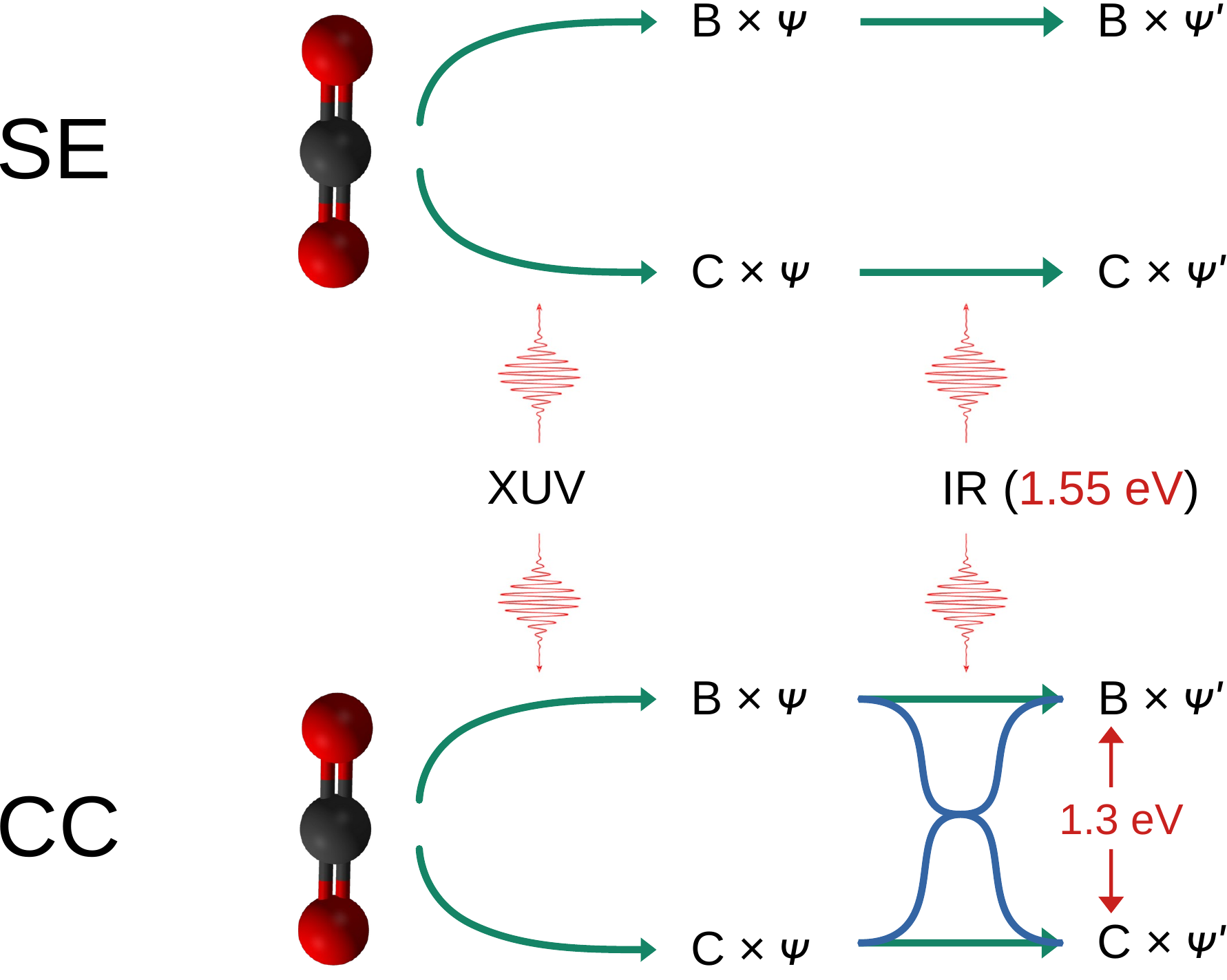}
    \caption{Absorption of XUV and IR photons in CO$_2$, indicating possible absorption pathways involving residual ion states B~${}^2\Sigma_u$ and C~${}^2\Sigma_g$, including the dipolar channel coupling between these two states during interaction with the IR field.}
    \label{fig:diagram}
\end{figure}

\begin{figure}[htbp]
    \centering
    \includegraphics[width=0.95\columnwidth]{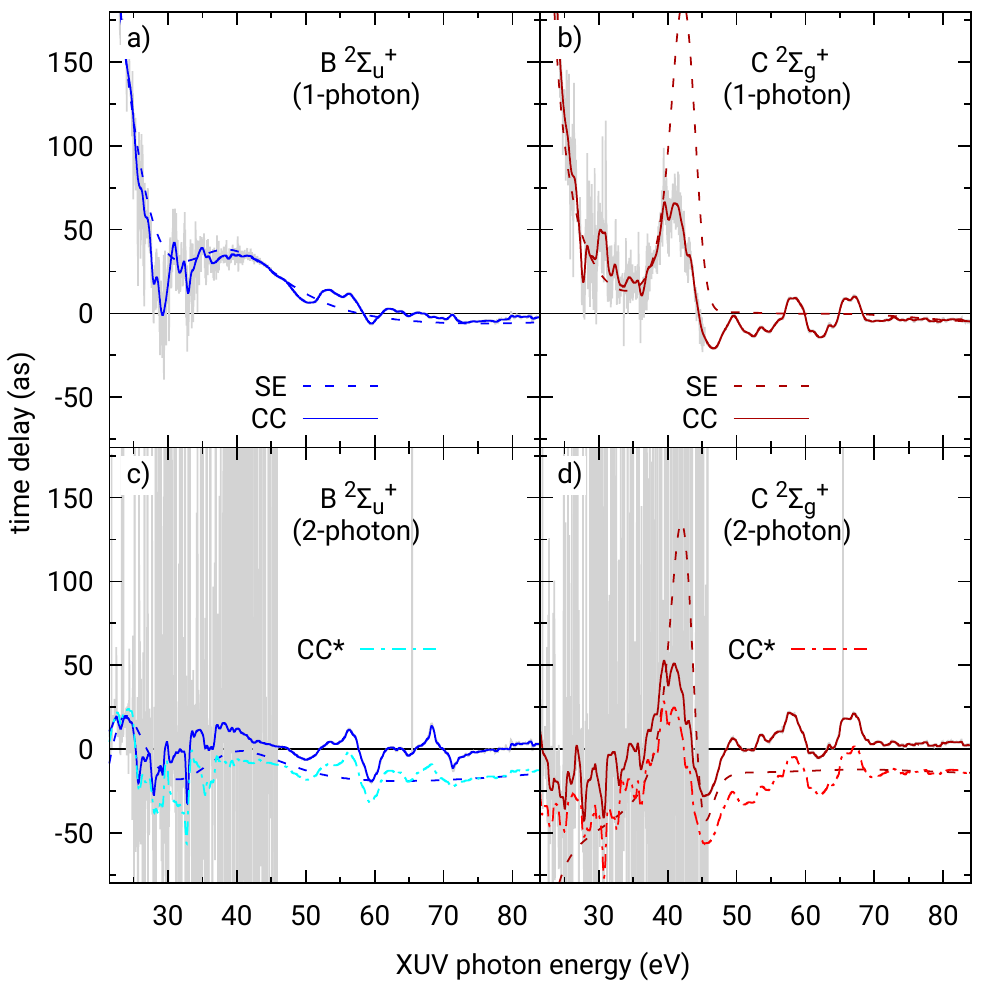}
    \caption{Comparison of one- and two-photon delays in ionization of unoriented CO$_2$ into states B~${}^2\Sigma_u^+$ and C~${}^2\Sigma_g^+$. Thicker solid curves correspond to smoothed CC results, dashed curves to SE results, chain curves to smoothed modified CC results (see text). Light grey curves are the unsmoothed CC delays.}
    \label{fig:CO2BC}
\end{figure}

As an example we investigate here the strongly coupled states B and C of CO$_2^+$, as illustrated in Fig.~\ref{fig:diagram}. When we inspect the 1-photon delays for ionization into state B and C, Fig.~\ref{fig:CO2BC}a--b, we see that the results are largely independent of the model, single-channel or coupled-channel. The only significant change is the shift to lower energies and change of size of the core-excited shape resonance in the C state, Fig.~\ref{fig:CO2BC}b, which is a known effect~\cite{CO2HHG}. This tells us that the absorption of the XUV photon alone is not sensitive to multi-electron effects, at least away from resonances. The wide structure in the B state with the centre at approximately 40~eV is related to onset of the \(p\) partial wave and is not of resonant character.

The similarity between the SE and CC calculations holds to some degree also for the 2-photon delays, Fig.~\ref{fig:CO2BC}c--d. However, here the CC results appear almost uniformly shifted from the SE results by approximately 20~as towards positive delays. The 2-photon SE results allow us to more easily identify the low-energy shape resonance in the B state~\cite{KamalovCO2}, which here appears centered at approximately 25~eV of XUV photon energy. In the CC model, it occurs at somewhat lower photon energies.

The calculated magnitude of the ion core transition dipole element \(D_{BC,z} = \langle \Phi_{\text{B}} | \sum_{i=1}^N z_i | \Phi_{\text{C}} \rangle \) that couples the residual ion states B and C together in the IR field is approximately equal to 1.0 atomic unit, which is a relatively strong coupling. For comparison, the only other lower CO$_2^+$ state that C is coupled to via a component of the dipole operator is the state A, with the transition dipole element magnitude of 0.09~a.u. Furthermore, the states B and C are separated by the energy 1.3~eV (see Tab.~\ref{tab:ionization-potentials}), which is very close to the energy 1.55~eV of the IR field used in the calculation. This means that in this case the IR field indeed very strongly couples the two states, producing a mixture of these two final states regardless of whether the initial XUV ionization resulted in one or the other. We verified this hypothesis by manually setting the $D_{BC,z}$ dipole coupling element to zero in the otherwise fully coupled calculation. The resulting time delays are marked as ``CC$^*$'' in Fig.~\ref{fig:CO2BC}c--d and clearly show much better agreement with the single-channel SE results than the delays obtained from the fully coupled original calculation.

The remaining discrepancies in the time delays for ionization into the C state below 30~eV can be ascribed to other couplings than the field-driven B-C ion core transition. The features in the energy range 50--70~eV, present also in the one-photon delays, are visible  in the one-photon cross sections and asymmetry parameters too~\cite{CO2HHG}. They might be unphysical pseudoresonances since at these energies the calculation is missing not only further singly ionic channels but also channels corresponding to double ionization of the molecule.

The effect of the field-driven coupling on the time delays can be quantified in the asymptotic theory, purely from the knowledge of the one-photon ionization amplitudes and the ion transition dipole element. As detailed in Appendix~\ref{sect:coupling}, the  two-photon ionization amplitude can be written as \(d^{(2)} = d_{\text{ion}}^{(2)} + d_{\text{pws}}^{(2)}\), see Eq.~\eqref{eq:matelemsplit}. Here the first term is the amplitude of absorption of the photon by the residual ion, proportional to the B--C transition dipole matrix element, while the second term corresponds to absorption by the photoelectron. The ion core transition is responsible only for a small correction of the dominant partial wave coupling term and \(d_{\text{ion}}^{(2)}\) can be regarded as a perturbation of \(d_{\text{pws}}^{(2)}\). Expanding Eq.~\eqref{eq:oavg2} to the first order in \(d_{\text{ion}}^{(2)}\) then yields
\begin{equation}
    \tau^{(2)} = \tau_{0}^{(2)} + \tau_{\text{coupl}}^{(2)} \,,
\end{equation}
where
\begin{equation}
    \tau_{0}^{(2)} = \frac{1}{2\omega} \arg Q_0 \,, \qquad
    \tau_{\text{coupl}}^{(2)} = \frac{1}{2\omega} \frac{\text{Im}\, \delta Q}{\text{Re}\, Q_0} \,,
\end{equation}
\begin{align}
    Q_0 &= \sum_{\substack{lm \\ abcd}} d_{\text{pws},+,flm,ab}^{(2)*} d_{\text{pws},-,flm,cd}^{(2)} A_{abcd} \,, \\
    \delta Q &= \sum_{\substack{lm \\ abcd}} d_{\text{ion},+,flm,ab}^{(2)*} d_{\text{pws},-,flm,cd}^{(2)} A_{abcd} \nonumber \\
    &+ \sum_{\substack{lm \\ abcd}} d_{\text{pws},+,flm,ab}^{(2)*} d_{\text{ion},-,flm,cd}^{(2)} A_{abcd} \,.
\end{align}
The partial two-photon ionization amplitudes \(d_{\text{xxx},\pm,flm,ab}^{(2)}\) and hence also the quantities \(Q_0\) and \(\delta Q_0\) can be approximately calculated from the asymptotic theory of Appendix~\ref{sect:coupling} that uses only the one-photon ionization amplitudes:
\begin{align}
    d_{\text{pws},\pm,flm,ab}^{(2)}
        &\approx -\mathrm{i}A_{\kappa_f k_f}^{\text{pws}} \sum_{l' m'} \langle l m | \hat{n}_a | l' m' \rangle d_{\pm,fl'm',b}^{(1)} \,, \\
    d_{\text{ion},\pm,flm,ab}^{(2)}
        &\approx -\sum_{n} A_{\kappa_n k_f}^{\text{ion}} D_{fn,a} d_{\pm,nlm,b}^{(1)} \,.
\end{align}
Here \(A_{\kappa k}\) are factors that depend only on the momentum \(\kappa\) and \(k\) of the photoelectron in the intermediate \(n\) and the final state \(f\) of the molecular ion, respectively, for the given absorption-absorption (\(+\)) or absorption-emission (\(-\)) ionization pathway. The quantities \(d_{nlm}^{(1)}\) are the terms of the partial wave expansion of ionization amplitude into the intermediate state \(n\). In the above two formulas, these one-photon amplitudes are recombined in terms of partial waves and final states, respectively. \(D_{fn,a}\) is the Cartesian \(a\)-component of the transition dipole between the states \(n\) and \(f\) of the residual ion.

\begin{figure}[htbp]
    \centering
    \includegraphics[width=0.95\columnwidth]{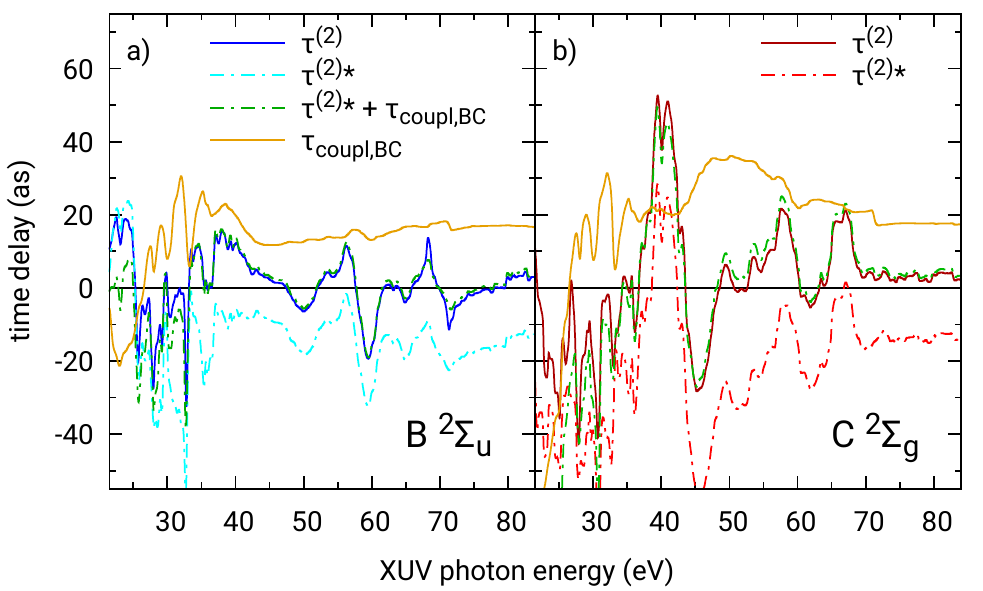}
    \caption{Contribution of the field-driven B--C coupling to the photoionization delays in CO$_2$ according to the asymptotic theory of Appendix~\ref{sect:coupling}. The curves labeled as \(\tau^{(2)}\) are obtained from the full two-photon method; asterisk marks results with excluded B--C residual ion coupling from the full theory. The quantity \(\tau_{\text{coupl,BC}}\) (yellow line in both panels) is the contribution to the time delay arising from the B--C coupling, as calculated in the asymptotic theory.}
    \label{fig:CO2-coupling-delays}
\end{figure}

In other words, the additional effect of the field-driven coupling between the two states extends the frequently used decomposition \(\tau^{(2)} \approx \tau_{\text{mol}} + \tau_{cc}\) with another term \(\tau_{\text{coupl}}\),
\begin{equation}
    \tau^{(2)} \approx \tau_{\text{mol}} + \tau_{cc} + \tau_{\text{coupl}} \,,
    \label{eq:tau2asy}
\end{equation}
or with more such terms when multiple residual ion states are field-coupled to the final state of interest. The effect of \(\tau_{\text{coupl}}\) on the results pertaining to B and C ion states of CO$_2^+$ is demonstrated in Fig.~\ref{fig:CO2-coupling-delays}. Other field-driven dipole couplings than this one are not considered. The figure shows that for sufficiently high photon energies this new term describes almost perfectly the difference between the CC and CC$^*$ datasets in Fig.~\ref{fig:CO2BC}, confirming consistency between the full second-order method discussed earlier and the asymptotic approach just introduced.

While the new term \(\tau_{\text{coupl}}\) is proportional to the transition dipole element between the final and the intermediate ion state, in the asymptotic approach this is the only way that the channel coupling explicitly enters the formulas~\eqref{eq:matelemdip}--\eqref{eq:Akkion}. If this transition dipole is obtained from some external source, \(\tau_{\text{coupl}}\) can be calculated even from an uncoupled photoionization model. In the studied case of CO$_2$, addition of  \(\tau_{\text{coupl}}\) to the SE results in Fig.~\ref{fig:CO2BC} would shift them upwards, making them compatible with the proper CC calculation. Only the  resonance features specific to the coupled model cannot be obtained in this way, since they are a manifestation of electron correlation.

\begin{figure}[htbp]
    \centering
    \includegraphics[width=0.95\columnwidth]{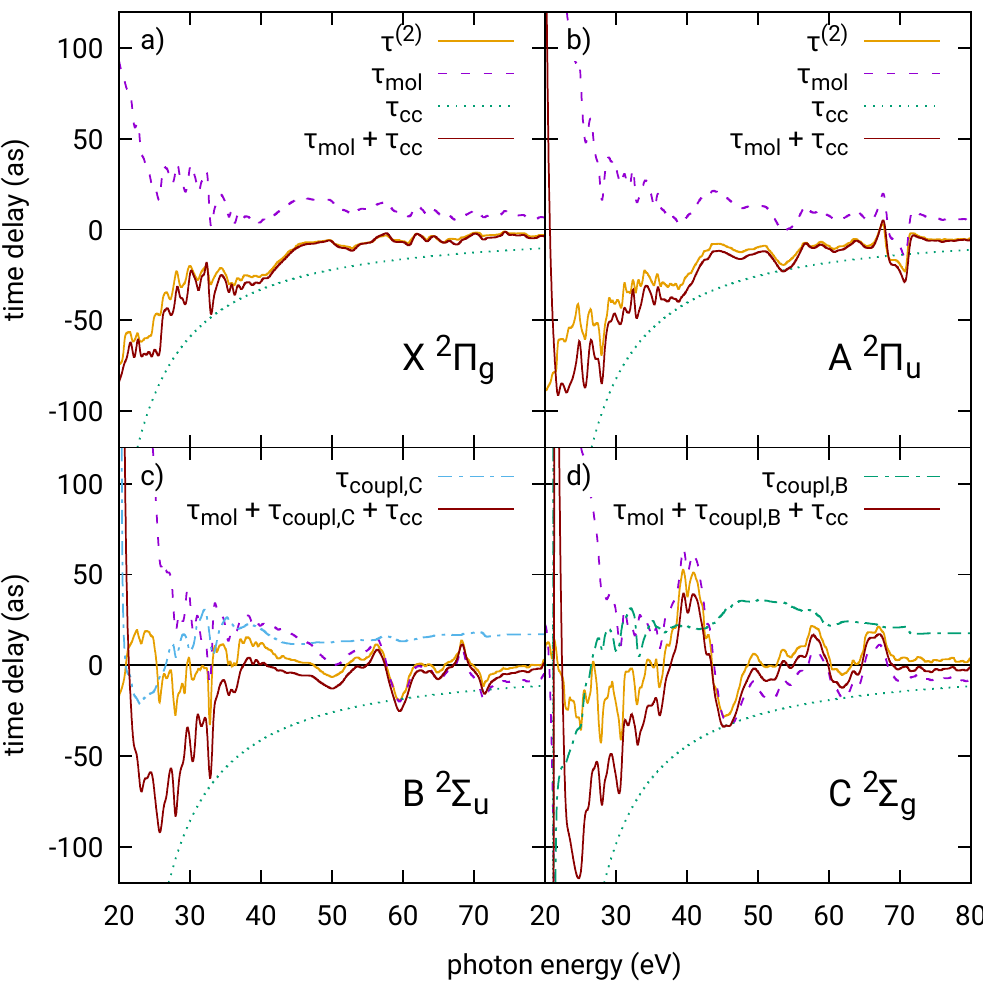}
    \caption{RABITT sideband delays for photoionization of CO$_2$ into its four lowest ion states (yellow light solid curve) compared to the asymptotic approximation (red dark solid curve). Individual terms of the asymptotic approximation are plotted as broken lines.}
    \label{fig:CO2-asy-delays}
\end{figure}

The complete asymptotic splitting of the delay given by Eq.~\eqref{eq:tau2asy} converges to the two-photon result in the limit of high energies. This is illustrated in Fig.~\ref{fig:CO2-asy-delays}. Inclusion of \(\tau_{\text{coupl}}\) for the states B and C is necessary due to their substantial dipolar coupling. However, even with the coupling delay accounted for the approximation for the excited states significantly deviates from the two-photon results at energies below 35~eV. At such energies only the complete multi-photon and multi-electron theory is appropriate and can reproduce the measurement in Fig.~\ref{fig:CO2-X-oavg}b.

%To reach higher energies, for the purposes of this section we replaced the original Gaussian continuum basis set with a set of 25 B-splines of order~6 (and used standard double precision). Other parameters of the calculation were left unchanged.

\section{Interference structures in time delays}
\label{sect:wpacket}

When compared to the time-dependent approach the time-independent R-matrix approach for calculation of 2-photon delays offers the advantage of an improved computational efficiency and a clearer interpretation in terms of channel-resolved amplitudes and their interferences. This allows for a detailed interpretation of some structures that are otherwise simply part of the total time-dependent ionization yield when a time-dependent method is used.

It was shown earlier~\cite{Lein,Fernandez} that---unlike atoms---in the case of molecules some partial wave components of oriented photoionization dipoles feature ``deep minima'' at specific energies. These minima are the result of destructive interference between electrons ionized from the individual atoms of the molecule. This picture is easily applicable to diatomics where it resembles a double-slit interference and leads to a formula for the minima relating interatomic distance to photoelectron energy or wavelength~\cite{NingH2p}.

As an example of use of the partial wave dipoles in assisting interpretation of the time-dependent picture, we investigate in detail the time evolution of parallel photoionization in H$_2$. Specifically, we focus on interpretation of the broad structure in the parallel photoionization time delays (Fig.~\ref{fig:H2-delays-SE}a) caused by the two-center interference.

\begin{figure*}
    \centering
    \includegraphics[width=0.85\textwidth]{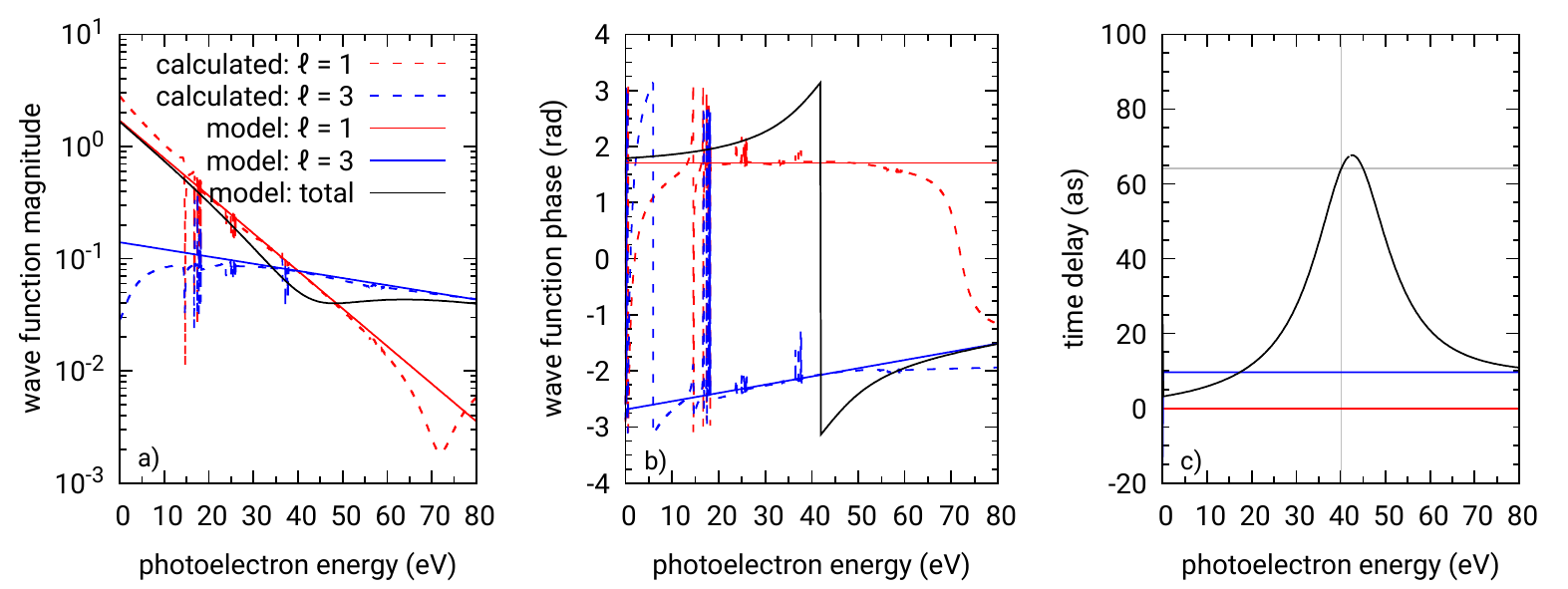}
    \caption{Left and center: Fit of the energy dependence of the magnitudes and phases respectively of the one-photon amplitudes of the axial ionization of H$_2$  by weak field polarized parallel with the molecular axis for the full CI model. Right: one-photon time delays calculated as the energy derivative of the combined transition dipole element.}
    \label{fig:fitting}
\end{figure*}

The \(p\)-wave component of the oriented photoionization dipole parallel with the molecular axis in the hydrogen molecule has a deep minimum at photoelectron energy of about 70~eV (see Fig.~\ref{fig:fitting}a), which means that at somewhat smaller kinetic energy it goes below the \(f\)-wave component that is otherwise smaller due to the higher centrifugal barrier that it has to overcome~\cite{CohenFano}. Consequently, these two partial wave dipoles have different phase-energy dependencies and so different corresponding partial wave photoionization delays. When the magnitude of the two components swaps, the phase of the total dipole element rapidly changes from being almost equal to that of the originally dominant \(p\)-wave to being very similar to that of the other partial wave. This swapping can be localized in a narrow region around the crossing energy and so the time delay, which is the energy derivative of the dipole element phase, will become very large. This is illustrated in Fig.~\ref{fig:fitting}f.

Note that in~\cite{NingH2p} where a similar structure in the parallel photoionization delays is discussed in H$_2^+$, it is ascribed to the minimum of the differential cross section. It is argued that ``\textit{the destructive interference suppresses the emission of the outgoing wave packet and the magnitude of the corresponding \(t_{\text{EWS}}\) delay is significantly increased.}''
While we agree that there is a connection to the minima in the differential cross sections, we feel this explanation is not accurate and not giving a detailed insight into the actual mechanism by which the interference could ``suppress'' a wave packet.

\begin{figure*}
    \centering
    \includegraphics[width=0.85\textwidth]{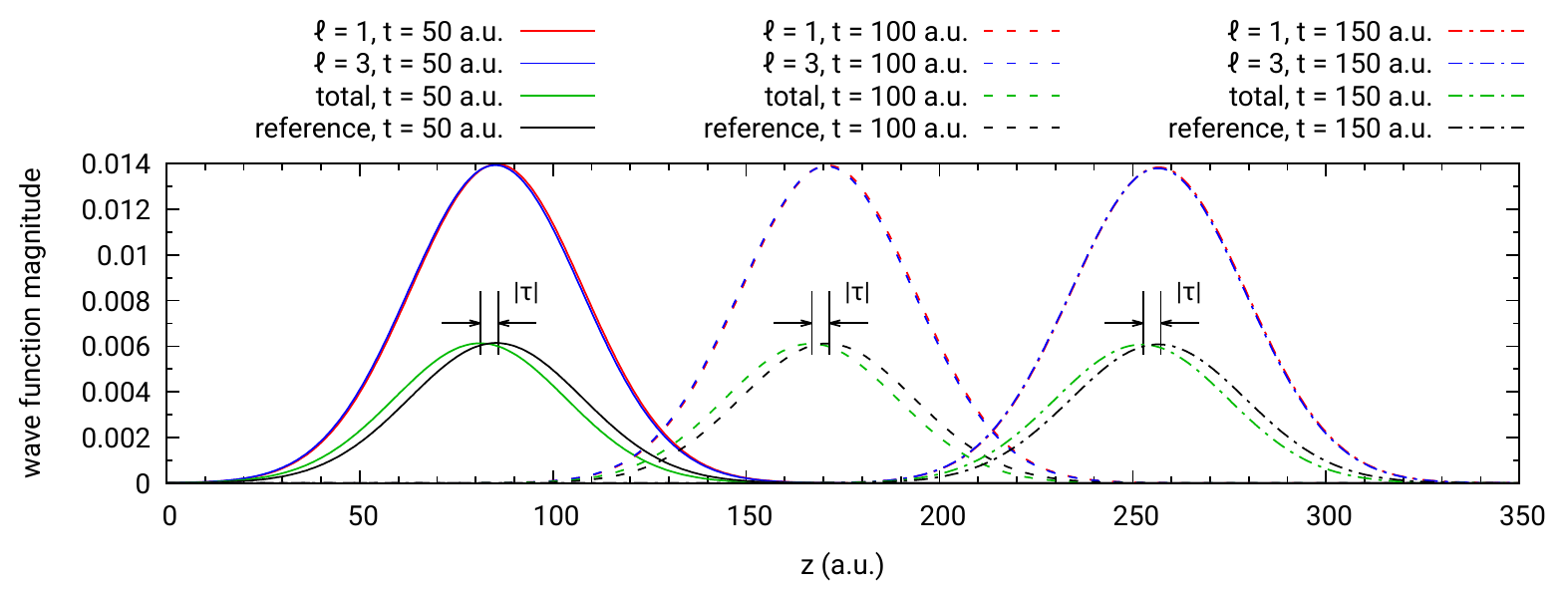}
    \caption{Ionized photoelectron wave packets at three times (\(t = 50\)~a.u., 100~a.u. and 150~a.u.), with spectral width parameter \(\lambda = 500\)~a.u. The magnitudes of the partial \(p\)-wave and \(f\)-wave components are plotted, as well as magnitude of their coherent and incoherent sum (``total'' vs.\ ``reference'', respectively, the latter not to scale). The apparent time delay arising in the asymmetric subtraction of the partial components corresponds to the time delay calculated in the time-independent approach.}
    \label{fig:wave-packets}
\end{figure*}

To further elaborate what this delay means in the time-dependent picture, we construct a simple model of the 1-photon ionization process. We irradiate the molecule with a spectrally narrow Gaussian pulse such that if the photoionization efficiency was constant, the resulting photoelectron wave packet would have the energy spectrum
\begin{equation}
    A(E_k) = \kappa \mathrm{e}^{-\lambda(k - k_0)^2} \,,
\end{equation}
where \(E_k = k^2/2\) is photoelectron kinetic energy, \(k_0\) is the central momentum and \(\kappa\) a suitable normalization constant. We know, however, that the photoionization has different efficiency at different energies, which is directly related to the photoionization amplitude (or dipole). We focus on the p- and f-components of the wave packet that propagates along the \(z\) axis
%The time-dependent wave packet that propagates along the axis \(z\) can  then be written as the combination
\begin{equation}
    \psi(z, t) = \psi_p(z, t) + \psi_f(z, t) \,,
\end{equation}
of the two contributing partial wave components
\begin{equation}
    \psi_l(z, t) = \int\limits_0^{+\infty} A(E_k) d_l^{(1)}(E_k) \mathrm{e}^{\mathrm{i}(kz - E_k t)} \mathrm{d}k. \label{eq:integp}
\end{equation}
Now, we could substitute for \(d_p^{(1)}(E_k)\) and \(d_f^{(1)}(E_k)\) the photoionization amplitudes (dipoles) calculated numerically. For simplicity, we will restrict ourselves to the narrow vicinity of the swapping energy (\(E_0\) = 40~eV) and perform a fit of the energy dependencies of the magnitude and phase of the dipoles; see Fig.~\ref{fig:fitting}. We end up with the following model
\begin{align}
    d_l^{(1)}(E_k) &= a_l \mathrm{e}^{\mathrm{i}\eta_l(E_k)} \mathrm{e}^{-\alpha_l E_k},
\end{align}
where the energy-dependent phase is
\begin{align}
    \eta_l(E_k) &= \eta_{l}^0 + (E - E_0)\tau_l
\end{align}
and the values of the constants are \(a_p = 1.70\), \(a_f = -2.1\), \(\alpha_p = 2.1\), \(\alpha_f = 0.4\), \(\eta_p^0 = 1.71\), \(\eta_f^0 = -2.10\), \(E_0 = 1.47\) (= 40~eV), \(\tau_p = 0\) and \(\tau_f = 0.4\). This model reproduces qualitatively the interference structure in the time delays in Fig.~\ref{fig:H2-delays-FCI-z}, as shown in Fig.~\ref{fig:fitting}c. For a sufficiently narrow spectrum, that is, large values of \(\lambda\), the integrations in Eq.~\eqref{eq:integp} can be extended to minus infinity, because the Gaussian envelope of the source pulse will not allow any significant contribution from negative momenta. The integral in Eq.~\eqref{eq:integp} can then be calculated analytically. The explicit forms of the wave packet~\eqref{eq:integp} corresponding to the model parameters is
\begin{align}
    \psi_l(z, t) &= \sqrt{\frac{\pi}{A_l}} \frac{D_l B_l}{2A_l} \mathrm{e}^{-C_l+\frac{B_l^2}{4A_l}} \,.
\end{align}
Here
\begin{align}
    &A_l = \lambda + \frac{\alpha_l}{2} + \mathrm{i}\frac{t - \tau_l}{2} \,, \quad &&C_l = \lambda k_0^2 \,, \\
    &B_l = - 2\lambda k_0 - \mathrm{i}z \,, \quad &&D_l = -a_l\kappa \mathrm{e}^{\mathrm{i}(\eta_l^0 - \tau_l E_0)}.
\end{align}
We can then find out the effect of the partial wave interference on the time delay by comparing the time-dependent position of the actual wave packet including the interference
\begin{equation}
    \rho_{\text{total}}(z, t) = | \psi_p(z,t) + \psi_f(z,t) |^2
\end{equation}
to the position of a reference wave packet, whose components have the same phase,
\begin{equation}
    \rho_{\text{reference}}(z, t) = \left| |\psi_p(z,t)| + |\psi_f(z,t)| \right|^2 \,.
\end{equation}
The comparison of the wave packets (for \(\lambda = 500\)) is plotted in Fig.~\ref{fig:wave-packets}. Even though the centres of the partial wave packets are very close to each other, the maximum of the total combined wave packet has its centre significantly shifted towards the origin and hence the wave packet appears retarded on the detector when compared to the reference wave packet whose components do not interfere. This apparent retardation can be directly calculated in this model case by evaluating the position of the peak of \(\rho_{\text{total}}(z,t)\) for a given time. This is then compared to the expected position \(z_0(t) = k_0 t\), where \(k_0 = \sqrt{2E_0}\). Here we assume that the peak is located at \(z(t) = z_0(t) + \Delta z\), with \(\Delta z\) being a small offset with respect to \(z_0(t)\). Simplification of the expression \(\rho_{\text{total}}(z,t)\) to the first order in \(\lambda^{-1}\) and localization of its maximum with respect to \(\Delta z\) leads to a \(\lambda\)-independent result
\begin{equation}
    \Delta z \simeq -\frac{1}{2} k_0 (\tau_p + \tau_f) + \frac{1}{2} k_0 (\alpha_p - \alpha_f) \tan \frac{\eta_p^0 - \eta_f^0}{2} \,.
\end{equation}
This is related to the apparent time delay \(\tau \simeq -\Delta z/k_0\) of the combined wave packet:
\begin{equation}
    \tau \simeq \frac{1}{2} (\tau_p + \tau_f) - \frac{1}{2} (\alpha_p - \alpha_f) \tan \frac{\eta_p^0 - \eta_f^0}{2} \,.
    \label{eq:tauwpkt}
\end{equation}
For the above-given parameters of the model, we get \(\tau \doteq 66\)~as, in good agreement with the value of 64~as obtained directly from the time-independent approach (highlighted by the grey cross-hair in Fig.~\ref{fig:fitting}c). We see from Eq.~\eqref{eq:tauwpkt} that when the phases \(\eta_p^0\) and \(\eta_f^0\) of the partial dipoles are the same at the crossing energy \(E_0\), the resulting time delay corresponds to a simple mean of the partial time delays of the two wave packets. The explanation is that in this case the wave packets do not interfere, the magnitudes of the wave packets are additive and the centre of the combined wave packet corresponds to a simple mean of the two constituent wave packets' centres.

The same is true when the local energy dependence of the magnitude of the partial dipoles at \(E_0\) are equal, \(\alpha_p = \alpha_f\). In the wave packet picture this also makes sense: The overall phase of the sum of the two partial wave packet contributions of the same magnitude will be the mean of their individual phases, \(\arg (\exp \mathrm{i}\eta_p + \exp \mathrm{i}\eta_f) = (\eta_p + \eta_f)/2\). And this in turn leads to the mean time delay \(\tau = (\tau_p + \tau_f)/2\) for the combined wave packet.

To sum up, the partial wave interference is responsible for asymmetric subtraction of the resulting wave packet, rather than ``suppression''~\cite{NingH2p} of the wave packet as a whole. It is now obvious that the same mechanism can also result in \textit{advancing} the wave packet, i.e.\ in negative time delays, which is what occurs in the perpendicular ionization of N$_2$ in Fig.~\ref{fig:N2-delays-SE}b. Nevertheless, as commented in~\cite{Pazourek}, observation of these interference delay peaks is experimentally challenging, because they are located angularly close to minima in the differential cross section.

\section{Conclusion}

In this article we investigated RABITT time delays calculated from full 2-photon amplitudes by the time-independent multi-photon R-matrix method. This method offers significant advantages compared to typical methods used in the field: it is computationally more efficient than the solution of the time-dependent Schrödinger equation and provides accurate second-order results for the time delays. We have explicitly verified that the conventional asymptotic theory (\(\tau^{(2)} \approx \tau_{\text{mol}} + \tau_{cc}\)) based on separability of the individual contributions is insufficient at low energies. As expected, the asymptotic theory is accurate at higher photoelectron energies but the threshold for its applicability depends on the target and ranges from a few eV of photoelectron energy for H$_2$ up to a few dozen for CO$_2$. Depending on the target and the IR photon energy, the RABITT delay may contain an additional contribution, \(\tau_{\text{coupl}}\), coming from field-induced ion-state coupling. In contrast to that, as expected, the second-order time-independent theory provides time delays in agreement with results obtained by solution of the time-dependent Schrödinger equation all the way to the threshold.

We calculated orientationally averaged absolute and/or relative time delays for N$_2$, CO$_2$, H$_2$O and N$_2$O, finding a very good agreement with published measurements for N$_2$ and CO$_2$, a fair agreement for H$_2$O, but a very poor one for N$_2$O, despite reproducing the experimental one-photon ionization cross sections very well. This implies that either the measurements~\cite{Huppert} should be revisited or our theory is failing to account for other effects that might play a significant role such as nuclear motion.

The comparison between single-channel and close-coupling calculations suggests that electron correlation significantly alters the time delays at photon energies even as high as 40~eV for some of the studied molecules. This is particularly noticeable around shape resonances in ionization of N$_2$ and CO$_2$.

We further discussed in detail the effect of field-driven coupling of ionization channels. We demonstrated that in the case of CO$_2$ the ion states B and C are strongly coupled by typical IR fields, resulting in a significant deviation of the calculated (and measurable) time delays in the coupled model from the results of the single-channel model. Based on this observation we have extended the widely used asymptotic approximation~\cite{BaykushevaWorner} for the two-photon time delays \(\tau^{(2)} = \tau_{\text{mol}} + \tau_{cc}\) to include an additional delay \(\tau_{\text{coupl}}\) that describes the effect of the field-driven coupling of the final residual ionic states. This correction can be used together with the single-channel models, provided that the  residual ion dipole transition element is obtained from some external source.

Finally, we inspected the origin of the structures in the time delays for oriented molecules. We demonstrate the connection between the 1-photon time delay and the shape of the photoelectron wave packet: the destructive partial wave interference alters the shape of the wave packet, resulting in its apparent retardation or advancement. In H$_2$, this interference is a coincidence caused by a deep minimum in the \(p\)-wave partial photoionization cross section. In N$_2$ it is caused by convergence of the \(p\) and \(f\) partial wave cross sections, with additional effect of the \(l = 5\) partial wave.

\section{Acknowledgements}

JB would like to thank Prof. Hugo van der Hart for the initial impulse into this research. Also, his advice---as well as that of Dr. Andrew Brown and of Dr. Jack Wragg from Queen's University Belfast---on running time-dependent simulations of RABITT in RMT is gratefully acknowledged.

JB and ZM further acknowledge support of the Czech Science Foundation as the project GA CR 20-15548Y and support of the PRIMUS program of Charles University as the project PRIMUS/20/SCI/003. Some computational resources were supplied by the project ``e-Infrastruktura CZ'' (e-INFRA LM2018140) provided within the program Projects of Large Research, Development and Innovations Infrastructures. This work was also supported by the Ministry of Education, Youth and Sports of the Czech Republic through the e-INFRA CZ (ID:90140).

JB and JDG were supported by EPSRC under grant No.\ EP/P022146/1. JDG also acknowledges support from the UK-AMOR consortium funded by EPSRC (EP/R029342/1).

\begin{appendix}

\section{Levin quadrature}
\label{sect:levin}

Levin~\cite{Levin} introduced an efficient quadrature method for highly oscillatory integrands that takes advantage of some additional knowledge about the integrated function. The method can be used to numerically evaluate  integrals of the form
\begin{equation}
    \mathcal{I} = \int\limits_a^b \bm{f}(r) \cdot \bm{w}(r) \mathrm{d}r \,,
\end{equation}
where components of \(\bm{w}(r)\) strongly oscillate, but components of \(\bm{f}(r)\) do not. Moreover, there has to be a known rectangular matrix \(\mathsf{A}(r)\) that relates the oscillating part of the integrand to its derivative, \(\bm{w}'(r) = \mathsf{A}(r)\bm{w}(r)\). Then the result of the integral can be written as a difference of surface terms
\begin{equation}
    \mathcal{I} = \bm{p}(b) \cdot \bm{w}(b) - \bm{p}(a) \cdot \bm{w}(a) \,,
    \label{eq:surface}
\end{equation}
provided that the auxiliary set of functions \(\bm{p}(r)\) satisfies the coupled differential equations
\begin{equation}
    \bm{p}'(r) + \mathsf{A}^\top(r) \bm{p}(r) = \bm{f}(r)
    \label{eq:levindiffeq}
\end{equation}
with any boundary conditions; the homogeneous solution does not contribute to Eq.~\eqref{eq:surface}.
In our RABITT calculations, the integrals of interest have always the form
\begin{equation}
    \mathcal{I} = \int\limits_a^b r^m H_{l_1}^{s_1}(\eta_1, k_1 r) H_{l_2}^{s_2}(\eta_2, k_2 r) \mathrm{d}r \,,
    \label{eq:integ}
\end{equation}
where \(H_l^{\pm}(\eta,\rho)\) is the Coulomb-Hankel function and \(\eta_i = -1/k_i\) for singly charged cations. Coulomb functions are generally expensive to evaluate numerically, so finding a quadrature scheme that avoids evaluating them too often is very beneficial. Choosing
\begin{equation}
    \bm{f}(r) =
    \left(\begin{matrix}
        r^m \\
        0 \\
        0 \\
        0
    \end{matrix}\right),
    \ \bm{w}(r) =
    \left(\begin{matrix}
        H_{l_1}^{s_1}(\eta_1, k_1 r)   H_{l_2}^{s_2}(\eta_2, k_2 r) \\
        H_{l_1}^{s_1}(\eta_1, k_1 r)   H_{l_2+1}^{s_2}(\eta_2, k_2 r) \\
        H_{l_1+1}^{s_1}(\eta_1, k_1 r) H_{l_2}^{s_2}(\eta_2, k_2 r) \\
        H_{l_1+1}^{s_1}(\eta_1, k_1 r) H_{l_2+1}^{s_2}(\eta_2, k_2 r)
    \end{matrix}\right)
\end{equation}
makes it possible to construct the matrix \(\mathsf{A}(r)\) from the known recurrence relations for the Coulomb-Hankel functions~\cite{Powell}.

To make the solution of the differential equation~\eqref{eq:levindiffeq} fast, we expand the unknown components of \(\bm{p}(r)\) in terms of Chebyshev polynomials of some given order and solve the set of equations as a ``collocation condition''. That is, we require that the equations hold in some chosen set of discrete points. In this case the Chebyshev nodes of the same order as the interpolating polynomials were used as the collocation points. The differential equations become a small set of algebraic equations, which is solved using standard LAPACK routines. The use of Chebyshev polynomials as the interpolating basis set is advantageous because it avoids the Runge phenomenon~\cite{olver_2007} by a dense distribution of nodes towards the edges of the interval. For all calculations in this article we used the Chebyshev order of~5.

For transitions from closed to open channels, one of the oscillating Coulomb-Hankel functions in~\eqref{eq:integ} becomes the exponentially decreasing real-valued Whittaker function. In such a case, it is used as the factor \(\bm{f}(r)\), leaving \(\bm{w}(r)\) with two independent components only. This reduces the rank of the linear system by half.

The recurrence relations for Coulomb functions always diverge in some direction. Levin quadrature employs both directions, so the solution of Eq.~\eqref{eq:levindiffeq} will always contain an exponentially increasing component, potentially making numerical calculations unstable. However, this is easy to avoid by using a fixed-order \textit{adaptive} variant of the method. The integration interval is always divided in half, quadrature estimates are calculated in both half-intervals and compared to the estimate for the whole interval. When the difference is not significant, the result is considered converged. Otherwise, recursive subdivisions are done until the fixed order becomes accurate enough and convergence is reached. Overall, for the present application Levin quadrature achieves comparable accuracy to Romberg quadrature with at least 100 times fewer evaluations of the Coulomb functions. The special functions need to be explicitly calculated only in the endpoints of all sub-intervals.

\section{Molecular delay vs one-photon delay}
\label{sect:taumol}

\begin{figure}[htbp]
    \centering
    \includegraphics[width=0.95\columnwidth]{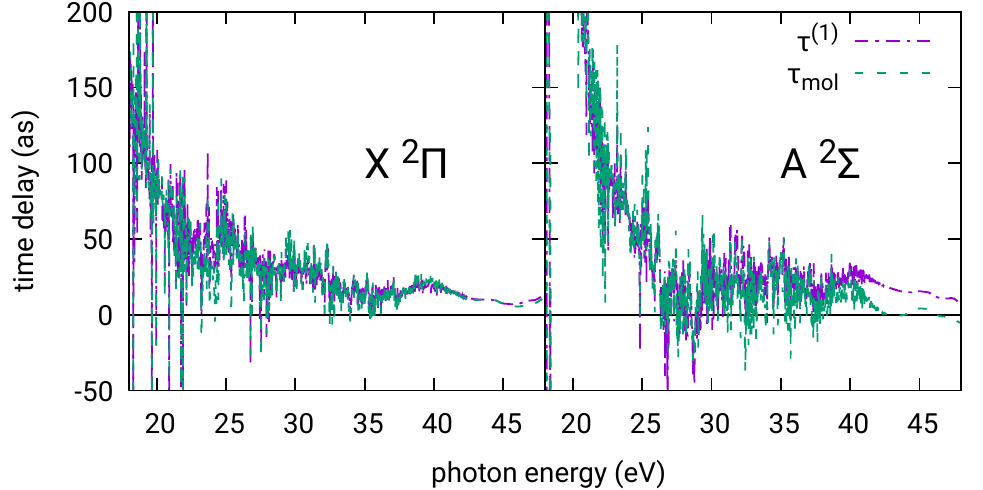}
    \caption{Comparison of unsmoothed unoriented one-photon delays \(\tau^{(1)}\) and molecular delays \(\tau_{\text{mol}}\) for ionization of N$_2$O into the ground and the first excited state of N$_2$O$^+$.}
    \label{fig:taumol-vs-tau1}
\end{figure}

Bakusheva and Wörner~\cite{BaykushevaWorner} define the two-photon molecular delay in their equation (25) as
\begin{equation}
    \tau_{\text{mol}}(2q,\hat{\bm{k}},\hat{R}_\gamma) = \frac{1}{2\omega} \arg \left[ b_{2q-1}^* b_{2q+1} \right]
    \label{eq:tmol}
\end{equation}
by means of the quantity
\begin{equation}
    b_{2q\pm1}(\hat{\bm{k}}, \hat{R}_\gamma) = \sum_{LM} b_{2q\pm1,LM}(\hat{R}_\gamma) Y_{LM}(\hat{\bm{k}}) \,,
    \label{eq:bdef}
\end{equation}
where, by their equation (19),
\begin{align}
    &b_{2q\pm1,LM}(\hat{R}_\gamma) = \sqrt{\frac{4\pi}{3}} (-1)^{m_2 + 1} E_{-m_2}^{\text{IR}} \nonumber \\
    &\times \sum_{\rho\rho'\lambda\mu} I_{\lambda \mu\rho'} \langle Y_{LM} | Y_{1\rho} | Y_{\lambda\mu} \rangle
    \mathcal{D}_{\rho'm_1}^{(1)}(\hat{R}_\gamma) \mathcal{D}_{\rho m_2}^{(1)}(\hat{R}_\gamma) \,.
\end{align}
Here the quantum numbers \(m_1\) and \(m_2\) are the laboratory-frame components of the XUV and IR fields, respectively. They are both set to zero for fields with identical linear polarization.
The sum over \(L\) and \(M\) in Eq.~\eqref{eq:bdef} invokes the resolution of identity. Combined with equations (3) and (4) from~\cite{BaykushevaWorner}, it yields
\begin{equation}
    b_{2q\pm1}(\hat{\bm{k}}, \hat{R}_\gamma) = -(\hat{\bm{k}} \cdot \bm{\epsilon}^{\text{IR}}) I_{i,f}(2q\pm1, \hat{\bm{k}}, \hat{R}_\gamma) \,,
    \label{eq:bsimple}
\end{equation}
where \(I_{i,f}\) is the matrix element of one-photon ionization from the initial state \(i\) to final state \(f\). In the present notation in the molecular frame it is
\begin{equation}
    I_{i,f}(2q \pm 1, \hat{\bm{k}}) = \bm{\epsilon}^{\text{XUV}} \cdot \sum_{lmq} \bm{d}_{\mp,flm}^{(1)} X_{lm}(\hat{\bm{k}}) \,.
\end{equation}
Apart from the irrelevant sign factor in Eq.~\eqref{eq:bsimple}, the complex phase of the quantity \(b\) is identical to the complex phase of the one-photon ionization matrix element. For a fixed orientation of the molecule with respect to the field and for a fixed photoelectron emission direction, the formula~\eqref{eq:tmol} then necessarily yields the discrete-derivative approximation to the one-photon delays, Eq.~\eqref{eq:tau1}. However, due to the additional angular weight factor \(\hat{\bm{k}} \cdot \bm{\epsilon}^{\text{IR}}\) arising from the asymptotic theory, the emission- and orientation-averaged \(\tau_{\text{mol}}\) may contain some features of the full second-order expression, Eq.~\eqref{eq:oavg2}, in addition to the plain one-photon delays given by Eq.~\eqref{eq:oavg1}. If we integrate the product \(b_{2q-1}^* b_{2q+1}\) in Eq.~\eqref{eq:tmol} over emission directions \(\hat{\bm{k}}\) and average over polarization directions \(\bm{\epsilon} = \bm{\epsilon}^{\text{XUV}} = \bm{\epsilon}^{\text{IR}}\), we obtain
\begin{equation}
    \tau_{\text{mol}} = \frac{1}{2\omega} \arg \sum_{\substack{lmp\\ l'm'p'}} d_{+,flmp}^{(1)*} d_{-,fl'm'p'}^{(1)}
    \sum_{qq'} A_{pp'qq'} B_{ll'mm'qq'} \,,
    \label{eq:tmolavg}
\end{equation}
where \(A_{pp'qq'}\) was given in Eq.~\eqref{eq:Aabcd}, while
\begin{align}
    B_{ll'mm'qq'}
    &= \int \hat{k}_q \hat{k}_{q'} X_{lm} X_{l'm'} \, \mathrm{d}^2\hat{\bm{k}} \nonumber \\
    &= \frac{4\pi}{3} \sum_{\lambda \mu} \mathcal{G}^{11\lambda}_{qq'\mu} \mathcal{G}^{\lambda ll'}_{\mu mm'} \,.
\end{align}
The symbol \(\mathcal{G}_{m_1 m_2 m_3}^{l_1 l_2 l_3}\) denotes the full angular integral of a product of three real spherical harmonics.
Equation~\eqref{eq:tmolavg} illustrates that the averaged \(\tau_{\text{mol}}\) sits, in terms of complexity, somewhere between \(\tau^{(1)}\) and \(\tau^{(2)}\), cf.\ Eqs.~\eqref{eq:oavg1} and~\eqref{eq:oavg2}. However, this difference alone does not seem to account for the disagreement in Fig.~\ref{fig:N2O-delays-CAS}. When we use the same 1-photon ionization dipoles \(d_{\pm,lmq}^{(1)}\) to calculate \(\tau_{\text{mol}}\), using Eq.~\eqref{eq:tmolavg}, and \(\tau^{(1)}\), we obtain very similar results, see Fig.~\ref{fig:taumol-vs-tau1}. The only discernible difference is visible for the A-state around 40~eV of photon energy, but not as massive as Fig.~\ref{fig:N2O-delays-CAS}a,c suggest, so the disagreement between our results and those of Huppert~\textit{et al.}~\cite{Huppert} must be coming from a different molecular description.

\section{Channel coupling in asymptotic theory}
\label{sect:coupling}

In their theoretical exposition of molecular delays, Kamalov \textit{et al.}~\cite{KamalovCO2} reach Eq.~\eqref{eq:tmol} too but with a formula for \(b\) that involves the \(S\)-matrix, at variance with Baykusheva and Wörner~\cite{BaykushevaWorner}. However, their starting point is the boundary condition
\begin{align}
    &\langle \bm{r} | \Psi_{\bm{k},n}^{(-)} \rangle \overset{r \rightarrow \infty}{\longrightarrow} \sum_{LM} \mathrm{i}^L \mathrm{e}^{-\mathrm{i}\sigma_L}
    \sum_{plm} Y_{l}^{m}(\hat{\bm{k}}) \frac{\mathrm{i}}{r\sqrt{\pi k_p}} \nonumber \\
    &\times 
    \left(\mathrm{e}^{-\mathrm{i}\phi_p(r)} \delta_{\substack{nLM\\ plm}} - \mathrm{e}^{+\mathrm{i}\phi_p(r)} S_{\substack{nLM\\ plm}}^*\right)
    Y_{l}^{m}(\hat{\bm{r}}) | \Phi_p \rangle \,,
\end{align}
which is incorrect. Additionally, note that their $h^\pm$ are actually asymptotically proportional to $e^{\mp i \phi_p(r)}$. The correct  stationary photoionization boundary condition given by Burke~\cite{Burke} is
\begin{align}
    &\langle \bm{r} | \Psi_{\bm{k},n}^{(-)} \rangle \overset{r \rightarrow \infty}{\longrightarrow} \sum_{LM} \mathrm{i}^L \mathrm{e}^{-\mathrm{i}\sigma_L} Y_{L}^{M*}(\hat{\bm{k}})
    \sum_{plm} \frac{-\mathrm{i}}{r\sqrt{2\pi k_p}} \nonumber \\
    &\times
    \left(\mathrm{e}^{+\mathrm{i}\phi_p(r)} \delta_{\substack{plm\\ nLM}} - \mathrm{e}^{-\mathrm{i}\phi_p(r)} S_{\substack{plm\\ nLM}}^*\right)
    Y_{l}^{m}(\hat{\bm{r}}) | \Phi_p \rangle \,.
    \label{eq:psiasy}
\end{align}
In both equations \(\phi_p(r) = k_p r + k_p^{-1} \ln 2k_pr - \pi l_p /2 + \sigma_{l_p}(k_p)\) is the total asymptotic phase of a Coulomb wave function and \(\Phi_p\) are the bound states of the ion. The wave function of the intermediate state after absorption of the first photon is
\begin{equation*}
    | \Psi_{i+\Omega}^{(+)} \rangle \approx
    \sum_{i_n} \int \frac{| \Psi_{\bm{\kappa}_n,i_n}^{(-)}\rangle \langle \Psi_{\bm{\kappa}_n,i_n}^{(-)} | D(\bm{\epsilon}^{\text{XUV}}) | \Psi_i \rangle}{E_i + \Omega - E_{\kappa_n} + \mathrm{i}0} \mathrm{d}^2 \hat{\bm{\kappa}}_n \mathrm{d}E_{\kappa_n} \,,
\end{equation*}
where we neglected the contribution of the bound spectrum to the resolution of the Green's operator of Eq.~\eqref{eq:schrinterm}. When we perform the integral over  photoelectron emission directions \(\hat{\bm{\kappa}}_n\), we obtain an  approximation for \(\langle \bm{r} | \Psi_{i + \Omega}^{(+)} \rangle\):
\begin{gather}
    \sum_{npq} \int
    \frac{\frac{1}{r}F_{pn}^{(-)} Y_{l_p}^{m_p} |\Phi_{i_p}\rangle \langle \frac{1}{r} F_{qn}^{(-)} Y_{l_q}^{m_q} \Phi_{i_q} | D(\bm{\epsilon}^{\text{XUV}}) | \Psi_i \rangle}{E_i + \Omega - E_{\kappa_n} + \mathrm{i}0} \mathrm{d}E_{\kappa_n} \,, \nonumber \\
    F_{qn}^{(-)}(r) = \frac{-\mathrm{i}}{\sqrt{2\pi \kappa_q}}
    \left(\mathrm{e}^{+\mathrm{i}\phi_q(r)} \delta_{qn} - \mathrm{e}^{-\mathrm{i}\phi_q(r)} S_{qn}^*\right) \,.
    \label{eq:Fqn}
\end{gather}
Due to Eq.~\eqref{eq:Fqn}, the above integral over channel energies \(E_{\kappa_n}\) consists of two terms with opposite signs in the exponentials. Denoting \(E_{\kappa_0} = \kappa_0^2/2 = E_i + \Omega\), the individual integrals can be schematically written as
\begin{align}
    &\int\limits_0^{+\infty} \frac{\mathrm{e}^{\pm\mathrm{i}\phi(r)} h(E_\kappa) \mathrm{d}E_\kappa}{E_i + \Omega - E_\kappa + \mathrm{i}0}
    =
    2\int\limits_0^{+\infty} \frac{\mathrm{e}^{\pm\mathrm{i}\phi(r)} h(E_\kappa)}{\kappa_0^2 - \kappa^2 + \mathrm{i}0} \mathrm{d}E_\kappa
    \nonumber \\
    &= \int\limits_0^{+\infty} \left(
    \frac{\mathrm{e}^{\pm\mathrm{i}\phi(r)} h(E_\kappa)}{\kappa_0 - \kappa + \mathrm{i}0}
    + 
    \frac{\mathrm{e}^{\pm\mathrm{i}\phi(r)} h(E_\kappa)}{\kappa_0 + \kappa + \mathrm{i}0}
    \right) \mathrm{d}\kappa
    \,,
    \label{eq:integrals}
\end{align}
where the decomposition into partial fractions has been used and \(h(E_\kappa)\) denotes omitted factors. As we are operating in the limit \(r \rightarrow +\infty\), the exponentials oscillate extremely fast in \(\kappa\) and the only non-negligible contributions to the integral must be coming from intervals where the remaining part of the integrand varies sufficiently quickly as well, i.e. around the simple pole. This means that of the two integrals in Eq.~\eqref{eq:integrals} only the first one, with \(\kappa_0 - \kappa\) in the denominator, is of interest, whereas the other one with the always positive denominator can be neglected. For the same reason, we can extend the lower integration bound of the first integral to minus infinity. Then it is possible to invoke the residue theorem for contour integration in the \(\kappa\)-plane, giving eventually
\begin{equation}
    \int\limits_0^{+\infty} \frac{\mathrm{e}^{\pm\mathrm{i}\phi(r)} h(E_\kappa) \mathrm{d}E_\kappa}{E_i + \Omega - E_\kappa + \mathrm{i}0}
    \approx \begin{cases}
    -2\pi\mathrm{i} \mathrm{e}^{+\mathrm{i}\phi(r)} h(E_{\kappa_0}) \\
    0
    \end{cases}
\end{equation}
for the positive and the negative exponential, respectively. Thus, of the two terms in \eqref{eq:Fqn} only the one with the positive exponential, diagonal in channels and partial waves, remains after the integration over \(E_\kappa\). Due to the channel diagonality, the resulting form of the intermediate state wave function greatly simplifies to
\begin{gather}
    \langle \bm{r} | \Psi_{i+\Omega}^{(+)} \rangle \approx \sum_{i_n l_n m_n} \frac{-2\pi \mathrm{e}^{+\mathrm{i}\phi_n(r)}}{r\sqrt{2\pi\kappa_n}} Y_{l_n}^{m_n}(\hat{\bm{r}}) | \Phi_{i_n} \rangle
    f_{i,n}
    \label{eq:interm} \\
    f_{i,n} = \sum_{i_q l_q m_q} \langle \frac{1}{r} F_{qn}^{(-)} Y_{l_q}^{m_q} \Phi_{i_q} | D(\bm{\epsilon}^{\text{XUV}}) | \Psi_i \rangle \,,
\end{gather}
consistently with Eq.~(10) of~\cite{multiphoton} where the channel amplitude becomes \(a_n = -\sqrt{2\pi/\kappa_n} f_{i,n}\).

The approximate two-photon transition matrix element after absorption or emission of the second photon by the photoelectron then follows from Eq.~\eqref{eq:dip} as
\begin{equation}
    \langle \Psi_{\bm{k}, i_f}^{(-)} | \bm{\epsilon}^{\text{IR}} \cdot \bm{r}_N  | \Psi_{i+\Omega}^{(+)} \rangle
    \approx A_{\kappa k}^{\text{pws}} \sum_{l_f m_f} b_{i_f l_f m_f} Y_{l_f}^{m_f}(\hat{\bm{k}}) \,,
    \label{eq:matelem}
\end{equation}
where
\begin{align}
    b_{i_f l_f m_f} &= -\mathrm{i} \sum_{l_p m_p} \bm{\epsilon}^{\text{IR}} \cdot  \langle l_f m_f | \hat{\bm{r}} | l_p m_p \rangle
    \bm{\epsilon}^{\text{XUV}} \cdot \bm{d}_{i_f l_p m_p}^{(1)} \,, \label{eq:blfmf} \\
    A_{\kappa k}^{\text{pws}} &\approx \frac{\mathrm{e}^{-\pi/2\kappa + \pi/2k}}{\sqrt{k \kappa} |\kappa - k|^2}
    \frac{(2\kappa)^{\mathrm{i}/\kappa}}{(2k)^{\mathrm{i}/k}}
    \frac{\Gamma(2 + \mathrm{i}/\kappa - \mathrm{i}/k)}{(\kappa - k)^{\mathrm{i}(1/\kappa - 1/k)}} \,.
\end{align}
In this second absorption step we followed the asymptotic theory of Dahlström \textit{et al.}~\cite{Dahlstrom}. Only the slowly oscillating part of the product \(\Psi_{\bm{k},i_f}^{(-)*} \Psi_{i+\Omega}^{(+)}\) contributes significantly to the matrix element~\eqref{eq:matelem}. Because \(\Psi_{i+\Omega}^{(+)}\) contains an exponential with a positive phase, the conjugated \(\Psi_{\bm{k},i_f}^{(-)*}\) has to contribute an exponential with a negative phase. This once again discards the term containing the \(S\)-matrix. The missing factor ``i'' in Eq.~\eqref{eq:bsimple} with respect to Eq.~\eqref{eq:blfmf} is consistent with a different choice of the overall phase of the second-order matrix element in~\cite{BaykushevaWorner}.

To sum up, using the correct multichannel photoionization stationary state leads to the same results as obtained by Baykusheva and Wörner, who took a shortcut in their derivation by neglecting the proper asymptotic condition~\eqref{eq:psiasy} of the final state as needed, even in their one-electron theory, for a non-spherical (and thus multi-channel) system. In any case, in the asymptotic approximation, there is no additional channel coupling by means of elements of the \(S\)-matrix as proposed by Kamalov~\textit{et al.}.

The only relevant coupling for the asymptotic theory is the field-driven dipole coupling between residual ion states, which is normally disregarded in the hope that the IR absorption is not resonant with any transition within the residual ion. In Eq.~\eqref{eq:matelem} we also implicitly disregarded ion core transitions. However, the matrix element has generally both contributions: in the residual ion (electrons \(1, \dots,  N - 1\)) and in the photoelectron (electron \(N\)). Concretely,
\begin{align}
    \langle \Psi_{\bm{k}, i_f}^{(-)} | D(\bm{\epsilon}^{\text{IR}}) | \Psi_{i+\Omega}^{(+)} \rangle
    &= \bm{\epsilon}^{\text{IR}} \cdot \langle \Psi_{\bm{k}, i_f}^{(-)} | \sum_{i=1}^{N-1} \bm{r}_i  | \Psi_{i+\Omega}^{(+)} \rangle \nonumber \\
    &+ \bm{\epsilon}^{\text{IR}} \cdot \langle \Psi_{\bm{k}, i_f}^{(-)} | \bm{r}_{N}  | \Psi_{i+\Omega}^{(+)} \rangle \,.
    \label{eq:matelemsplit}
\end{align}
The first, ion core transition term of the amplitude can be evaluated analogously to the second, continuum-continuum transition given by Eq.~\eqref{eq:matelem}, yielding a very similar formula
\begin{align}
    d_{\text{ion}}^{(2)} &= \bm{\epsilon}^{\text{IR}} \cdot \langle \Psi_{\bm{k}, i_f}^{(-)} | \sum_{i=1}^{N-1} \bm{r}_i  | \Psi_{i+\Omega}^{(+)} \rangle
    \nonumber \\
    &\approx \sum_{i_n} A_{\kappa_n k}^{\text{ion}} \sum_{l_f m_f} b_{i_n i_f l_f m_f}^{\text{ion}} Y_{l_f}^{m_f}(\hat{\bm{k}}) \,,
    \label{eq:matelemdip}
\end{align}
where
\begin{align}
    b_{i_n i_f l_f m_f}^{\text{ion}} &= -\bm{\epsilon}^{\text{IR}}\cdot \langle \Phi_{i_f} | \sum_{i = 1}^{N-1} \bm{r}_i | \Phi_{i_n} \rangle
        \ \bm{\epsilon}^{\text{XUV}} \cdot \bm{d}_{i_n l_f m_f}^{(1)}
    \label{eq:bion} \\
    A_{\kappa k}^{\mathrm{ion}} &\approx \frac{\mathrm{e}^{-\pi/2\kappa + \pi/2k}}{\sqrt{k \kappa} (\kappa - k)}
    \frac{(2\kappa)^{\mathrm{i}/\kappa}}{(2k)^{\mathrm{i}/k}}
    \frac{\Gamma(1 + \mathrm{i}/\kappa - \mathrm{i}/k)}{(\kappa - k)^{\mathrm{i}(1/\kappa - 1/k)}}
    \label{eq:Akkion}
\end{align}
and the index \(i_n\) in Eq.~\eqref{eq:matelemdip} runs over all relevant intermediate ionic states that are dipole-coupled to the final ion state \(i_f\).

%In the special case when the IR frequency is close to the energy difference between a pair of residual ion states, it is enough to consider a single term of the sum as it dominates over the remaining ones due to the properties of \(A_{\kappa k}^{\text{ion}}\). If the ionization thresholds of the two states are similar, at high photoelectron energies it is not necessary to distinguish the intermediate momenta \(\kappa_n\) in \(A_{\kappa_n k}^\text{ion}\) the associated channels. Finally, note that when \(|k - \kappa| \ll \kappa k\), which satisfied when \(\omega \ll \Omega \rightarrow +\infty\), then \(A_{\kappa k}^{\text{ion}} \approx (\kappa - k) A_{\kappa k}\) due to properties of the Gamma function. Altogether, in this special case one can use a simpler approximate formula
%\begin{align}
%    \langle \Psi_{\bm{k}, i_f}^{(-)} | D(\bm{\epsilon}^{\text{IR}}) | \Psi_{i+\Omega}^{(+)} \rangle
%    \approx A_{\kappa k} \sum_{l_f m_f} b_{i_n i_f l_f m_f}^{\text{tot}} Y_{l_f}^{m_f}(\hat{\bm{k}}) \,,
%\end{align}
%with \(b_{i_f l_f m_f}^{\text{tot}} = b_{i_f l_f m_f} + (\kappa - k) b_{i_n i_f l_f m_f}^{\text{ion}}\). From the energy conservation we obtain \(|\kappa - k| \approx \omega/k\). So, for fixed IR frequency \(\omega\) and large photoelectron momenta \(k\), the second term is a small perturbation of the first one.

\end{appendix}

\bibliography{bibliography}

\begin{thebibliography}{66}%
\makeatletter
\providecommand \@ifxundefined [1]{%
 \@ifx{#1\undefined}
}%
\providecommand \@ifnum [1]{%
 \ifnum #1\expandafter \@firstoftwo
 \else \expandafter \@secondoftwo
 \fi
}%
\providecommand \@ifx [1]{%
 \ifx #1\expandafter \@firstoftwo
 \else \expandafter \@secondoftwo
 \fi
}%
\providecommand \natexlab [1]{#1}%
\providecommand \enquote  [1]{``#1''}%
\providecommand \bibnamefont  [1]{#1}%
\providecommand \bibfnamefont [1]{#1}%
\providecommand \citenamefont [1]{#1}%
\providecommand \href@noop [0]{\@secondoftwo}%
\providecommand \href [0]{\begingroup \@sanitize@url \@href}%
\providecommand \@href[1]{\@@startlink{#1}\@@href}%
\providecommand \@@href[1]{\endgroup#1\@@endlink}%
\providecommand \@sanitize@url [0]{\catcode `\\12\catcode `\$12\catcode
  `\&12\catcode `\#12\catcode `\^12\catcode `\_12\catcode `\%12\relax}%
\providecommand \@@startlink[1]{}%
\providecommand \@@endlink[0]{}%
\providecommand \url  [0]{\begingroup\@sanitize@url \@url }%
\providecommand \@url [1]{\endgroup\@href {#1}{\urlprefix }}%
\providecommand \urlprefix  [0]{URL }%
\providecommand \Eprint [0]{\href }%
\providecommand \doibase [0]{http://dx.doi.org/}%
\providecommand \selectlanguage [0]{\@gobble}%
\providecommand \bibinfo  [0]{\@secondoftwo}%
\providecommand \bibfield  [0]{\@secondoftwo}%
\providecommand \translation [1]{[#1]}%
\providecommand \BibitemOpen [0]{}%
\providecommand \bibitemStop [0]{}%
\providecommand \bibitemNoStop [0]{.\EOS\space}%
\providecommand \EOS [0]{\spacefactor3000\relax}%
\providecommand \BibitemShut  [1]{\csname bibitem#1\endcsname}%
\let\auto@bib@innerbib\@empty
%</preamble>
\bibitem [{\citenamefont {Paul}\ \emph {et~al.}(2001)\citenamefont {Paul},
  \citenamefont {Toma}, \citenamefont {Breger}, \citenamefont {Mullot},
  \citenamefont {Augé}, \citenamefont {Balcou}, \citenamefont {Muller},\ and\
  \citenamefont {Agostini}}]{Paul}%
  \BibitemOpen
  \bibfield  {author} {\bibinfo {author} {\bibfnamefont {P.~M.}\ \bibnamefont
  {Paul}}, \bibinfo {author} {\bibfnamefont {E.~S.}\ \bibnamefont {Toma}},
  \bibinfo {author} {\bibfnamefont {P.}~\bibnamefont {Breger}}, \bibinfo
  {author} {\bibfnamefont {G.}~\bibnamefont {Mullot}}, \bibinfo {author}
  {\bibfnamefont {F.}~\bibnamefont {Augé}}, \bibinfo {author} {\bibfnamefont
  {P.}~\bibnamefont {Balcou}}, \bibinfo {author} {\bibfnamefont {H.~G.}\
  \bibnamefont {Muller}}, \ and\ \bibinfo {author} {\bibfnamefont
  {P.}~\bibnamefont {Agostini}},\ }\href {\doibase 10.1126/science.1059413}
  {\bibfield  {journal} {\bibinfo  {journal} {Science}\ }\textbf {\bibinfo
  {volume} {292}},\ \bibinfo {pages} {1689} (\bibinfo {year}
  {2001})}\BibitemShut {NoStop}%
\bibitem [{\citenamefont {Muller}(2002)}]{RABITT}%
  \BibitemOpen
  \bibfield  {author} {\bibinfo {author} {\bibfnamefont {H.~G.}\ \bibnamefont
  {Muller}},\ }\href {\doibase 10.1007/s00340-002-0894-8} {\bibfield  {journal}
  {\bibinfo  {journal} {Appl. Phys. B}\ }\textbf {\bibinfo {volume} {74}},\
  \bibinfo {pages} {S17} (\bibinfo {year} {2002})}\BibitemShut {NoStop}%
\bibitem [{\citenamefont {Guénot}\ \emph {et~al.}(2014)\citenamefont {Guénot}
  \emph {et~al.}}]{ArHeNe}%
  \BibitemOpen
  \bibfield  {author} {\bibinfo {author} {\bibfnamefont {D.}~\bibnamefont
  {Guénot}} \emph {et~al.},\ }\href {\doibase 10.1088/0953-4075/47/24/245602}
  {\bibfield  {journal} {\bibinfo  {journal} {J. Phys. B: At. Mol. Opt. Phys.}\
  }\textbf {\bibinfo {volume} {47}},\ \bibinfo {pages} {245602} (\bibinfo
  {year} {2014})}\BibitemShut {NoStop}%
\bibitem [{\citenamefont {Cattaneo}\ \emph {et~al.}(2016)\citenamefont
  {Cattaneo}, \citenamefont {Vos}, \citenamefont {Lucchini}, \citenamefont
  {Gallmann}, \citenamefont {Cirelli},\ and\ \citenamefont {Keller}}]{Compar}%
  \BibitemOpen
  \bibfield  {author} {\bibinfo {author} {\bibfnamefont {L.}~\bibnamefont
  {Cattaneo}}, \bibinfo {author} {\bibfnamefont {J.}~\bibnamefont {Vos}},
  \bibinfo {author} {\bibfnamefont {M.}~\bibnamefont {Lucchini}}, \bibinfo
  {author} {\bibfnamefont {L.}~\bibnamefont {Gallmann}}, \bibinfo {author}
  {\bibfnamefont {C.}~\bibnamefont {Cirelli}}, \ and\ \bibinfo {author}
  {\bibfnamefont {U.}~\bibnamefont {Keller}},\ }\href {\doibase
  10.1364/OE.24.029066} {\bibfield  {journal} {\bibinfo  {journal} {Opt.
  Express}\ }\textbf {\bibinfo {volume} {24}},\ \bibinfo {pages} {29060}
  (\bibinfo {year} {2016})}\BibitemShut {NoStop}%
\bibitem [{\citenamefont {Alexandridi}\ \emph {et~al.}(2021)\citenamefont
  {Alexandridi}, \citenamefont {Platzer}, \citenamefont {Barreau},
  \citenamefont {Busto}, \citenamefont {Zhong}, \citenamefont {Turconi},
  \citenamefont {Neoričić}, \citenamefont {Laurell}, \citenamefont {Arnold},
  \citenamefont {Borot}, \citenamefont {Hergott}, \citenamefont {Tcherbakoff},
  \citenamefont {Lejman}, \citenamefont {Gisselbrecht}, \citenamefont
  {Lindroth}, \citenamefont {L’Huillier}, \citenamefont {Dahlström},\ and\
  \citenamefont {Salières}}]{ArgonCooper}%
  \BibitemOpen
  \bibfield  {author} {\bibinfo {author} {\bibfnamefont {C.}~\bibnamefont
  {Alexandridi}}, \bibinfo {author} {\bibfnamefont {D.}~\bibnamefont
  {Platzer}}, \bibinfo {author} {\bibfnamefont {L.}~\bibnamefont {Barreau}},
  \bibinfo {author} {\bibfnamefont {D.}~\bibnamefont {Busto}}, \bibinfo
  {author} {\bibfnamefont {S.}~\bibnamefont {Zhong}}, \bibinfo {author}
  {\bibfnamefont {M.}~\bibnamefont {Turconi}}, \bibinfo {author} {\bibfnamefont
  {L.}~\bibnamefont {Neoričić}}, \bibinfo {author} {\bibfnamefont
  {H.}~\bibnamefont {Laurell}}, \bibinfo {author} {\bibfnamefont {C.~L.}\
  \bibnamefont {Arnold}}, \bibinfo {author} {\bibfnamefont {A.}~\bibnamefont
  {Borot}}, \bibinfo {author} {\bibfnamefont {J.-F.}\ \bibnamefont {Hergott}},
  \bibinfo {author} {\bibfnamefont {O.}~\bibnamefont {Tcherbakoff}}, \bibinfo
  {author} {\bibfnamefont {M.}~\bibnamefont {Lejman}}, \bibinfo {author}
  {\bibfnamefont {M.}~\bibnamefont {Gisselbrecht}}, \bibinfo {author}
  {\bibfnamefont {E.}~\bibnamefont {Lindroth}}, \bibinfo {author}
  {\bibfnamefont {A.}~\bibnamefont {L’Huillier}}, \bibinfo {author}
  {\bibfnamefont {J.~M.}\ \bibnamefont {Dahlström}}, \ and\ \bibinfo {author}
  {\bibfnamefont {P.}~\bibnamefont {Salières}},\ }\href {\doibase
  10.1103/PhysRevResearch.3.L012012} {\bibfield  {journal} {\bibinfo  {journal}
  {Phys. Rev. Research}\ }\textbf {\bibinfo {volume} {3}},\ \bibinfo {pages}
  {L012012} (\bibinfo {year} {2021})}\BibitemShut {NoStop}%
\bibitem [{\citenamefont {Cattaneo}\ \emph {et~al.}(2018)\citenamefont
  {Cattaneo}, \citenamefont {Vos}, \citenamefont {Bello}, \citenamefont
  {Palacios}, \citenamefont {Heuser}, \citenamefont {Pedrelli}, \citenamefont
  {Lucchini}, \citenamefont {Cirelli}, \citenamefont {Martín},\ and\
  \citenamefont {Keller}}]{cattaneo2018}%
  \BibitemOpen
  \bibfield  {author} {\bibinfo {author} {\bibfnamefont {L.}~\bibnamefont
  {Cattaneo}}, \bibinfo {author} {\bibfnamefont {J.}~\bibnamefont {Vos}},
  \bibinfo {author} {\bibfnamefont {R.~Y.}\ \bibnamefont {Bello}}, \bibinfo
  {author} {\bibfnamefont {A.}~\bibnamefont {Palacios}}, \bibinfo {author}
  {\bibfnamefont {S.}~\bibnamefont {Heuser}}, \bibinfo {author} {\bibfnamefont
  {L.}~\bibnamefont {Pedrelli}}, \bibinfo {author} {\bibfnamefont
  {M.}~\bibnamefont {Lucchini}}, \bibinfo {author} {\bibfnamefont
  {C.}~\bibnamefont {Cirelli}}, \bibinfo {author} {\bibfnamefont
  {F.}~\bibnamefont {Martín}}, \ and\ \bibinfo {author} {\bibfnamefont
  {U.}~\bibnamefont {Keller}},\ }\href {\doibase 10.1038/s41567-018-0103-2}
  {\bibfield  {journal} {\bibinfo  {journal} {Nat. Phys.}\ }\textbf {\bibinfo
  {volume} {14}},\ \bibinfo {pages} {733} (\bibinfo {year} {2018})}\BibitemShut
  {NoStop}%
\bibitem [{\citenamefont {Huppert}\ \emph {et~al.}(2016)\citenamefont
  {Huppert}, \citenamefont {Jordan}, \citenamefont {Baykusheva}, \citenamefont
  {von Conta},\ and\ \citenamefont {Wörner}}]{Huppert}%
  \BibitemOpen
  \bibfield  {author} {\bibinfo {author} {\bibfnamefont {M.}~\bibnamefont
  {Huppert}}, \bibinfo {author} {\bibfnamefont {I.}~\bibnamefont {Jordan}},
  \bibinfo {author} {\bibfnamefont {D.}~\bibnamefont {Baykusheva}}, \bibinfo
  {author} {\bibfnamefont {A.}~\bibnamefont {von Conta}}, \ and\ \bibinfo
  {author} {\bibfnamefont {H.~J.}\ \bibnamefont {Wörner}},\ }\href {\doibase
  10.1103/PhysRevLett.117.093001} {\bibfield  {journal} {\bibinfo  {journal}
  {Phys. Rev. Lett.}\ }\textbf {\bibinfo {volume} {117}},\ \bibinfo {pages}
  {093001} (\bibinfo {year} {2016})}\BibitemShut {NoStop}%
\bibitem [{\citenamefont {Loriot}\ \emph {et~al.}(2020)\citenamefont {Loriot},
  \citenamefont {Marciniak}, \citenamefont {Nandi}, \citenamefont {Karras},
  \citenamefont {Hervé}, \citenamefont {Constant}, \citenamefont {Plésiat},
  \citenamefont {Palacios}, \citenamefont {Martín},\ and\ \citenamefont
  {Lépine}}]{LoriotN2}%
  \BibitemOpen
  \bibfield  {author} {\bibinfo {author} {\bibfnamefont {V.}~\bibnamefont
  {Loriot}}, \bibinfo {author} {\bibfnamefont {A.}~\bibnamefont {Marciniak}},
  \bibinfo {author} {\bibfnamefont {S.}~\bibnamefont {Nandi}}, \bibinfo
  {author} {\bibfnamefont {G.}~\bibnamefont {Karras}}, \bibinfo {author}
  {\bibfnamefont {M.}~\bibnamefont {Hervé}}, \bibinfo {author} {\bibfnamefont
  {E.}~\bibnamefont {Constant}}, \bibinfo {author} {\bibfnamefont
  {E.}~\bibnamefont {Plésiat}}, \bibinfo {author} {\bibfnamefont
  {A.}~\bibnamefont {Palacios}}, \bibinfo {author} {\bibfnamefont
  {F.}~\bibnamefont {Martín}}, \ and\ \bibinfo {author} {\bibfnamefont
  {F.}~\bibnamefont {Lépine}},\ }\href {\doibase 10.1088/2515-7647/ab7b10}
  {\bibfield  {journal} {\bibinfo  {journal} {J. Phys.: Photonics}\ }\textbf
  {\bibinfo {volume} {2}},\ \bibinfo {pages} {024003} (\bibinfo {year}
  {2020})}\BibitemShut {NoStop}%
\bibitem [{\citenamefont {Nandi}\ \emph {et~al.}(2020)\citenamefont {Nandi},
  \citenamefont {Plésiat}, \citenamefont {Zhong}, \citenamefont {Palacios},
  \citenamefont {Busto}, \citenamefont {Isinger}, \citenamefont {Neoričić},
  \citenamefont {Arnold}, \citenamefont {Squibb}, \citenamefont {Feifel},
  \citenamefont {Decleva}, \citenamefont {l'Huillier}, \citenamefont
  {Martín},\ and\ \citenamefont {Gisselbrecht}}]{Nandi}%
  \BibitemOpen
  \bibfield  {author} {\bibinfo {author} {\bibfnamefont {S.}~\bibnamefont
  {Nandi}}, \bibinfo {author} {\bibfnamefont {E.}~\bibnamefont {Plésiat}},
  \bibinfo {author} {\bibfnamefont {S.}~\bibnamefont {Zhong}}, \bibinfo
  {author} {\bibfnamefont {A.}~\bibnamefont {Palacios}}, \bibinfo {author}
  {\bibfnamefont {D.}~\bibnamefont {Busto}}, \bibinfo {author} {\bibfnamefont
  {M.}~\bibnamefont {Isinger}}, \bibinfo {author} {\bibfnamefont
  {L.}~\bibnamefont {Neoričić}}, \bibinfo {author} {\bibfnamefont {C.~L.}\
  \bibnamefont {Arnold}}, \bibinfo {author} {\bibfnamefont {R.~J.}\
  \bibnamefont {Squibb}}, \bibinfo {author} {\bibfnamefont {R.}~\bibnamefont
  {Feifel}}, \bibinfo {author} {\bibfnamefont {P.}~\bibnamefont {Decleva}},
  \bibinfo {author} {\bibfnamefont {A.}~\bibnamefont {l'Huillier}}, \bibinfo
  {author} {\bibfnamefont {F.}~\bibnamefont {Martín}}, \ and\ \bibinfo
  {author} {\bibfnamefont {M.}~\bibnamefont {Gisselbrecht}},\ }\href {\doibase
  10.1126/sciadv.aba7762} {\bibfield  {journal} {\bibinfo  {journal} {Sci.
  Adv.}\ }\textbf {\bibinfo {volume} {6}},\ \bibinfo {pages} {eaba7762}
  (\bibinfo {year} {2020})}\BibitemShut {NoStop}%
\bibitem [{\citenamefont {Vos}\ \emph {et~al.}(2018)\citenamefont {Vos},
  \citenamefont {Cattaneo}, \citenamefont {Patchkovskii}, \citenamefont
  {Zimmermann}, \citenamefont {Cirelli}, \citenamefont {Lucchini},
  \citenamefont {Kheifets}, \citenamefont {Landsman},\ and\ \citenamefont
  {Keller}}]{vos_2018}%
  \BibitemOpen
  \bibfield  {author} {\bibinfo {author} {\bibfnamefont {J.}~\bibnamefont
  {Vos}}, \bibinfo {author} {\bibfnamefont {L.}~\bibnamefont {Cattaneo}},
  \bibinfo {author} {\bibfnamefont {S.}~\bibnamefont {Patchkovskii}}, \bibinfo
  {author} {\bibfnamefont {T.}~\bibnamefont {Zimmermann}}, \bibinfo {author}
  {\bibfnamefont {C.}~\bibnamefont {Cirelli}}, \bibinfo {author} {\bibfnamefont
  {M.}~\bibnamefont {Lucchini}}, \bibinfo {author} {\bibfnamefont
  {A.}~\bibnamefont {Kheifets}}, \bibinfo {author} {\bibfnamefont {A.~S.}\
  \bibnamefont {Landsman}}, \ and\ \bibinfo {author} {\bibfnamefont
  {U.}~\bibnamefont {Keller}},\ }\href {\doibase 10.1126/science.aao4731}
  {\bibfield  {journal} {\bibinfo  {journal} {Science}\ }\textbf {\bibinfo
  {volume} {360}},\ \bibinfo {pages} {1326} (\bibinfo {year}
  {2018})}\BibitemShut {NoStop}%
\bibitem [{\citenamefont {Kamalov}\ \emph {et~al.}(2020)\citenamefont
  {Kamalov}, \citenamefont {Wang}, \citenamefont {Bucksbaum}, \citenamefont
  {Haxton},\ and\ \citenamefont {Cryan}}]{KamalovCO2}%
  \BibitemOpen
  \bibfield  {author} {\bibinfo {author} {\bibfnamefont {A.}~\bibnamefont
  {Kamalov}}, \bibinfo {author} {\bibfnamefont {A.~L.}\ \bibnamefont {Wang}},
  \bibinfo {author} {\bibfnamefont {P.~H.}\ \bibnamefont {Bucksbaum}}, \bibinfo
  {author} {\bibfnamefont {D.~J.}\ \bibnamefont {Haxton}}, \ and\ \bibinfo
  {author} {\bibfnamefont {J.~P.}\ \bibnamefont {Cryan}},\ }\href {\doibase
  10.1103/PhysRevA.102.023118} {\bibfield  {journal} {\bibinfo  {journal}
  {Phys. Rev. A}\ }\textbf {\bibinfo {volume} {102}},\ \bibinfo {pages}
  {023118} (\bibinfo {year} {2020})}\BibitemShut {NoStop}%
\bibitem [{\citenamefont {Dahlström}\ \emph {et~al.}(2013)\citenamefont
  {Dahlström}, \citenamefont {Guénot}, \citenamefont {Klünder},
  \citenamefont {Gisselbrecht}, \citenamefont {Mauritsson}, \citenamefont
  {L'Huillier}, \citenamefont {Maquet},\ and\ \citenamefont
  {Taïeb}}]{Dahlstrom}%
  \BibitemOpen
  \bibfield  {author} {\bibinfo {author} {\bibfnamefont {J.~M.}\ \bibnamefont
  {Dahlström}}, \bibinfo {author} {\bibfnamefont {D.}~\bibnamefont {Guénot}},
  \bibinfo {author} {\bibfnamefont {K.}~\bibnamefont {Klünder}}, \bibinfo
  {author} {\bibfnamefont {M.}~\bibnamefont {Gisselbrecht}}, \bibinfo {author}
  {\bibfnamefont {J.}~\bibnamefont {Mauritsson}}, \bibinfo {author}
  {\bibfnamefont {A.}~\bibnamefont {L'Huillier}}, \bibinfo {author}
  {\bibfnamefont {A.}~\bibnamefont {Maquet}}, \ and\ \bibinfo {author}
  {\bibfnamefont {R.}~\bibnamefont {Taïeb}},\ }\href {\doibase
  10.1016/j.chemphys.2012.01.017} {\bibfield  {journal} {\bibinfo  {journal}
  {Chem. Phys.}\ }\textbf {\bibinfo {volume} {414}},\ \bibinfo {pages} {53}
  (\bibinfo {year} {2013})}\BibitemShut {NoStop}%
\bibitem [{\citenamefont {Dahlström}\ \emph {et~al.}(2012)\citenamefont
  {Dahlström}, \citenamefont {L'Huillier},\ and\ \citenamefont
  {Maquet}}]{Dahlstrom_JPB}%
  \BibitemOpen
  \bibfield  {author} {\bibinfo {author} {\bibfnamefont {J.~M.}\ \bibnamefont
  {Dahlström}}, \bibinfo {author} {\bibfnamefont {A.}~\bibnamefont
  {L'Huillier}}, \ and\ \bibinfo {author} {\bibfnamefont {A.}~\bibnamefont
  {Maquet}},\ }\href {\doibase 10.1088/0953-4075/45/18/183001} {\bibfield
  {journal} {\bibinfo  {journal} {J. Phys. B: At., Mol. Opt. Phys.}\ }\textbf
  {\bibinfo {volume} {45}},\ \bibinfo {pages} {183001} (\bibinfo {year}
  {2012})}\BibitemShut {NoStop}%
\bibitem [{\citenamefont {Serov}\ and\ \citenamefont
  {Kheifets}(2017)}]{SerovKheifets}%
  \BibitemOpen
  \bibfield  {author} {\bibinfo {author} {\bibfnamefont {V.~V.}\ \bibnamefont
  {Serov}}\ and\ \bibinfo {author} {\bibfnamefont {A.~S.}\ \bibnamefont
  {Kheifets}},\ }\href {\doibase 10.1063/1.4993493} {\bibfield  {journal}
  {\bibinfo  {journal} {J. Chem. Phys.}\ }\textbf {\bibinfo {volume} {147}},\
  \bibinfo {pages} {204303} (\bibinfo {year} {2017})}\BibitemShut {NoStop}%
\bibitem [{\citenamefont {Baykusheva}\ and\ \citenamefont
  {Wörner}(2017)}]{BaykushevaWorner}%
  \BibitemOpen
  \bibfield  {author} {\bibinfo {author} {\bibfnamefont {D.}~\bibnamefont
  {Baykusheva}}\ and\ \bibinfo {author} {\bibfnamefont {H.~J.}\ \bibnamefont
  {Wörner}},\ }\href {\doibase 10.1063/1.4977933} {\bibfield  {journal}
  {\bibinfo  {journal} {J. Chem. Phys.}\ }\textbf {\bibinfo {volume} {146}},\
  \bibinfo {pages} {124306} (\bibinfo {year} {2017})}\BibitemShut {NoStop}%
\bibitem [{\citenamefont {Jordan}\ \emph {et~al.}(2020)\citenamefont {Jordan},
  \citenamefont {Huppert}, \citenamefont {Rattenbacher}, \citenamefont {Peper},
  \citenamefont {Jelovina}, \citenamefont {Perry}, \citenamefont {von Conta},
  \citenamefont {Schild},\ and\ \citenamefont {Wörner}}]{LiquidDelays}%
  \BibitemOpen
  \bibfield  {author} {\bibinfo {author} {\bibfnamefont {I.}~\bibnamefont
  {Jordan}}, \bibinfo {author} {\bibfnamefont {M.}~\bibnamefont {Huppert}},
  \bibinfo {author} {\bibfnamefont {D.}~\bibnamefont {Rattenbacher}}, \bibinfo
  {author} {\bibfnamefont {M.}~\bibnamefont {Peper}}, \bibinfo {author}
  {\bibfnamefont {D.}~\bibnamefont {Jelovina}}, \bibinfo {author}
  {\bibfnamefont {C.}~\bibnamefont {Perry}}, \bibinfo {author} {\bibfnamefont
  {A.}~\bibnamefont {von Conta}}, \bibinfo {author} {\bibfnamefont
  {A.}~\bibnamefont {Schild}}, \ and\ \bibinfo {author} {\bibfnamefont {H.~J.}\
  \bibnamefont {Wörner}},\ }\href {\doibase 10.1126/science.abb0979}
  {\bibfield  {journal} {\bibinfo  {journal} {Science}\ }\textbf {\bibinfo
  {volume} {369}},\ \bibinfo {pages} {974} (\bibinfo {year}
  {2020})}\BibitemShut {NoStop}%
\bibitem [{\citenamefont {Gallmann}\ \emph {et~al.}(2017)\citenamefont
  {Gallmann}, \citenamefont {Jordan}, \citenamefont {Wörner}, \citenamefont
  {Castiglioni}, \citenamefont {Hengsberger}, \citenamefont {Osterwalder},
  \citenamefont {Arrell}, \citenamefont {Chergui}, \citenamefont {Liberatore},
  \citenamefont {Rothlisberger},\ and\ \citenamefont {Keller}}]{Gallmann}%
  \BibitemOpen
  \bibfield  {author} {\bibinfo {author} {\bibfnamefont {L.}~\bibnamefont
  {Gallmann}}, \bibinfo {author} {\bibfnamefont {I.}~\bibnamefont {Jordan}},
  \bibinfo {author} {\bibfnamefont {H.~J.}\ \bibnamefont {Wörner}}, \bibinfo
  {author} {\bibfnamefont {L.}~\bibnamefont {Castiglioni}}, \bibinfo {author}
  {\bibfnamefont {M.}~\bibnamefont {Hengsberger}}, \bibinfo {author}
  {\bibfnamefont {J.}~\bibnamefont {Osterwalder}}, \bibinfo {author}
  {\bibfnamefont {C.~A.}\ \bibnamefont {Arrell}}, \bibinfo {author}
  {\bibfnamefont {M.}~\bibnamefont {Chergui}}, \bibinfo {author} {\bibfnamefont
  {E.}~\bibnamefont {Liberatore}}, \bibinfo {author} {\bibfnamefont
  {U.}~\bibnamefont {Rothlisberger}}, \ and\ \bibinfo {author} {\bibfnamefont
  {U.}~\bibnamefont {Keller}},\ }\href {\doibase 10.1063/1.4997175} {\bibfield
  {journal} {\bibinfo  {journal} {Struct. Dyn.}\ }\textbf {\bibinfo {volume}
  {4}},\ \bibinfo {pages} {061502} (\bibinfo {year} {2017})}\BibitemShut
  {NoStop}%
\bibitem [{\citenamefont {Eckart}(2020)}]{HASE}%
  \BibitemOpen
  \bibfield  {author} {\bibinfo {author} {\bibfnamefont {S.}~\bibnamefont
  {Eckart}},\ }\href {\doibase 10.1103/PhysRevResearch.2.033248} {\bibfield
  {journal} {\bibinfo  {journal} {Phys. Rev. Research}\ }\textbf {\bibinfo
  {volume} {2}},\ \bibinfo {pages} {033248} (\bibinfo {year}
  {2020})}\BibitemShut {NoStop}%
\bibitem [{\citenamefont {Trabert}\ \emph {et~al.}(2021)\citenamefont
  {Trabert}, \citenamefont {Brennecke}, \citenamefont {Fehre}, \citenamefont
  {Anders}, \citenamefont {Geyer}, \citenamefont {Grundmann}, \citenamefont
  {Schöffler}, \citenamefont {Schmidt}, \citenamefont {Jahnke}, \citenamefont
  {Dörner}, \citenamefont {Kunitski},\ and\ \citenamefont {Eckart}}]{Eckart}%
  \BibitemOpen
  \bibfield  {author} {\bibinfo {author} {\bibfnamefont {D.}~\bibnamefont
  {Trabert}}, \bibinfo {author} {\bibfnamefont {S.}~\bibnamefont {Brennecke}},
  \bibinfo {author} {\bibfnamefont {K.}~\bibnamefont {Fehre}}, \bibinfo
  {author} {\bibfnamefont {N.}~\bibnamefont {Anders}}, \bibinfo {author}
  {\bibfnamefont {A.}~\bibnamefont {Geyer}}, \bibinfo {author} {\bibfnamefont
  {S.}~\bibnamefont {Grundmann}}, \bibinfo {author} {\bibfnamefont {M.~S.}\
  \bibnamefont {Schöffler}}, \bibinfo {author} {\bibfnamefont {L.~P.~H.}\
  \bibnamefont {Schmidt}}, \bibinfo {author} {\bibfnamefont {T.}~\bibnamefont
  {Jahnke}}, \bibinfo {author} {\bibfnamefont {R.}~\bibnamefont {Dörner}},
  \bibinfo {author} {\bibfnamefont {M.}~\bibnamefont {Kunitski}}, \ and\
  \bibinfo {author} {\bibfnamefont {S.}~\bibnamefont {Eckart}},\ }\href
  {\doibase 10.1038/s41467-021-21845-6} {\bibfield  {journal} {\bibinfo
  {journal} {Nat. Commun.}\ }\textbf {\bibinfo {volume} {12}},\ \bibinfo
  {pages} {1697} (\bibinfo {year} {2021})}\BibitemShut {NoStop}%
\bibitem [{\citenamefont {Biswas}\ \emph {et~al.}(2020)\citenamefont {Biswas},
  \citenamefont {Förg}, \citenamefont {Ortmann}, \citenamefont {Schötz},
  \citenamefont {Schweinberger}, \citenamefont {Zimmermann}, \citenamefont
  {Pi}, \citenamefont {Baykusheva}, \citenamefont {Masood}, \citenamefont
  {Liontos}, \citenamefont {Kamal}, \citenamefont {Kling}, \citenamefont
  {Alharbi}, \citenamefont {Alharbi}, \citenamefont {Azzeer}, \citenamefont
  {Hartmann}, \citenamefont {Wörner}, \citenamefont {Landsmann},\ and\
  \citenamefont {Kling}}]{Biswas}%
  \BibitemOpen
  \bibfield  {author} {\bibinfo {author} {\bibfnamefont {S.}~\bibnamefont
  {Biswas}}, \bibinfo {author} {\bibfnamefont {B.}~\bibnamefont {Förg}},
  \bibinfo {author} {\bibfnamefont {L.}~\bibnamefont {Ortmann}}, \bibinfo
  {author} {\bibfnamefont {J.}~\bibnamefont {Schötz}}, \bibinfo {author}
  {\bibfnamefont {W.}~\bibnamefont {Schweinberger}}, \bibinfo {author}
  {\bibfnamefont {T.}~\bibnamefont {Zimmermann}}, \bibinfo {author}
  {\bibfnamefont {L.}~\bibnamefont {Pi}}, \bibinfo {author} {\bibfnamefont
  {D.}~\bibnamefont {Baykusheva}}, \bibinfo {author} {\bibfnamefont
  {A.}~\bibnamefont {Masood}}, \bibinfo {author} {\bibfnamefont
  {I.}~\bibnamefont {Liontos}}, \bibinfo {author} {\bibfnamefont {A.~M.}\
  \bibnamefont {Kamal}}, \bibinfo {author} {\bibfnamefont {N.~G.}\ \bibnamefont
  {Kling}}, \bibinfo {author} {\bibfnamefont {A.~F.}\ \bibnamefont {Alharbi}},
  \bibinfo {author} {\bibfnamefont {M.}~\bibnamefont {Alharbi}}, \bibinfo
  {author} {\bibfnamefont {A.~M.}\ \bibnamefont {Azzeer}}, \bibinfo {author}
  {\bibfnamefont {G.}~\bibnamefont {Hartmann}}, \bibinfo {author}
  {\bibfnamefont {H.~J.}\ \bibnamefont {Wörner}}, \bibinfo {author}
  {\bibfnamefont {A.~S.}\ \bibnamefont {Landsmann}}, \ and\ \bibinfo {author}
  {\bibfnamefont {M.~F.}\ \bibnamefont {Kling}},\ }\href {\doibase
  10.1038/s41567-020-0887-8} {\bibfield  {journal} {\bibinfo  {journal} {Nat.
  Phys.}\ }\textbf {\bibinfo {volume} {16}},\ \bibinfo {pages} {778} (\bibinfo
  {year} {2020})}\BibitemShut {NoStop}%
\bibitem [{\citenamefont {Joseph}\ \emph {et~al.}(2020)\citenamefont {Joseph},
  \citenamefont {Holzmeier}, \citenamefont {Bresteau}, \citenamefont
  {Spezzani}, \citenamefont {Ruchon}, \citenamefont {Hergott}, \citenamefont
  {Tcherbakoff}, \citenamefont {D'Oliveira}, \citenamefont {Houver},\ and\
  \citenamefont {Dowek}}]{joseph2020}%
  \BibitemOpen
  \bibfield  {author} {\bibinfo {author} {\bibfnamefont {J.}~\bibnamefont
  {Joseph}}, \bibinfo {author} {\bibfnamefont {F.}~\bibnamefont {Holzmeier}},
  \bibinfo {author} {\bibfnamefont {D.}~\bibnamefont {Bresteau}}, \bibinfo
  {author} {\bibfnamefont {C.}~\bibnamefont {Spezzani}}, \bibinfo {author}
  {\bibfnamefont {T.}~\bibnamefont {Ruchon}}, \bibinfo {author} {\bibfnamefont
  {J.~F.}\ \bibnamefont {Hergott}}, \bibinfo {author} {\bibfnamefont
  {O.}~\bibnamefont {Tcherbakoff}}, \bibinfo {author} {\bibfnamefont
  {P.}~\bibnamefont {D'Oliveira}}, \bibinfo {author} {\bibfnamefont {J.~C.}\
  \bibnamefont {Houver}}, \ and\ \bibinfo {author} {\bibfnamefont
  {D.}~\bibnamefont {Dowek}},\ }\href {\doibase 10.1088/1361-6455/ab9f0d}
  {\bibfield  {journal} {\bibinfo  {journal} {J. Phys. B: At. Mol. Opt. Phys.}\
  }\textbf {\bibinfo {volume} {53}},\ \bibinfo {pages} {184007} (\bibinfo
  {year} {2020})}\BibitemShut {NoStop}%
\bibitem [{\citenamefont {Fuchs}\ \emph {et~al.}(2020)\citenamefont {Fuchs},
  \citenamefont {Douguet}, \citenamefont {Donsa}, \citenamefont {Martin},
  \citenamefont {Burgdörfer}, \citenamefont {Argenti}, \citenamefont
  {Cattaneo},\ and\ \citenamefont {Keller}}]{fuchs2020}%
  \BibitemOpen
  \bibfield  {author} {\bibinfo {author} {\bibfnamefont {J.}~\bibnamefont
  {Fuchs}}, \bibinfo {author} {\bibfnamefont {N.}~\bibnamefont {Douguet}},
  \bibinfo {author} {\bibfnamefont {S.}~\bibnamefont {Donsa}}, \bibinfo
  {author} {\bibfnamefont {F.}~\bibnamefont {Martin}}, \bibinfo {author}
  {\bibfnamefont {J.}~\bibnamefont {Burgdörfer}}, \bibinfo {author}
  {\bibfnamefont {L.}~\bibnamefont {Argenti}}, \bibinfo {author} {\bibfnamefont
  {L.}~\bibnamefont {Cattaneo}}, \ and\ \bibinfo {author} {\bibfnamefont
  {U.}~\bibnamefont {Keller}},\ }\href {\doibase 10.1364/OPTICA.378639}
  {\bibfield  {journal} {\bibinfo  {journal} {Optica}\ }\textbf {\bibinfo
  {volume} {7}},\ \bibinfo {pages} {154} (\bibinfo {year} {2020})}\BibitemShut
  {NoStop}%
\bibitem [{\citenamefont {Loriot}\ \emph {et~al.}(2017)\citenamefont {Loriot},
  \citenamefont {Marciniak}, \citenamefont {Karras}, \citenamefont {Schindler},
  \citenamefont {Renois-Predelus}, \citenamefont {Compagnon}, \citenamefont
  {Concina}, \citenamefont {Brédy}, \citenamefont {Celep}, \citenamefont
  {Bordas}, \citenamefont {Constant},\ and\ \citenamefont
  {Lépine}}]{Loriot2017}%
  \BibitemOpen
  \bibfield  {author} {\bibinfo {author} {\bibfnamefont {V.}~\bibnamefont
  {Loriot}}, \bibinfo {author} {\bibfnamefont {A.}~\bibnamefont {Marciniak}},
  \bibinfo {author} {\bibfnamefont {G.}~\bibnamefont {Karras}}, \bibinfo
  {author} {\bibfnamefont {B.}~\bibnamefont {Schindler}}, \bibinfo {author}
  {\bibfnamefont {G.}~\bibnamefont {Renois-Predelus}}, \bibinfo {author}
  {\bibfnamefont {I.}~\bibnamefont {Compagnon}}, \bibinfo {author}
  {\bibfnamefont {B.}~\bibnamefont {Concina}}, \bibinfo {author} {\bibfnamefont
  {R.}~\bibnamefont {Brédy}}, \bibinfo {author} {\bibfnamefont
  {G.}~\bibnamefont {Celep}}, \bibinfo {author} {\bibfnamefont
  {C.}~\bibnamefont {Bordas}}, \bibinfo {author} {\bibfnamefont
  {E.}~\bibnamefont {Constant}}, \ and\ \bibinfo {author} {\bibfnamefont
  {F.}~\bibnamefont {Lépine}},\ }\href {\doibase 10.1088/2040-8986/aa8e10}
  {\bibfield  {journal} {\bibinfo  {journal} {J. Opt.}\ }\textbf {\bibinfo
  {volume} {19}},\ \bibinfo {pages} {114003} (\bibinfo {year}
  {2017})}\BibitemShut {NoStop}%
\bibitem [{\citenamefont {Bharti}\ \emph {et~al.}(2021)\citenamefont {Bharti},
  \citenamefont {Atri-Schuller}, \citenamefont {Menning}, \citenamefont
  {Hamilton}, \citenamefont {Moshammer}, \citenamefont {Pfeifer}, \citenamefont
  {Doguet}, \citenamefont {Bartschat},\ and\ \citenamefont
  {Harth}}]{Multisideband}%
  \BibitemOpen
  \bibfield  {author} {\bibinfo {author} {\bibfnamefont {D.}~\bibnamefont
  {Bharti}}, \bibinfo {author} {\bibfnamefont {D.}~\bibnamefont
  {Atri-Schuller}}, \bibinfo {author} {\bibfnamefont {G.}~\bibnamefont
  {Menning}}, \bibinfo {author} {\bibfnamefont {K.~R.}\ \bibnamefont
  {Hamilton}}, \bibinfo {author} {\bibfnamefont {R.}~\bibnamefont {Moshammer}},
  \bibinfo {author} {\bibfnamefont {T.}~\bibnamefont {Pfeifer}}, \bibinfo
  {author} {\bibfnamefont {N.}~\bibnamefont {Doguet}}, \bibinfo {author}
  {\bibfnamefont {K.}~\bibnamefont {Bartschat}}, \ and\ \bibinfo {author}
  {\bibfnamefont {A.}~\bibnamefont {Harth}},\ }\href {\doibase
  10.1103/PhysRevA.103.022834} {\bibfield  {journal} {\bibinfo  {journal}
  {Phys. Rev. A}\ }\textbf {\bibinfo {volume} {103}},\ \bibinfo {pages}
  {022834} (\bibinfo {year} {2021})}\BibitemShut {NoStop}%
\bibitem [{\citenamefont {Ivanov}\ and\ \citenamefont
  {Smirnova}(2011)}]{IvanovSmirnova}%
  \BibitemOpen
  \bibfield  {author} {\bibinfo {author} {\bibfnamefont {M.}~\bibnamefont
  {Ivanov}}\ and\ \bibinfo {author} {\bibfnamefont {O.}~\bibnamefont
  {Smirnova}},\ }\href {\doibase 10.1103/PhysRevLett.107.213605} {\bibfield
  {journal} {\bibinfo  {journal} {Phys. Rev. Lett.}\ }\textbf {\bibinfo
  {volume} {107}},\ \bibinfo {pages} {213605} (\bibinfo {year}
  {2011})}\BibitemShut {NoStop}%
\bibitem [{\citenamefont {Pazourek}\ \emph {et~al.}(2015)\citenamefont
  {Pazourek}, \citenamefont {Nagele},\ and\ \citenamefont
  {Burgdörfer}}]{Pazourek}%
  \BibitemOpen
  \bibfield  {author} {\bibinfo {author} {\bibfnamefont {R.}~\bibnamefont
  {Pazourek}}, \bibinfo {author} {\bibfnamefont {S.}~\bibnamefont {Nagele}}, \
  and\ \bibinfo {author} {\bibfnamefont {J.}~\bibnamefont {Burgdörfer}},\
  }\href {\doibase 10.1103/RevModPhys.87.765} {\bibfield  {journal} {\bibinfo
  {journal} {Rev. Mod. Phys.}\ }\textbf {\bibinfo {volume} {87}},\ \bibinfo
  {pages} {765} (\bibinfo {year} {2015})}\BibitemShut {NoStop}%
\bibitem [{\citenamefont {Haessler}\ \emph {et~al.}(2009)\citenamefont
  {Haessler}, \citenamefont {Fabre}, \citenamefont {Higuet}, \citenamefont
  {Caillat}, \citenamefont {Ruchon}, \citenamefont {Breger}, \citenamefont
  {Carré}, \citenamefont {Constant}, \citenamefont {Maquet}, \citenamefont
  {Mével}, \citenamefont {Salières}, \citenamefont {Taïeb},\ and\
  \citenamefont {Mairesse}}]{Haessler}%
  \BibitemOpen
  \bibfield  {author} {\bibinfo {author} {\bibfnamefont {S.}~\bibnamefont
  {Haessler}}, \bibinfo {author} {\bibfnamefont {B.}~\bibnamefont {Fabre}},
  \bibinfo {author} {\bibfnamefont {J.}~\bibnamefont {Higuet}}, \bibinfo
  {author} {\bibfnamefont {J.}~\bibnamefont {Caillat}}, \bibinfo {author}
  {\bibfnamefont {T.}~\bibnamefont {Ruchon}}, \bibinfo {author} {\bibfnamefont
  {P.}~\bibnamefont {Breger}}, \bibinfo {author} {\bibfnamefont
  {B.}~\bibnamefont {Carré}}, \bibinfo {author} {\bibfnamefont
  {E.}~\bibnamefont {Constant}}, \bibinfo {author} {\bibfnamefont
  {A.}~\bibnamefont {Maquet}}, \bibinfo {author} {\bibfnamefont
  {E.}~\bibnamefont {Mével}}, \bibinfo {author} {\bibfnamefont
  {P.}~\bibnamefont {Salières}}, \bibinfo {author} {\bibfnamefont
  {R.}~\bibnamefont {Taïeb}}, \ and\ \bibinfo {author} {\bibfnamefont
  {Y.}~\bibnamefont {Mairesse}},\ }\href {\doibase 10.1103/PhysRevA.80.011404}
  {\bibfield  {journal} {\bibinfo  {journal} {Phys. Rev. A}\ }\textbf {\bibinfo
  {volume} {80}},\ \bibinfo {pages} {011404(R)} (\bibinfo {year}
  {2009})}\BibitemShut {NoStop}%
\bibitem [{\citenamefont {Chacón}\ and\ \citenamefont
  {Ruiz}(2018)}]{ChaconCO}%
  \BibitemOpen
  \bibfield  {author} {\bibinfo {author} {\bibfnamefont {A.}~\bibnamefont
  {Chacón}}\ and\ \bibinfo {author} {\bibfnamefont {C.}~\bibnamefont {Ruiz}},\
  }\href {\doibase 10.1364/OE.26.004548} {\bibfield  {journal} {\bibinfo
  {journal} {Opt. Express}\ }\textbf {\bibinfo {volume} {26}},\ \bibinfo
  {pages} {4548} (\bibinfo {year} {2018})}\BibitemShut {NoStop}%
\bibitem [{\citenamefont {Kheifets}\ and\ \citenamefont
  {Bray}(2021)}]{KheifetsBray}%
  \BibitemOpen
  \bibfield  {author} {\bibinfo {author} {\bibfnamefont {A.~S.}\ \bibnamefont
  {Kheifets}}\ and\ \bibinfo {author} {\bibfnamefont {A.~W.}\ \bibnamefont
  {Bray}},\ }\href {\doibase 10.1103/PhysRevA.103.L011101} {\bibfield
  {journal} {\bibinfo  {journal} {Phys. Rev. A}\ }\textbf {\bibinfo {volume}
  {103}},\ \bibinfo {pages} {L011101} (\bibinfo {year} {2021})}\BibitemShut
  {NoStop}%
\bibitem [{\citenamefont {Hockett}\ \emph {et~al.}(2016)\citenamefont
  {Hockett}, \citenamefont {Frumker}, \citenamefont {Villeneuve},\ and\
  \citenamefont {Corkum}}]{Hockett}%
  \BibitemOpen
  \bibfield  {author} {\bibinfo {author} {\bibfnamefont {P.}~\bibnamefont
  {Hockett}}, \bibinfo {author} {\bibfnamefont {E.}~\bibnamefont {Frumker}},
  \bibinfo {author} {\bibfnamefont {D.~M.}\ \bibnamefont {Villeneuve}}, \ and\
  \bibinfo {author} {\bibfnamefont {P.~B.}\ \bibnamefont {Corkum}},\ }\href
  {\doibase 10.1088/0953-4075/49/9/095602} {\bibfield  {journal} {\bibinfo
  {journal} {J. Phys. B: At. Mol. Opt. Phys.}\ }\textbf {\bibinfo {volume}
  {49}},\ \bibinfo {pages} {095602} (\bibinfo {year} {2016})}\BibitemShut
  {NoStop}%
\bibitem [{\citenamefont {Feng}\ and\ \citenamefont {van~der
  Hart}(2003)}]{Feng}%
  \BibitemOpen
  \bibfield  {author} {\bibinfo {author} {\bibfnamefont {L.}~\bibnamefont
  {Feng}}\ and\ \bibinfo {author} {\bibfnamefont {H.~W.}\ \bibnamefont {van~der
  Hart}},\ }\href {\doibase 10.1088/0953-4075/36/1/101} {\bibfield  {journal}
  {\bibinfo  {journal} {J. Phys. B: At. Mol. Opt. Phys.}\ }\textbf {\bibinfo
  {volume} {36}},\ \bibinfo {pages} {L1} (\bibinfo {year} {2003})}\BibitemShut
  {NoStop}%
\bibitem [{\citenamefont {Shakeshaft}(2007)}]{Shakeshaft}%
  \BibitemOpen
  \bibfield  {author} {\bibinfo {author} {\bibfnamefont {R.}~\bibnamefont
  {Shakeshaft}},\ }\href {\doibase 10.1103/PhysRevA.76.063405} {\bibfield
  {journal} {\bibinfo  {journal} {Phys. Rev. A}\ }\textbf {\bibinfo {volume}
  {76}},\ \bibinfo {pages} {063405} (\bibinfo {year} {2007})}\BibitemShut
  {NoStop}%
\bibitem [{\citenamefont {Vinbladh}\ \emph {et~al.}(2019)\citenamefont
  {Vinbladh}, \citenamefont {Dahlström},\ and\ \citenamefont
  {Lindroth}}]{vinbladh2019}%
  \BibitemOpen
  \bibfield  {author} {\bibinfo {author} {\bibfnamefont {J.}~\bibnamefont
  {Vinbladh}}, \bibinfo {author} {\bibfnamefont {J.~M.}\ \bibnamefont
  {Dahlström}}, \ and\ \bibinfo {author} {\bibfnamefont {E.}~\bibnamefont
  {Lindroth}},\ }\href {\doibase 10.1103/PhysRevA.100.043424} {\bibfield
  {journal} {\bibinfo  {journal} {Phys. Rev. A}\ }\textbf {\bibinfo {volume}
  {100}},\ \bibinfo {pages} {043424} (\bibinfo {year} {2019})}\BibitemShut
  {NoStop}%
\bibitem [{\citenamefont {Benda}\ and\ \citenamefont
  {Ma\v{s}\'{i}n}(2021)}]{multiphoton}%
  \BibitemOpen
  \bibfield  {author} {\bibinfo {author} {\bibfnamefont {J.}~\bibnamefont
  {Benda}}\ and\ \bibinfo {author} {\bibfnamefont {Z.}~\bibnamefont
  {Ma\v{s}\'{i}n}},\ }\href {\doibase 10.1038/s41598-021-89733-z} {\bibfield
  {journal} {\bibinfo  {journal} {Sci. Rep.}\ }\textbf {\bibinfo {volume}
  {11}},\ \bibinfo {pages} {11686} (\bibinfo {year} {2021})}\BibitemShut
  {NoStop}%
\bibitem [{\citenamefont {Brown}\ \emph {et~al.}(2020)\citenamefont {Brown},
  \citenamefont {Armstrong}, \citenamefont {Benda}, \citenamefont {Clarke},
  \citenamefont {Wragg}, \citenamefont {Hamilton}, \citenamefont {Mašín},
  \citenamefont {Gorfinkiel},\ and\ \citenamefont {van~der Hart}}]{RMT}%
  \BibitemOpen
  \bibfield  {author} {\bibinfo {author} {\bibfnamefont {A.~C.}\ \bibnamefont
  {Brown}}, \bibinfo {author} {\bibfnamefont {G.~S.~J.}\ \bibnamefont
  {Armstrong}}, \bibinfo {author} {\bibfnamefont {J.}~\bibnamefont {Benda}},
  \bibinfo {author} {\bibfnamefont {D.~D.~A.}\ \bibnamefont {Clarke}}, \bibinfo
  {author} {\bibfnamefont {J.}~\bibnamefont {Wragg}}, \bibinfo {author}
  {\bibfnamefont {K.~R.}\ \bibnamefont {Hamilton}}, \bibinfo {author}
  {\bibfnamefont {Z.}~\bibnamefont {Mašín}}, \bibinfo {author} {\bibfnamefont
  {J.}~\bibnamefont {Gorfinkiel}}, \ and\ \bibinfo {author} {\bibfnamefont
  {H.~W.}\ \bibnamefont {van~der Hart}},\ }\href {\doibase
  10.1016/j.cpc.2019.107062} {\bibfield  {journal} {\bibinfo  {journal}
  {Comput. Phys. Commun.}\ }\textbf {\bibinfo {volume} {250}},\ \bibinfo
  {pages} {107062} (\bibinfo {year} {2020})}\BibitemShut {NoStop}%
\bibitem [{\citenamefont {Benda}\ \emph {et~al.}(2020)\citenamefont {Benda},
  \citenamefont {Gorfinkiel}, \citenamefont {Mašín}, \citenamefont
  {Armstrong}, \citenamefont {Brown}, \citenamefont {Clarke}, \citenamefont
  {van~der Hart},\ and\ \citenamefont {Wragg}}]{MRMT}%
  \BibitemOpen
  \bibfield  {author} {\bibinfo {author} {\bibfnamefont {J.}~\bibnamefont
  {Benda}}, \bibinfo {author} {\bibfnamefont {J.~D.}\ \bibnamefont
  {Gorfinkiel}}, \bibinfo {author} {\bibfnamefont {Z.}~\bibnamefont {Mašín}},
  \bibinfo {author} {\bibfnamefont {G.~S.~J.}\ \bibnamefont {Armstrong}},
  \bibinfo {author} {\bibfnamefont {A.~C.}\ \bibnamefont {Brown}}, \bibinfo
  {author} {\bibfnamefont {D.~D.~A.}\ \bibnamefont {Clarke}}, \bibinfo {author}
  {\bibfnamefont {H.~W.}\ \bibnamefont {van~der Hart}}, \ and\ \bibinfo
  {author} {\bibfnamefont {J.}~\bibnamefont {Wragg}},\ }\href {\doibase
  10.1103/PhysRevA.102.052826} {\bibfield  {journal} {\bibinfo  {journal}
  {Phys. Rev. A}\ }\textbf {\bibinfo {volume} {102}},\ \bibinfo {pages}
  {052826} (\bibinfo {year} {2020})}\BibitemShut {NoStop}%
\bibitem [{\citenamefont {Cohen-Tannoudji}\ \emph {et~al.}(1998)\citenamefont
  {Cohen-Tannoudji}, \citenamefont {Dupont-Roc},\ and\ \citenamefont
  {Grynberg}}]{cohen-tannoudji1998}%
  \BibitemOpen
  \bibfield  {author} {\bibinfo {author} {\bibfnamefont {C.}~\bibnamefont
  {Cohen-Tannoudji}}, \bibinfo {author} {\bibfnamefont {J.}~\bibnamefont
  {Dupont-Roc}}, \ and\ \bibinfo {author} {\bibfnamefont {G.}~\bibnamefont
  {Grynberg}},\ }\href {\doibase 10.1002/9783527617197.fmatter} {\emph
  {\bibinfo {title} {Atom—Photon Interactions}}}\ (\bibinfo  {publisher}
  {John Wiley \& Sons, Ltd, New York},\ \bibinfo {year} {1998})\BibitemShut
  {NoStop}%
\bibitem [{\citenamefont {Harvey}\ \emph {et~al.}(2014)\citenamefont {Harvey},
  \citenamefont {Brambila}, \citenamefont {Morales},\ and\ \citenamefont
  {Smirnova}}]{Harvey}%
  \BibitemOpen
  \bibfield  {author} {\bibinfo {author} {\bibfnamefont {A.}~\bibnamefont
  {Harvey}}, \bibinfo {author} {\bibfnamefont {D.~S.}\ \bibnamefont
  {Brambila}}, \bibinfo {author} {\bibfnamefont {F.}~\bibnamefont {Morales}}, \
  and\ \bibinfo {author} {\bibfnamefont {O.}~\bibnamefont {Smirnova}},\ }\href
  {\doibase 10.1088/0953-4075/47/21/215005} {\bibfield  {journal} {\bibinfo
  {journal} {J. Phys. B: At. Mol. Opt. Phys.}\ }\textbf {\bibinfo {volume}
  {47}},\ \bibinfo {pages} {215005} (\bibinfo {year} {2014})}\BibitemShut
  {NoStop}%
\bibitem [{\citenamefont {Aymar}\ and\ \citenamefont
  {Crance}(1980)}]{AymarCrance}%
  \BibitemOpen
  \bibfield  {author} {\bibinfo {author} {\bibfnamefont {M.}~\bibnamefont
  {Aymar}}\ and\ \bibinfo {author} {\bibfnamefont {M.}~\bibnamefont {Crance}},\
  }\href {\doibase 10.1088/0022-3700/13/9/002} {\bibfield  {journal} {\bibinfo
  {journal} {J. Phys. B: At. Mol. Phys.}\ }\textbf {\bibinfo {volume} {13}},\
  \bibinfo {pages} {L287} (\bibinfo {year} {1980})}\BibitemShut {NoStop}%
\bibitem [{\citenamefont {Levin}(1996)}]{Levin}%
  \BibitemOpen
  \bibfield  {author} {\bibinfo {author} {\bibfnamefont {D.}~\bibnamefont
  {Levin}},\ }\href {\doibase 10.1016/0377-0427(94)00118-9} {\bibfield
  {journal} {\bibinfo  {journal} {J. Comput. Appl. Math.}\ }\textbf {\bibinfo
  {volume} {67}},\ \bibinfo {pages} {95} (\bibinfo {year} {1996})}\BibitemShut
  {NoStop}%
\bibitem [{\citenamefont {Powell}(1947)}]{Powell}%
  \BibitemOpen
  \bibfield  {author} {\bibinfo {author} {\bibfnamefont {J.~L.}\ \bibnamefont
  {Powell}},\ }\href {\doibase 10.1103/PhysRev.72.626} {\bibfield  {journal}
  {\bibinfo  {journal} {Phys. Rev.}\ }\textbf {\bibinfo {volume} {72}},\
  \bibinfo {pages} {626} (\bibinfo {year} {1947})}\BibitemShut {NoStop}%
\bibitem [{\citenamefont {Zamastil}\ and\ \citenamefont
  {Benda}(2017)}]{Zamastil}%
  \BibitemOpen
  \bibfield  {author} {\bibinfo {author} {\bibfnamefont {J.}~\bibnamefont
  {Zamastil}}\ and\ \bibinfo {author} {\bibfnamefont {J.}~\bibnamefont
  {Benda}},\ }\href {\doibase 10.1007/978-3-319-65780-6} {\emph {\bibinfo
  {title} {{Quantum Mechanics and Electrodynamics}}}}\ (\bibinfo  {publisher}
  {Springer, Cham},\ \bibinfo {year} {2017})\BibitemShut {NoStop}%
\bibitem [{\citenamefont {Chen}\ \emph {et~al.}(1974)\citenamefont {Chen},
  \citenamefont {Smith},\ and\ \citenamefont {Simons}}]{N2IP}%
  \BibitemOpen
  \bibfield  {author} {\bibinfo {author} {\bibfnamefont {T.-T.}\ \bibnamefont
  {Chen}}, \bibinfo {author} {\bibfnamefont {W.~D.}\ \bibnamefont {Smith}}, \
  and\ \bibinfo {author} {\bibfnamefont {J.}~\bibnamefont {Simons}},\ }\href
  {\doibase 10.1016/0009-2614(74)85419-9} {\bibfield  {journal} {\bibinfo
  {journal} {Chem. Phys. Lett.}\ }\textbf {\bibinfo {volume} {26}},\ \bibinfo
  {pages} {296} (\bibinfo {year} {1974})}\BibitemShut {NoStop}%
\bibitem [{\citenamefont {Kelkensberg}\ \emph {et~al.}(2011)\citenamefont
  {Kelkensberg}, \citenamefont {Rouzeé}, \citenamefont {Siu}, \citenamefont
  {Gademann}, \citenamefont {Jonsson}, \citenamefont {Lucchini}, \citenamefont
  {Lucchese},\ and\ \citenamefont {Vrakking}}]{CO2IP}%
  \BibitemOpen
  \bibfield  {author} {\bibinfo {author} {\bibfnamefont {F.}~\bibnamefont
  {Kelkensberg}}, \bibinfo {author} {\bibfnamefont {A.}~\bibnamefont
  {Rouzeé}}, \bibinfo {author} {\bibfnamefont {W.}~\bibnamefont {Siu}},
  \bibinfo {author} {\bibfnamefont {G.}~\bibnamefont {Gademann}}, \bibinfo
  {author} {\bibfnamefont {P.}~\bibnamefont {Jonsson}}, \bibinfo {author}
  {\bibfnamefont {M.}~\bibnamefont {Lucchini}}, \bibinfo {author}
  {\bibfnamefont {R.~R.}\ \bibnamefont {Lucchese}}, \ and\ \bibinfo {author}
  {\bibfnamefont {M.~J.~J.}\ \bibnamefont {Vrakking}},\ }\href {\doibase
  10.1103/PhysRevA.84.051404} {\bibfield  {journal} {\bibinfo  {journal} {Phys.
  Rev. A}\ }\textbf {\bibinfo {volume} {84}},\ \bibinfo {pages} {051404(R)}
  (\bibinfo {year} {2011})}\BibitemShut {NoStop}%
\bibitem [{\citenamefont {Potts}\ and\ \citenamefont {Price}(1972)}]{H2OIP}%
  \BibitemOpen
  \bibfield  {author} {\bibinfo {author} {\bibfnamefont {A.~W.}\ \bibnamefont
  {Potts}}\ and\ \bibinfo {author} {\bibfnamefont {W.~C.}\ \bibnamefont
  {Price}},\ }\href {\doibase 10.1098/rspa.1972.0004} {\bibfield  {journal}
  {\bibinfo  {journal} {Proc. R. Soc. London Ser. A}\ }\textbf {\bibinfo
  {volume} {326}},\ \bibinfo {pages} {181} (\bibinfo {year}
  {1972})}\BibitemShut {NoStop}%
\bibitem [{\citenamefont {Truesdale}\ \emph {et~al.}(1983)\citenamefont
  {Truesdale}, \citenamefont {Southworth}, \citenamefont {Kobrin},
  \citenamefont {Lindle},\ and\ \citenamefont {Shirley}}]{Truesdale}%
  \BibitemOpen
  \bibfield  {author} {\bibinfo {author} {\bibfnamefont {C.~M.}\ \bibnamefont
  {Truesdale}}, \bibinfo {author} {\bibfnamefont {S.}~\bibnamefont
  {Southworth}}, \bibinfo {author} {\bibfnamefont {P.~H.}\ \bibnamefont
  {Kobrin}}, \bibinfo {author} {\bibfnamefont {D.~W.}\ \bibnamefont {Lindle}},
  \ and\ \bibinfo {author} {\bibfnamefont {D.~A.}\ \bibnamefont {Shirley}},\
  }\href {\doibase 10.1063/1.444742} {\bibfield  {journal} {\bibinfo  {journal}
  {J. Chem. Phys.}\ }\textbf {\bibinfo {volume} {78}},\ \bibinfo {pages} {7117}
  (\bibinfo {year} {1983})}\BibitemShut {NoStop}%
\bibitem [{\citenamefont {Smith}\ \emph {et~al.}(2020)\citenamefont {Smith},
  \citenamefont {Burns}, \citenamefont {Simmonett}, \citenamefont {Parrish},
  \citenamefont {Schieber}, \citenamefont {Galvelis}, \citenamefont {Kraus},
  \citenamefont {Kruse}, \citenamefont {{Di Remigio}}, \citenamefont
  {Alenaizan}, \citenamefont {James}, \citenamefont {Lehtola}, \citenamefont
  {Misiewicz}, \citenamefont {Scheurer}, \citenamefont {Shaw}, \citenamefont
  {Schriber}, \citenamefont {Xie}, \citenamefont {Glick}, \citenamefont
  {Sirianni}, \citenamefont {O’Brien}, \citenamefont {Waldrop}, \citenamefont
  {Kumar}, \citenamefont {Hohenstein}, \citenamefont {Pritchard}, \citenamefont
  {Brooks}, \citenamefont {Schaefer}, \citenamefont {Sokolov}, \citenamefont
  {Patkowski}, \citenamefont {DePrince}, \citenamefont {Bozkaya}, \citenamefont
  {King}, \citenamefont {Evangelista}, \citenamefont {Turney}, \citenamefont
  {Crawford},\ and\ \citenamefont {Sherrill}}]{PSI4}%
  \BibitemOpen
  \bibfield  {author} {\bibinfo {author} {\bibfnamefont {D.~G.~A.}\
  \bibnamefont {Smith}}, \bibinfo {author} {\bibfnamefont {L.~A.}\ \bibnamefont
  {Burns}}, \bibinfo {author} {\bibfnamefont {A.~C.}\ \bibnamefont
  {Simmonett}}, \bibinfo {author} {\bibfnamefont {R.~M.}\ \bibnamefont
  {Parrish}}, \bibinfo {author} {\bibfnamefont {M.~C.}\ \bibnamefont
  {Schieber}}, \bibinfo {author} {\bibfnamefont {R.}~\bibnamefont {Galvelis}},
  \bibinfo {author} {\bibfnamefont {P.}~\bibnamefont {Kraus}}, \bibinfo
  {author} {\bibfnamefont {H.}~\bibnamefont {Kruse}}, \bibinfo {author}
  {\bibfnamefont {R.}~\bibnamefont {{Di Remigio}}}, \bibinfo {author}
  {\bibfnamefont {A.}~\bibnamefont {Alenaizan}}, \bibinfo {author}
  {\bibfnamefont {A.~M.}\ \bibnamefont {James}}, \bibinfo {author}
  {\bibfnamefont {S.}~\bibnamefont {Lehtola}}, \bibinfo {author} {\bibfnamefont
  {J.~P.}\ \bibnamefont {Misiewicz}}, \bibinfo {author} {\bibfnamefont
  {M.}~\bibnamefont {Scheurer}}, \bibinfo {author} {\bibfnamefont {R.~A.}\
  \bibnamefont {Shaw}}, \bibinfo {author} {\bibfnamefont {J.~B.}\ \bibnamefont
  {Schriber}}, \bibinfo {author} {\bibfnamefont {Y.}~\bibnamefont {Xie}},
  \bibinfo {author} {\bibfnamefont {Z.~L.}\ \bibnamefont {Glick}}, \bibinfo
  {author} {\bibfnamefont {D.~A.}\ \bibnamefont {Sirianni}}, \bibinfo {author}
  {\bibfnamefont {J.~S.}\ \bibnamefont {O’Brien}}, \bibinfo {author}
  {\bibfnamefont {J.~M.}\ \bibnamefont {Waldrop}}, \bibinfo {author}
  {\bibfnamefont {A.}~\bibnamefont {Kumar}}, \bibinfo {author} {\bibfnamefont
  {E.~G.}\ \bibnamefont {Hohenstein}}, \bibinfo {author} {\bibfnamefont
  {B.~P.}\ \bibnamefont {Pritchard}}, \bibinfo {author} {\bibfnamefont {B.~R.}\
  \bibnamefont {Brooks}}, \bibinfo {author} {\bibfnamefont {H.~F.}\
  \bibnamefont {Schaefer}}, \bibinfo {author} {\bibfnamefont {A.~Y.}\
  \bibnamefont {Sokolov}}, \bibinfo {author} {\bibfnamefont {K.}~\bibnamefont
  {Patkowski}}, \bibinfo {author} {\bibfnamefont {A.~E.}\ \bibnamefont
  {DePrince}}, \bibinfo {author} {\bibfnamefont {U.}~\bibnamefont {Bozkaya}},
  \bibinfo {author} {\bibfnamefont {R.~A.}\ \bibnamefont {King}}, \bibinfo
  {author} {\bibfnamefont {F.~A.}\ \bibnamefont {Evangelista}}, \bibinfo
  {author} {\bibfnamefont {J.~M.}\ \bibnamefont {Turney}}, \bibinfo {author}
  {\bibfnamefont {T.~D.}\ \bibnamefont {Crawford}}, \ and\ \bibinfo {author}
  {\bibfnamefont {C.~D.}\ \bibnamefont {Sherrill}},\ }\href {\doibase
  10.1063/5.0006002} {\bibfield  {journal} {\bibinfo  {journal} {J. Chem.
  Phys.}\ }\textbf {\bibinfo {volume} {152}},\ \bibinfo {pages} {184108}
  (\bibinfo {year} {2020})}\BibitemShut {NoStop}%
\bibitem [{\citenamefont {Mašín}\ \emph {et~al.}(2020)\citenamefont
  {Mašín}, \citenamefont {Benda}, \citenamefont {Gorfinkiel}, \citenamefont
  {Harvey},\ and\ \citenamefont {Tennyson}}]{UKRmolp}%
  \BibitemOpen
  \bibfield  {author} {\bibinfo {author} {\bibfnamefont {Z.}~\bibnamefont
  {Mašín}}, \bibinfo {author} {\bibfnamefont {J.}~\bibnamefont {Benda}},
  \bibinfo {author} {\bibfnamefont {J.~D.}\ \bibnamefont {Gorfinkiel}},
  \bibinfo {author} {\bibfnamefont {A.~G.}\ \bibnamefont {Harvey}}, \ and\
  \bibinfo {author} {\bibfnamefont {J.}~\bibnamefont {Tennyson}},\ }\href
  {\doibase 10.1016/j.cpc.2019.107092} {\bibfield  {journal} {\bibinfo
  {journal} {Comput. Phys. Commun.}\ }\textbf {\bibinfo {volume} {249}},\
  \bibinfo {pages} {107092} (\bibinfo {year} {2020})}\BibitemShut {NoStop}%
\bibitem [{\citenamefont {Little}\ and\ \citenamefont
  {Tennyson}(2013)}]{DuncanN2a}%
  \BibitemOpen
  \bibfield  {author} {\bibinfo {author} {\bibfnamefont {D.~A.}\ \bibnamefont
  {Little}}\ and\ \bibinfo {author} {\bibfnamefont {J.}~\bibnamefont
  {Tennyson}},\ }\href {\doibase 10.1088/0953-4075/46/14/145102} {\bibfield
  {journal} {\bibinfo  {journal} {J. Phys. B: At. Mol. Opt. Phys.}\ }\textbf
  {\bibinfo {volume} {46}},\ \bibinfo {pages} {145102} (\bibinfo {year}
  {2013})}\BibitemShut {NoStop}%
\bibitem [{\citenamefont {Little}\ and\ \citenamefont
  {Tennyson}(2014)}]{DuncanN2b}%
  \BibitemOpen
  \bibfield  {author} {\bibinfo {author} {\bibfnamefont {D.~A.}\ \bibnamefont
  {Little}}\ and\ \bibinfo {author} {\bibfnamefont {J.}~\bibnamefont
  {Tennyson}},\ }\href {\doibase 10.1088/0953-4075/47/10/105204} {\bibfield
  {journal} {\bibinfo  {journal} {J. Phys. B: At. Mol. Opt. Phys.}\ }\textbf
  {\bibinfo {volume} {47}},\ \bibinfo {pages} {105204} (\bibinfo {year}
  {2014})}\BibitemShut {NoStop}%
\bibitem [{\citenamefont {Werner}\ \emph {et~al.}(2012)\citenamefont {Werner},
  \citenamefont {Knowles}, \citenamefont {Knizia}, \citenamefont {Manby},\ and\
  \citenamefont {Sch{\"u}tz}}]{Molpro}%
  \BibitemOpen
  \bibfield  {author} {\bibinfo {author} {\bibfnamefont {H.-J.}\ \bibnamefont
  {Werner}}, \bibinfo {author} {\bibfnamefont {P.~J.}\ \bibnamefont {Knowles}},
  \bibinfo {author} {\bibfnamefont {G.}~\bibnamefont {Knizia}}, \bibinfo
  {author} {\bibfnamefont {F.~R.}\ \bibnamefont {Manby}}, \ and\ \bibinfo
  {author} {\bibfnamefont {M.}~\bibnamefont {Sch{\"u}tz}},\ }\href {\doibase
  10.1002/wcms.82} {\bibfield  {journal} {\bibinfo  {journal} {WIREs Comput.
  Mol. Sci.}\ }\textbf {\bibinfo {volume} {2}},\ \bibinfo {pages} {242}
  (\bibinfo {year} {2012})}\BibitemShut {NoStop}%
\bibitem [{\citenamefont {Plummer}\ \emph {et~al.}(1977)\citenamefont
  {Plummer}, \citenamefont {Gustafsson}, \citenamefont {Gudat},\ and\
  \citenamefont {Eastman}}]{PlummerN2}%
  \BibitemOpen
  \bibfield  {author} {\bibinfo {author} {\bibfnamefont {E.~W.}\ \bibnamefont
  {Plummer}}, \bibinfo {author} {\bibfnamefont {T.}~\bibnamefont {Gustafsson}},
  \bibinfo {author} {\bibfnamefont {W.}~\bibnamefont {Gudat}}, \ and\ \bibinfo
  {author} {\bibfnamefont {D.~E.}\ \bibnamefont {Eastman}},\ }\href {\doibase
  10.1103/PhysRevA.15.2339} {\bibfield  {journal} {\bibinfo  {journal} {Phys.
  Rev. A}\ }\textbf {\bibinfo {volume} {15}},\ \bibinfo {pages} {2339}
  (\bibinfo {year} {1977})}\BibitemShut {NoStop}%
\bibitem [{\citenamefont {Samson}\ \emph {et~al.}(1977)\citenamefont {Samson},
  \citenamefont {Haddad},\ and\ \citenamefont {Gardner}}]{SamsonN2}%
  \BibitemOpen
  \bibfield  {author} {\bibinfo {author} {\bibfnamefont {J.~A.~R.}\
  \bibnamefont {Samson}}, \bibinfo {author} {\bibfnamefont {G.~N.}\
  \bibnamefont {Haddad}}, \ and\ \bibinfo {author} {\bibfnamefont {J.~L.}\
  \bibnamefont {Gardner}},\ }\href {\doibase 10.1088/0022-3700/10/9/024}
  {\bibfield  {journal} {\bibinfo  {journal} {J. Phys. B: At. Mol. Opt. Phys.}\
  }\textbf {\bibinfo {volume} {10}},\ \bibinfo {pages} {1749} (\bibinfo {year}
  {1977})}\BibitemShut {NoStop}%
\bibitem [{\citenamefont {Woodruff}\ and\ \citenamefont
  {Marr}(1977)}]{WoodruffMarrN2}%
  \BibitemOpen
  \bibfield  {author} {\bibinfo {author} {\bibfnamefont {P.~A.}\ \bibnamefont
  {Woodruff}}\ and\ \bibinfo {author} {\bibfnamefont {G.~V.}\ \bibnamefont
  {Marr}},\ }\href {\doibase 10.1098/rspa.1977.0188} {\bibfield  {journal}
  {\bibinfo  {journal} {Proc. R. Soc. Lond. A}\ }\textbf {\bibinfo {volume}
  {358}},\ \bibinfo {pages} {87} (\bibinfo {year} {1977})}\BibitemShut
  {NoStop}%
\bibitem [{\citenamefont {Marr}\ \emph {et~al.}(1979)\citenamefont {Marr},
  \citenamefont {Morton}, \citenamefont {Holmes},\ and\ \citenamefont
  {McCoy}}]{MarrN2}%
  \BibitemOpen
  \bibfield  {author} {\bibinfo {author} {\bibfnamefont {G.~V.}\ \bibnamefont
  {Marr}}, \bibinfo {author} {\bibfnamefont {J.~M.}\ \bibnamefont {Morton}},
  \bibinfo {author} {\bibfnamefont {R.~M.}\ \bibnamefont {Holmes}}, \ and\
  \bibinfo {author} {\bibfnamefont {D.~G.}\ \bibnamefont {McCoy}},\ }\href
  {\doibase 10.1088/0022-3700/12/1/013} {\bibfield  {journal} {\bibinfo
  {journal} {J. Phys. B: At. Mol. Opt. Phys.}\ }\textbf {\bibinfo {volume}
  {12}},\ \bibinfo {pages} {43} (\bibinfo {year} {1979})}\BibitemShut {NoStop}%
\bibitem [{\citenamefont {Brion}\ and\ \citenamefont {Tan}(1979)}]{BrionTan}%
  \BibitemOpen
  \bibfield  {author} {\bibinfo {author} {\bibfnamefont {C.~E.}\ \bibnamefont
  {Brion}}\ and\ \bibinfo {author} {\bibfnamefont {K.~H.}\ \bibnamefont
  {Tan}},\ }\href {\doibase 10.1016/0368-2048(79)87039-5} {\bibfield  {journal}
  {\bibinfo  {journal} {J. Electron Spectrosc. Relat. Phenomena}\ }\textbf
  {\bibinfo {volume} {15}},\ \bibinfo {pages} {241} (\bibinfo {year}
  {1979})}\BibitemShut {NoStop}%
\bibitem [{\citenamefont {Carlson}\ \emph {et~al.}(1983)\citenamefont
  {Carlson}, \citenamefont {Keller}, \citenamefont {Taylor}, \citenamefont
  {Whitley},\ and\ \citenamefont {Grimm}}]{Carlson}%
  \BibitemOpen
  \bibfield  {author} {\bibinfo {author} {\bibfnamefont {T.~A.}\ \bibnamefont
  {Carlson}}, \bibinfo {author} {\bibfnamefont {P.~R.}\ \bibnamefont {Keller}},
  \bibinfo {author} {\bibfnamefont {J.~W.}\ \bibnamefont {Taylor}}, \bibinfo
  {author} {\bibfnamefont {T.}~\bibnamefont {Whitley}}, \ and\ \bibinfo
  {author} {\bibfnamefont {F.~A.}\ \bibnamefont {Grimm}},\ }\href {\doibase
  10.1063/1.445519} {\bibfield  {journal} {\bibinfo  {journal} {J. Chem.
  Phys.}\ }\textbf {\bibinfo {volume} {79}},\ \bibinfo {pages} {97} (\bibinfo
  {year} {1983})}\BibitemShut {NoStop}%
\bibitem [{\citenamefont {Braunstein}\ and\ \citenamefont
  {McKoy}(1987)}]{BraunsteinMcKoy}%
  \BibitemOpen
  \bibfield  {author} {\bibinfo {author} {\bibfnamefont {M.}~\bibnamefont
  {Braunstein}}\ and\ \bibinfo {author} {\bibfnamefont {V.}~\bibnamefont
  {McKoy}},\ }\href {\doibase 10.1063/1.453620} {\bibfield  {journal} {\bibinfo
   {journal} {J. Chem. Phys.}\ }\textbf {\bibinfo {volume} {87}},\ \bibinfo
  {pages} {224} (\bibinfo {year} {1987})}\BibitemShut {NoStop}%
\bibitem [{\citenamefont {Mašín}\ \emph {et~al.}(2018)\citenamefont
  {Mašín}, \citenamefont {Harvey}, \citenamefont {Spanner}, \citenamefont
  {Patchkovskii}, \citenamefont {Ivanov},\ and\ \citenamefont
  {Smirnova}}]{CO2HHG}%
  \BibitemOpen
  \bibfield  {author} {\bibinfo {author} {\bibfnamefont {Z.}~\bibnamefont
  {Mašín}}, \bibinfo {author} {\bibfnamefont {A.~G.}\ \bibnamefont {Harvey}},
  \bibinfo {author} {\bibfnamefont {M.}~\bibnamefont {Spanner}}, \bibinfo
  {author} {\bibfnamefont {S.}~\bibnamefont {Patchkovskii}}, \bibinfo {author}
  {\bibfnamefont {M.}~\bibnamefont {Ivanov}}, \ and\ \bibinfo {author}
  {\bibfnamefont {O.}~\bibnamefont {Smirnova}},\ }\href {\doibase
  10.1088/1361-6455/aac598} {\bibfield  {journal} {\bibinfo  {journal} {J.
  Phys. B: At. Mol. Opt. Phys.}\ }\textbf {\bibinfo {volume} {51}},\ \bibinfo
  {pages} {134006} (\bibinfo {year} {2018})}\BibitemShut {NoStop}%
\bibitem [{\citenamefont {Kheifets}(2021)}]{ResonantRABITT}%
  \BibitemOpen
  \bibfield  {author} {\bibinfo {author} {\bibfnamefont {A.}~\bibnamefont
  {Kheifets}},\ }\href {\doibase 10.1103/PhysRevA.104.L021103} {\bibfield
  {journal} {\bibinfo  {journal} {Phys. Rev. A}\ }\textbf {\bibinfo {volume}
  {104}},\ \bibinfo {pages} {L021103} (\bibinfo {year} {2021})}\BibitemShut
  {NoStop}%
\bibitem [{\citenamefont {Lein}\ \emph {et~al.}(2002)\citenamefont {Lein},
  \citenamefont {Hay}, \citenamefont {Velotta}, \citenamefont {Marangos},\ and\
  \citenamefont {Knight}}]{Lein}%
  \BibitemOpen
  \bibfield  {author} {\bibinfo {author} {\bibfnamefont {M.}~\bibnamefont
  {Lein}}, \bibinfo {author} {\bibfnamefont {N.}~\bibnamefont {Hay}}, \bibinfo
  {author} {\bibfnamefont {R.}~\bibnamefont {Velotta}}, \bibinfo {author}
  {\bibfnamefont {J.~P.}\ \bibnamefont {Marangos}}, \ and\ \bibinfo {author}
  {\bibfnamefont {P.~L.}\ \bibnamefont {Knight}},\ }\href {\doibase
  10.1103/PhysRevA.66.023805} {\bibfield  {journal} {\bibinfo  {journal} {Phys.
  Rev. A}\ }\textbf {\bibinfo {volume} {66}},\ \bibinfo {pages} {023805}
  (\bibinfo {year} {2002})}\BibitemShut {NoStop}%
\bibitem [{\citenamefont {Fernández}\ \emph {et~al.}(2009)\citenamefont
  {Fernández}, \citenamefont {Fojón},\ and\ \citenamefont
  {Martín}}]{Fernandez}%
  \BibitemOpen
  \bibfield  {author} {\bibinfo {author} {\bibfnamefont {J.}~\bibnamefont
  {Fernández}}, \bibinfo {author} {\bibfnamefont {O.}~\bibnamefont {Fojón}},
  \ and\ \bibinfo {author} {\bibfnamefont {F.}~\bibnamefont {Martín}},\ }\href
  {\doibase 10.1103/PhysRevA.79.023420} {\bibfield  {journal} {\bibinfo
  {journal} {Phys. Rev. A}\ }\textbf {\bibinfo {volume} {79}},\ \bibinfo
  {pages} {023420} (\bibinfo {year} {2009})}\BibitemShut {NoStop}%
\bibitem [{\citenamefont {Ning}\ \emph {et~al.}(2014)\citenamefont {Ning},
  \citenamefont {Peng}, \citenamefont {Song}, \citenamefont {Jiang},
  \citenamefont {Nagele}, \citenamefont {Pazourek}, \citenamefont
  {Burgdörfer},\ and\ \citenamefont {Gong}}]{NingH2p}%
  \BibitemOpen
  \bibfield  {author} {\bibinfo {author} {\bibfnamefont {Q.-C.}\ \bibnamefont
  {Ning}}, \bibinfo {author} {\bibfnamefont {L.-Y.}\ \bibnamefont {Peng}},
  \bibinfo {author} {\bibfnamefont {S.~N.}\ \bibnamefont {Song}}, \bibinfo
  {author} {\bibfnamefont {W.-C.}\ \bibnamefont {Jiang}}, \bibinfo {author}
  {\bibfnamefont {S.}~\bibnamefont {Nagele}}, \bibinfo {author} {\bibfnamefont
  {R.}~\bibnamefont {Pazourek}}, \bibinfo {author} {\bibfnamefont
  {J.}~\bibnamefont {Burgdörfer}}, \ and\ \bibinfo {author} {\bibfnamefont
  {Q.}~\bibnamefont {Gong}},\ }\href {\doibase 10.1103/PhysRevA.90.013423}
  {\bibfield  {journal} {\bibinfo  {journal} {Phys. Rev. A}\ }\textbf {\bibinfo
  {volume} {90}},\ \bibinfo {pages} {013423} (\bibinfo {year}
  {2014})}\BibitemShut {NoStop}%
\bibitem [{\citenamefont {Cohen}\ and\ \citenamefont {Fano}(1966)}]{CohenFano}%
  \BibitemOpen
  \bibfield  {author} {\bibinfo {author} {\bibfnamefont {H.~D.}\ \bibnamefont
  {Cohen}}\ and\ \bibinfo {author} {\bibfnamefont {U.}~\bibnamefont {Fano}},\
  }\href {\doibase 10.1103/PhysRev.150.30} {\bibfield  {journal} {\bibinfo
  {journal} {Phys. Rev.}\ }\textbf {\bibinfo {volume} {150}},\ \bibinfo {pages}
  {30} (\bibinfo {year} {1966})}\BibitemShut {NoStop}%
\bibitem [{\citenamefont {Olver}(2007)}]{olver_2007}%
  \BibitemOpen
  \bibfield  {author} {\bibinfo {author} {\bibfnamefont {S.}~\bibnamefont
  {Olver}},\ }\href {\doibase 10.1017/S0956792507007012} {\bibfield  {journal}
  {\bibinfo  {journal} {Eur. J. Appl. Math.}\ }\textbf {\bibinfo {volume}
  {18}},\ \bibinfo {pages} {435–447} (\bibinfo {year} {2007})}\BibitemShut
  {NoStop}%
\bibitem [{\citenamefont {Burke}(2011)}]{Burke}%
  \BibitemOpen
  \bibfield  {author} {\bibinfo {author} {\bibfnamefont {P.~G.}\ \bibnamefont
  {Burke}},\ }\href {\doibase 10.1007/978-3-642-15931-2} {\emph {\bibinfo
  {title} {{R-Matrix Theory of Atomic Collisions}}}}\ (\bibinfo  {publisher}
  {Springer-Verlag Berlin Heidelberg},\ \bibinfo {year} {2011})\BibitemShut
  {NoStop}%
\end{thebibliography}%

\end{document}